\documentclass[aps, pra, reprint, superscriptaddress]{revtex4-1}
\usepackage{header}

\begin{document}
\bibliographystyle{apsrev4-1}
\title{Quantum Linear System Solvers: A Survey of Algorithms and Applications}
\author{Mauro E. S. Morales}
\email{mauroms@umd.edu}
\thanks{}
\affiliation{Joint Center for Quantum Information and Computer Science, University of Maryland, USA}
\affiliation{Centre for Quantum Software and Information, University of Technology Sydney, Australia}
\author{Lirandë Pira}
\email{lpira@nus.edu.sg}
\thanks{}
\affiliation{Centre for Quantum Technologies, National University of Singapore, Singapore}
\affiliation{Centre for Quantum Software and Information, University of Technology Sydney, Australia}
\author{Philipp Schleich}
\email{philipps@cs.toronto.edu}
\thanks{\\\\\\\\\\}
\affiliation{Department of Computer Science, University of Toronto, Canada}
\affiliation{Vector Institute for Artificial Intelligence, Toronto, Canada}
\author{Kelvin Koor}
\affiliation{Centre for Quantum Technologies, National University of Singapore, Singapore}
\author{Pedro C. S. Costa}
\affiliation{ContinoQuantum, Sydney, Australia}
\author{Dong An}
\affiliation{Beijing International Center for Mathematical Research, Peking University, Beijing, China}
\affiliation{Joint Center for Quantum Information and Computer Science, University of Maryland, USA}
\author{Al\'an Aspuru-Guzik}
\affiliation{Department of Computer Science, University of Toronto, Canada}
\affiliation{Department of Chemistry, University of Toronto, Canada}
\affiliation{Vector Institute for Artificial Intelligence, Toronto, Canada}
\affiliation{Canadian Institute for Advanced Research, Toronto, Canada.}
\affiliation{Acceleration Consortium, Toronto, Canada.}
\affiliation{Department of Chemical Engineering \& Applied Chemistry, University of Toronto, Toronto, Canada.}
\affiliation{Department of Materials Science \& Engineering, University of Toronto, Toronto, Canada.}
\affiliation{Lebovic Fellow, Canadian Institute for Advanced Research, Toronto, Canada.}
\author{Lin Lin}
\affiliation{Department of Mathematics, University of California, Berkeley, USA}
\affiliation{Applied Mathematics and Computational Research Division,
Lawrence Berkeley National Laboratory, Berkeley, USA}
\author{Patrick Rebentrost}
\affiliation{Centre for Quantum Technologies, National University of Singapore, Singapore}
\affiliation{Department of Computer Science, National University of Singapore, Singapore}
\author{Dominic W. Berry}
\affiliation{School of Mathematical and Physical Sciences, Macquarie University, Sydney Australia}

\date{\today}

\begin{abstract}
Solving linear systems of equations plays a fundamental role in numerous computational problems from different fields of science. The widespread use of numerical methods to solve these systems motivates investigating the feasibility of solving linear systems problems using quantum computers. In this work, we provide a survey of the main advances in quantum linear systems algorithms, together with some applications. We summarize and analyze the main ideas behind some of the algorithms for the quantum linear systems problem in the literature. The analysis begins by examining the Harrow-Hassidim-Lloyd (HHL) solver. We note its limitations and reliance on computationally expensive quantum methods, then highlight subsequent research efforts which aimed to address these limitations and optimize runtime efficiency and precision via various paradigms. We focus in particular on the post-HHL enhancements which have paved the way towards optimal lower bounds with respect to error tolerance and condition number. By doing so, we propose a taxonomy that categorizes these studies. Furthermore, by contextualizing these developments within the broader landscape of quantum computing, we explore the foundational work that have inspired and informed their development, as well as subsequent refinements. Finally, we discuss the potential applications of these algorithms in differential equations, quantum machine learning, and many-body physics.
\end{abstract}

\maketitle
\onecolumngrid

\newpage
\tableofcontents

\newpage

\section{Introduction}\label{sec:introduction}
Linear systems of equations play a central role in many areas of science and engineering, with a wide range of applications involving large instances of these equations. This often implies that solving such problems require efficient computational methods. Classical methods for solving these systems, such as Gaussian elimination and iterative methods, have been long studied and optimized~\cite{wendland2017numerical}. However, as the size and complexity of these systems grow, classical computational models encounter computational bottlenecks which limit their efficiency and scalability. In recent years, quantum computing has emerged as a promising paradigm to tackle computationally intensive problems \cite{nielsen_quantum_2010}. Quantum computers have already demonstrated exponential speedups in solving specific problems --- the most prominent of which is Shor's algorithm for factoring~\cite{shor_polynomial_1997}. This groundbreaking result spurred researchers to also seek speed-ups for other computational tasks via quantum computation. To this end, the Harrow-Hassidim-Lloyd (HHL) quantum algorithm was the first to quantumly solve problems associated to linear systems \cite{harrow_quantum_2009}. The main contribution of this algorithm comes from the \textit{logarithmic} dependence on the dimension of the input matrix. Could the complexity of HHL, with respect to the other parameters such as condition number and output error, be improved? Indeed, the development of quantum algorithms dedicated toward such improvements constitute an important area of research in quantum algorithms \cite{montanaro_quantum_2016, dalzell2023quantum}. This work, is an attempt at providing a comprehensive overview of the significant algorithmic advancements for QLSP solvers.

Classically, the problem of solving linear systems is as follows. Given a matrix $A$ and a vector $\mathbf{b}$, find $\mathbf{x}$ such that $A\mathbf{x}=\mathbf{b}$. The \textit{quantum} version of this problem, although related to the classical version, has subtle distinctions. We call this the quantum linear systems problem (QLSP). In this problem we are given an $N \times N$ matrix $A$ and a vector $\mathbf{b}$ (appropriately encoded as a quantum state) and the task is to output the quantum state $\ket{\mathbf{x}}$. The key difference is that instead of outputting the vector $\mathbf{x}$ (as an array of numbers for example), it is the quantum state $\ket{\mathbf{x}}$ (which encodes $\mathbf{x}$)) that is outputted. The entries of $\ket{\mathbf{x}}$ are not directly accessible, rather they can only be acquired through measurements of the relevant observables, e.g. $\braket{\mathbf{x}|M|\mathbf{x}}$.

When constructing QLSP solvers, there are a few criteria to fulfill to ensure the efficiency of these solvers. One of these is that the matrix $A$ has to be sparse and well-conditioned. The well-conditioned criterion makes $A$ robustly invertible \cite{aaronson_read_2015}. Another important quantity to consider is the condition number $\kappa$ of $A$, which quantifies how close $A$ is to being singular. In terms of runtime, HHL has complexity $\text{poly}(\log N, \kappa)$. It is thus important that $\kappa$ is not too large, otherwise the $\log N$ advantage in the dimension over the classical methods may be spoiled by the higher cost in $\kappa$ .

Algorithmic enhancements to the efficiency of QLSP have  been introduced in Ref.~\cite{childs_quantum_2017}. These are ways to represent the inverse through a linear combination of unitaries, and furthermore the employment of variable-time amplitude amplification (VTAA) \cite{ambainis_variable_2010}, which improves the dependency on $\kappa$ from quadratic to linear. 
Another improvement comes via block-encoding and quantum singular value transformation (QSVT) \cite{gilyen2019quantum}, which results to a very similar approach as the quantum walk algorithm in \cite{childs_quantum_2017}. Roughly speaking, QSVT provides a framework and efficient way to invert matrices block-encoded within larger unitaries; it allows to implement polynomial functions of a block-encoding, thereby applying the inverse function through a polynomial approximation.

\begin{figure}[t]
    \centering    \includegraphics[width=.99\textwidth]{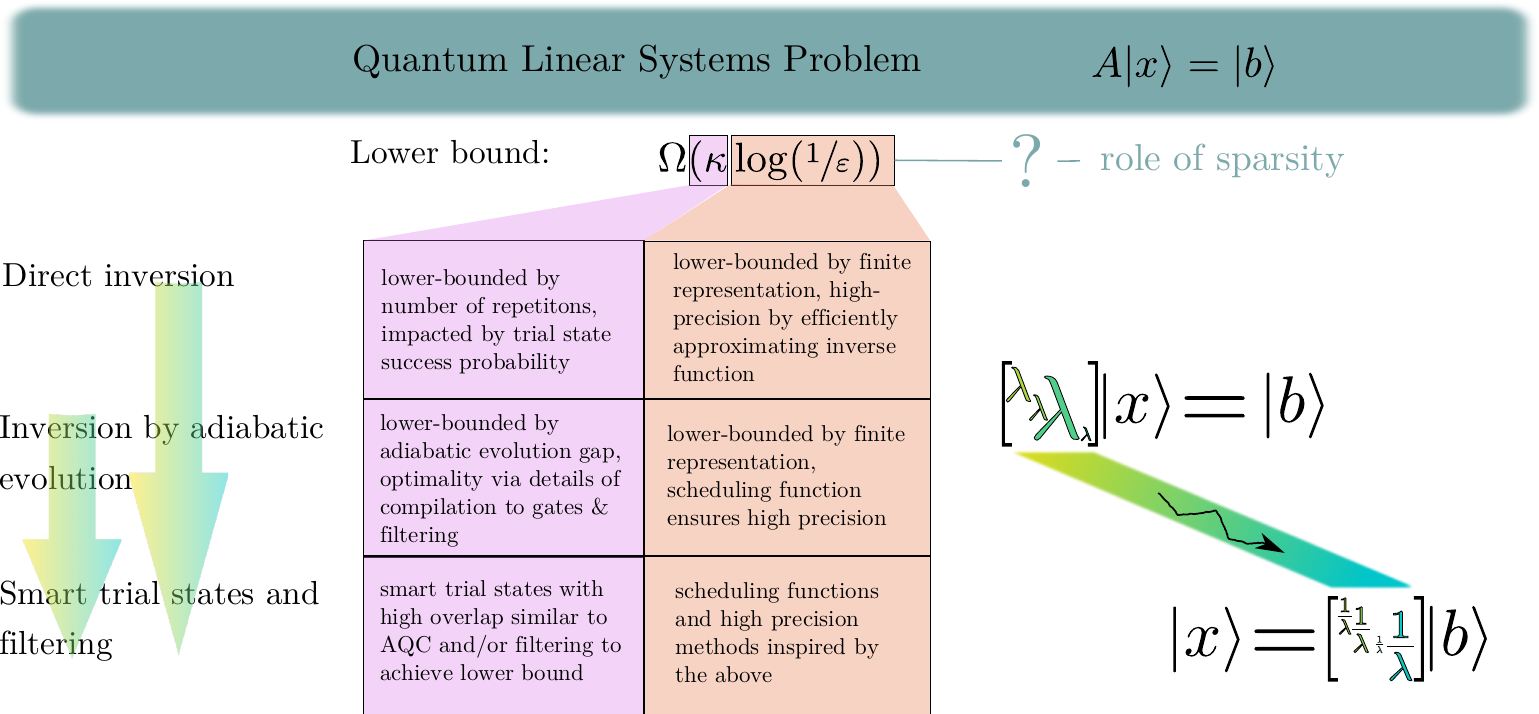}
    \caption{\textbf{Illustration of the quantum linear systems problem $A\ket{x}=\ket{b}$ and its classes of solutions.} Broadly speaking, there are three classes of solutions, i.e., direct inversion, inversion by adiabatic evolution, smart trial states and filtering. Each method's complexity is determined by factors such as the number of repetitions, adiabatic evolution gap, sparsity or finite representation precision. The role of sparsity influences efficient matrix inversion in achieving optimal results.}
    \label{fig:qlsp}
\end{figure}

On the other hand, there are also solutions based on adiabatic quantum computing (AQC), which in recent years have seen several improvements \cite{Subasi2019adiabatic, an_quantum_2020, Lin2020optimalpolynomial, costa_optimal_2021}. For instance, a slight modification of the quantum Zeno based algorithm in Ref.~\cite{Lin2020optimalpolynomial} leads to optimal scaling with respect to both $\kappa,\varepsilon$, up to a negligible factor of $\log\log \kappa$. 
Ref.~\cite{costa_optimal_2021} presents a different algorithm based on the discrete adiabatic theorem~\cite{dranov1998discrete} that achieves optimal scaling with respect to both $\kappa$ and $\varepsilon$, and achieves performance according to the known theoretical lower bound. More recently, Ref.~\cite{dalzell2024shortcutoptimalquantumlinear} departs from relying on the adiabatic theorems and augments the QLSP with an extra variable reaching the same complexity as Ref.~\cite{costa_optimal_2021}. Note that the improved AQC-based algorithms directly achieve a linear dependency on $\kappa$, as their setup allows starting with an initial state that does not impact the success probability. Ref.~\cite{dalzell2023quantum} achieves a similar result by introducing the kernel reflection method and an extended linear system that allows to use a similar better initial state. The taxonomy we introduce above is based on the strategy employed to solve the problem. In the first group we assemble direct inversion methods as mentioned above. In the second group, we detail adiabatic-inspired methods which we call inversion by adiabatic evolution. Finally, the third group highlights techniques which rely on trial state preparation and filtering. The main features of these methods are detailed in \cref{fig:qlsp} including their performance compared to the theoretical lower bound.

While this review focuses on the theoretical developments of QLSP solvers, for completeness we also briefly mention some aspects of experimental implementations, which at the time were all looking at the HHL algorithm. For instance, Ref.~\cite{cai_experimental_2013} implements the simple example of a system of $2\times 2$ linear equations on photonic qubits, where the fidelity remains fairly high, depending on the input vectors; the same system size and hardware combination was studied in Ref.~\cite{barz_solving_2013}.  Ref.~\cite{pan_experimental_2014} implements the QLSP on nuclear magnetic resonance experiments with four qubits, albeit also looking at a matrix of size $2\times 2$. 
These hardware implementations followed after the proposal of HHL and before algorithmic improvements; more recent implementations, as we mention in \cref{sec:nearterm}, tend to focus on near-term algorithms.

This review aims to serve as a comprehensive resource for researchers interested in understanding the current landscape of quantum algorithms for linear systems of equations, and their potential impact on computational science. An emphasis is placed on methods and advancements in algorithmic complexity and optimal scaling. We provide a succinct and an up-to-date overview of the main provable fault-tolerant quantum algorithms for linear systems, and their applications in quantum machine learning and differential equations. Additionally, we also briefly comment on the progress made in the development of the near-term and early fault-tolerant QLSP solvers which utilize variational and hybrid quantum-classical approaches. We are aware of older studies which also review the developments in QLSP solvers, namely Ref.~\cite{dervovic_quantum_2018}.

The rest of the manuscript is structured as follows. \cref{sec:preliminaries} sets the notation and algorithmic preliminaries. \cref{sec:qlsp} formulates the QLSP problem and the various input models. and the HHL algorithm focusing on its error analysis. \cref{sec:improv} provides the overview of works that improve the algorithmic complexity of the QLSP categorized as per the taxonomy we propose here. \cref{sec:optimal} discusses optimal scaling and constant factors. \cref{sec:nearterm} highlights near-term solutions to the QLSP. In \cref{sec:applications} we note the role of QLSP solvers in differential equations in \cref{subsec:de}, quantum machine learning in \cref{subsec:qml}, and in many-body physics in solving Green's function in fermionic systems in \ref{sec:greensfunction}. Finally, \cref{sec:concl} concludes this study and provides an outlook of open directions for further research.

\section{Preliminaries}\label{sec:preliminaries}
\subsection{Notation}\label{sec:notation}
For simplicity we assume $N=2^n$ for some $n \in \mathbb{N}$ throughout this paper. Logarithms are of base 2. Let $\mathbb{N} = \{1,2,\dots\}$ be the set of positive natural numbers. For an integer $d \in \mathbb{N}$, we define $[d] = \{1,2,\dots,d\}$. For a vector $v$, $\|v\|_1,\|v\|_2$ refer to its $\ell_1/\ell_2$-norms respectively, and we typically simply write $\|v\|$ for the $\ell_2$-norm. 
For a matrix $A$, we usually use the spectral norm $\|A\| $. By $\|A\|_1$, $\|A\|_2 $ we denote the induced $\ell_1$,~$\ell_2$-norms. The notation $A \succeq 0$ means $A$ is positive semi-definite~(PSD), i.e., the eigenvalues of $A$ are all non-negative. $I_n$ denotes the $n$-qubit identity matrix of size $2^n \times 2^n$. We say a Hermitian matrix $A$ is $s$-sparse if $A$ has at most $s$ nonzero entries per row/column. We use the following convention for binary representation up to $n$-bit precision: if $\alpha \geq 1$, $\alpha = \alpha_1\dots\alpha_n = \alpha_12^{n-1}+\dots+\alpha_n2^0$; if $0 \leq \alpha < 1$, $\alpha = 0.\alpha_1\dots\alpha_n = \alpha_12^{-1}+\dots+\alpha_n2^{-n}$. Here, $\alpha_i \in\{0,1\}$ for all $i$. Note that if $\alpha = \alpha_1\dots\alpha_n$, then $\alpha/2^n = 0.\alpha_1\dots\alpha_n$. For an invertible matrix $A$, the condition number 
$\kappa(A)$ quantifies the sensitivity of $A^{-1}$ toward errors in the input vector $b$ --- i.e., how perturbations $\varepsilon$ in the input $b$ are amplified by $A^{-1}$ relative to the input itself. Formally, this is the ratio of largest over smallest singular value, or eigenvalues respectively if $A$ is PSD:
\begin{align}
    \kappa(A) := \sup_{\|b\|=\|\varepsilon\|=1} \frac{\|A^{-1}\varepsilon\|}{\|A^{-1}b\|} = \|A^{-1}\|\|A\| = \frac{|\lambda|_{\max}(A)}{|\lambda|_{\min}(A)}.
\end{align}
The condition number can be defined over various norms; as above, we typically use the spectral norm. Finally, we use tildes in Big-Oh notation $\tilde{O}(\cdot)$ to hide polylogarithmic factors, i.e., $\tilde{O}(f(x)) := O(f(x)\cdot {\rm polylog}(f(x)))$.

\subsection{Quantum Algorithmic Primitives}\label{sec:QuantumAlgorithmPrimitives}
The following quantum algorithmic primitives are important components in quantum linear systems solvers. For ease of reference we describe the basics of their workings. Readers familiar with these may skip to \cref{sec:qlsp} and come back for reference.

\begin{enumerate}
    \item \textbf{Multi-Controlled Unitaries} Let $U$ be a unitary matrix. The unitary $\mathsf{ctrl}\text{-}U = \sum_{j=0}^{2^n-1} \ketbra{j}{j} \otimes U^j$ controlled on $n$ qubits can be implemented using $n$ single-qubit-controlled unitaries:
    \begin{align*}
        \mathsf{ctrl}\text{-}U = \sum_{j=0}^{2^n-1} \ketbra{j}{j} \otimes U^j &= \sum_{j_1,\dots,j_n=0}^{1} \ketbra{j_1}{j_1} \otimes \dots \otimes \ketbra{j_n}{j_n} \otimes U^{\sum_{i=1}^n j_i 2^{n-i}}\\
        &= \sum_{j_1,\dots,j_n=0}^{1} \left(\ketbra{j_1}{j_1} \otimes U^{j_1 2^{n-1}}\right) \cdots \left(\ketbra{j_n}{j_n} \otimes U^{j_n 2^{0}}\right)\\
        &= \prod_{i=0}^{n-1} \left( \ketbra{0}{0} \otimes I + \ketbra{1}{1} \otimes U^{2^i} \right).
    \end{align*}
    where $j=j_1\dots j_n = \sum_{i=1}^{n} j_i 2^{n-i}$ in binary. There is a slight abuse of notation in the last two lines where in essence we have written $(M_1 \otimes N_1)(M_2 \otimes N_2)$ for the term $(M_1 \otimes I \otimes N_1)(I \otimes M_2 \otimes N_2) = M_1 \otimes M_2 \otimes (N_1N_2)$. 

    \item \textbf{Quantum Fourier Transform (QFT)}
    The QFT is a key subroutine used in quantum phase estimation (and many other quantum algorithms). Let $\mathbf{x}\in \mathbb{C}^N$ (recall for simplicity we take $N=2^n$ for some $n \in \mathbb{N}$). The DFT (discrete Fourier transform) implements $\mathbf{x} \mapsto \mathbf{y}=\text{DFT}(\mathbf{x})$, where
    \begin{align*}
        DFT = \frac{1}{\sqrt{N}} \sum_{j,k=0}^{N-1} e^{2\pi ijk/N} \ketbra{j}{k}.
    \end{align*}
    The QFT implements the same transformation: for a basis state $\ket{j}$ we have
    \begin{align*}
        \mathsf{QFT}\ket{j} = \frac{1}{\sqrt{N}} \sum_{k=0}^{N-1} e^{2\pi ijk/N} \ket{k}.
    \end{align*}
    In binary, this is
    \begin{align*}
        \mathsf{QFT}\ket{j_1\dots j_n} = \frac{1}{\sqrt{2^n}}\left( \ket{0}+e^{2\pi i0.j_n}\ket{1} \right)\left( \ket{0}+e^{2\pi i0.j_{n-1}j_n}\ket{1} \right)\cdots \left( \ket{0}+e^{2\pi i0.j_1j_2\dots j_n}\ket{1}\right).
    \end{align*}
    For the implementation of $\mathsf{QFT}$ in terms of elementary gates, we refer the reader to \cite{nielsen_quantum_2010}. The gate complexity of $\mathsf{QFT}$ is $\Theta(\log^2 N) = \Theta(n^2)$, where $n$ is the number of qubits. The QFT is to the DFT what the QLSP is to the LSP, with the same caveat: the Fourier-transformed output $\mathsf{QFT}\ket{x}$ is not $\text{DFT}(\mathbf{x})$ itself, rather $\mathsf{QFT}\ket{x}$ has the entries of $\text{DFT}(\mathbf{x})$ encoded in its amplitudes.
    
    \item \textbf{Quantum Phase Estimation (QPE) \cite{cleve_quantum_1997}.} 
    Given a unitary $U \in \mathbb{C}^{N \times N}$ and an eigenvector $\ket{\psi}$ of $U$ with corresponding eigenvalue $e^{2\pi i \lambda}$, where $\lambda \in [0,1)$. QPE computes $\lambda$, or more precisely, a $t$-bit approximation of $\lambda$. For simplicity, first assume $\lambda$ can be exactly represented by $t=n$ bits, $\lambda = 0.\lambda_{1}\dots\lambda_{n}$. Then, the action of $\mathsf{QPE}$ on a $n$-qubit register initialized to $\ket{0^n}$ is
    \begin{align*}
        \mathsf{QPE}\ket{0^n}{\ket{\psi}} = \ket{2^n\lambda}\ket{\psi} = \ket{\lambda_{1}\dots\lambda_{n}}\ket{\psi},
    \end{align*}
    from which we extract the values $\lambda_1,\dots,\lambda_n$. Note that $\ket{\psi}$ is left invariant throughout. For a general $\lambda$ not representable by a finite number of bits, a more detailed analysis (see \cite{nielsen_quantum_2010}) shows that with $t=n+\lceil \log\left( 2+\frac{1}{2\delta} \right)  \rceil$,
    \begin{align*}
        \mathsf{QPE}\ket{0^t}{\ket{\psi}} = \ket{2^t\tilde\lambda}\ket{\psi}
    \end{align*}
    where the output is precise up to $n$-bits $|\tilde\lambda-\lambda|<\frac{1}{2^n}$  with probability $1-\delta$. This procedure requires $O(t^2)$ elementary gate operations (incurred from the Hadamards and $\mathsf{QFT}$) and makes one call to the oracle implementing the controlled-$U$. The circuit implementing $\mathsf{QPE}$ is shown in \cref{figure:QPE_circuit}.
    
    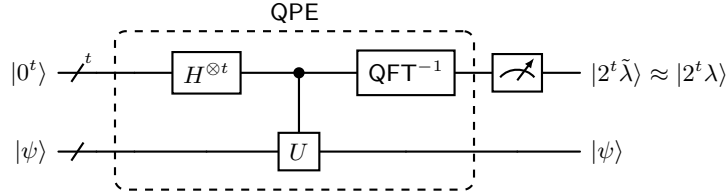
\begin{figure}[H]
        \centering
        \begin{quantikz}
        \lstick{$\ket{0^t}$} & \qwbundle{t} & \gategroup[2,steps=4,style={dashed,rounded corners}]{$\mathsf{QPE}$} & \gate{H^{\otimes t}} & \ctrl{1} & \gate{\mathsf{QFT}^{-1}} & \meter{} & \rstick{$\ket{2^t\tilde\lambda} \approx \ket{2^t\lambda}$}\\
        \lstick{$\ket{\psi}$} & \qwbundle{} &&& \gate{U} &&& \rstick{$\ket{\psi}$}
        \end{quantikz}
        \caption{Circuit implementing quantum phase estimation. The controlled-$U$ is shorthand for $\sum_{j=0}^{2^t-1} \ketbra{j}{j} \otimes U^j$.}
        \label{figure:QPE_circuit}
    \end{figure}

    \item \textbf{Amplitude Amplification (AA) \cite{brassard2002quantum}.} Amplitude amplification is a commonly used subroutine in many quantum algorithms. Let $f$ be a function $f: \{0,1\}^n \longrightarrow \{0,1\}$ marking the bases spanning the desired sector ($f(x)=1$) of Hilbert space and an oracle $O_f\ket{x,y} = \ket{x,y \oplus f(x)}$ implementing $f$ unitarily. Let $\ket{\psi} = U\ket{0^n}$ be the output of our quantum algorithm $U$. The goal of AA is to boost the probability amplitudes of $\ket{\psi}$ on the desired basis states. We can write
    \begin{align*}
        \ket{\psi} = \sum_{x=0}^{2^n-1} a_x \ket{x} &= \sum_{x: f(x)=0} a_x \ket{x} + \sum_{x: f(x)=1} a_x \ket{x}\\
        &= \sqrt{1-p}\ket{\alpha} + \sqrt{p}\ket{\beta},
    \end{align*}
    where without loss of generality the `success probability' $p = \sum_{x: f(x)=1} |a_x|^2 \leq 1/2$. Here $\ket{\alpha} = \frac{1}{\sqrt{1-p}} \sum_{x: f(x)=0} a_x \ket{x}$ and $\ket{\beta} = \frac{1}{\sqrt{p}} \sum_{x: f(x)=1} a_x \ket{x}$ specify respectively the undesired and desired sectors of Hilbert space. Our aim is to boost the success probability $p$ to $\beta$ such that $|\beta| \lessapprox 1$. The quantum circuit implementing this \textit{amplitude amplification} is given in \cref{AACircuit}, where the key component is the amplitude amplifier $G$ (for Grover, whose search algorithm is a precursor of amplitude amplification). As we see in ~\cref{AA}, this requires us to run our original algorithm $U$ twice for each $G$. In a nutshell, running $G$ for $k={O}(1/\sqrt{p})$ times gives $G^k \left( \sqrt{1-p}\ket{\alpha} + \sqrt{p}\ket{\beta} \right) = \sqrt{\alpha} \ket{\alpha} + \sqrt{\beta} \ket{\beta}$ so that $\abs{\alpha}$ is small and $\abs{\beta}$ is close to one. More details can be found in Ref.~\cite{brassard2002quantum}. 

    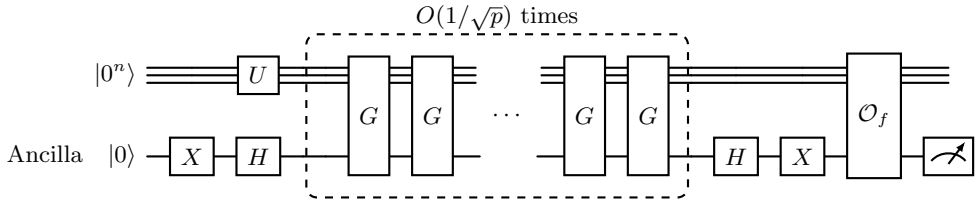
\begin{figure}[ht]
        \centering
        \begin{quantikz}[wire types = {b},classical
        gap=0.1cm, column sep=0.3cm]
        \lstick{$\ket{0^n}$} && \gate{U} && \gategroup[2,steps=6,style={dashed,rounded corners}]{${O}(1/\sqrt{p})$ times} & \gate[2]{G} & \gate[2]{G} & \midstick[2,brackets=none]{$\cdots$} & \gate[2]{G} & \gate[2]{G} &&&& \gate[2]{\soracle_f} &\\
        \lstick{\text{Ancilla}\quad$\ket{0}$} & \gate{X} & \gate{H} &&&&&&&&& \gate{H} & \gate{X} && \meter{}
        \end{quantikz}
        \caption{Quantum circuit implementing amplitude amplification.}
        \label{AACircuit}
    \end{figure}

    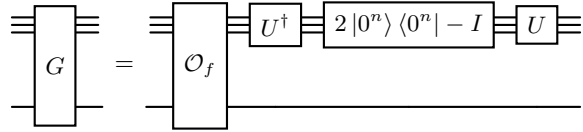
\begin{figure}[ht]
        \centering
        \begin{quantikz}[wire types = {b},classical
        gap=0.1cm, column sep=0.3cm]
        & \gate[2]{G} & \midstick[2,brackets=none]{=} & \gate[2]{\soracle_f} & \gate{U^{\dag}} & \gate{2\ket{0^n}\bra{0^n}-I} & \gate{U} &\\ 
        &&&&&&&
        \end{quantikz}
        \caption{The Amplitude Amplifier $G$.}
        \label{AA}
    \end{figure}
\end{enumerate}

\section{The Quantum Linear Systems Problem}\label{sec:qlsp}
In this section, we formally define the QLSP as it will be addressed throughout this review, along with the input models commonly employed in QLSP solvers.

\subsection{Problem Formulation}
First, we look at the linear systems problem that is typically known from linear algebra.
\begin{problem}[Linear System Problem (LSP)]\label{prob:LSP}
    Given an invertible matrix $A \in \mathbb{C}^{N \times N}$ and a vector $\mathbf{b} \in \mathbb{C}^N$, return a vector $\mathbf{x} \in \mathbb{C}^N$ satisfying $\mathbf{x}=A^{-1}\mathbf{b}$.
\end{problem}
That means that here, we expect a full classical description of the solution vector that allows inspection of every entry.

Next, we move to the quantum linear systems problem.
We are given an $N$-dimensional vector $\mathbf{v}=[v_1,\dots,v_N]^T \in \mathbb{C}^N$ which is encoded in a $\lceil\log(N)\rceil$-qubit quantum state: $\ket{v}=\sum_{i=1}^N v_i\ket{i}/\|\sum_{i=1}^N v_i\ket{i}\|$. Such an encoding is called the amplitude encoding of a vector, as the information is stored in the amplitude with indices labeled by basis state. The quantum version of the linear systems problem is defined as follows.
\begin{problem}[Quantum Linear System Problem (QLSP)]\label{prob:QLSP}
    We are given an invertible matrix $A \in \mathbb{C}^{N \times N}$ and a vector $\mathbf{b} \in \mathbb{C}^N$. Assume without loss of generality that (i.) $A$ is Hermitian, positive-semidefinite and has unit spectral norm $\|A\|=1$, so that $\|A^{-1}\|=\kappa(A)$ and (ii.) $\|b\|=1$.
    Let $\mathbf{x}=A^{-1}\mathbf{b}$ be as in the LSP in \cref{prob:LSP}. Denote the associated quantum states of $\mathbf{b,x}$ by
    \begin{align*}
        \ket{\mathbf b} = \sum_{i=1}^N b_i\ket{i} \quad \text{and} \quad \ket{\mathbf x} = \frac{\sum_{i=1}^N x_i\ket{i}}{\|\sum_{i=1}^N x_i\ket{i}\|}.
    \end{align*}
    Assuming oracle access to $A$ and an efficient oracle preparing $\ket{b}$, return a state $\ket{\tilde{\mathbf{x}}}$ such that {$\|\ket{\tilde{\mathbf{x}}} - \ket{\mathbf x}\| < \varepsilon$} for an allowed error tolerance $\varepsilon>0$.
\end{problem}
In this phrasing of the QLSP, the final solution is the output as a quantum state. This means we do not have direct ``human-readable'' access to it and will require some sort of subsequent measurement. In that spirit, solving the QLSP is a subroutine rather than a end-to-end algorithm.

\begin{remark}
    If $A$ is not Hermitian, we can (with an additional ancilla qubit) block-encode it in a larger Hermitian matrix 
    $\begin{pmatrix}
        0 & A\\
        A^\dag & 0
    \end{pmatrix}$.
    Solving
    $
    \begin{pmatrix}
        0 & A\\
        A^\dag & 0
    \end{pmatrix}
    \mathbf{y} = 
    \begin{pmatrix}
        \mathbf{b}\\
        0
    \end{pmatrix}
    $
    for $\mathbf{y}$ yields 
    $
    \mathbf{y} = 
    \begin{pmatrix}
        0\\
        \mathbf{x}
    \end{pmatrix},
    $
    where $\mathbf{x} = A^{-1}\mathbf{b}$. To simplify notation, we let $A\succeq 0$.
    Finally, suppose $A,\mathbf{b}$ were unnormalized, with normalized versions $A'=A/\|A\|$ and $\mathbf{b'}=\mathbf{b}/\|\mathbf{b}\|$. Then the new solution $\mathbf{x'}=A'^{-1}\mathbf{b'}=\frac{\|A\|}{\|b\|}\mathbf{x}$ is equivalent to the `original' solution $\mathbf{x}=A^{-1}\mathbf{b}$ up to normalization, and both are represented by $\ket{\mathbf{x}}$.
\end{remark}

\subsection{Input Model}
Specifying the input model is an important component in the design of quantum algorithms --- for QLSP, what is specifically meant is how to access the input matrix $A$ and the right-hand side vector $\mathbf{b}$ on a quantum computer by means of oracle queries.  We also say that $A$ and $\mathbf{b}$ are encoded in the oracle. In \cref{Definition:Block_encoding} an example of such an encoding in form of a block-encoding is given, where the matrix $A$ is encoded as a submatrix of a unitary $U$. In this case, the number of oracle queries is the number of times $U$ is applied in the execution of the algorithm. Another type of access is provided by the sparse matrix oracle access which we define below.

\begin{definition}[Sparse matrix oracles]\label{def:sparse_oracle}
    An $\spr$-sparse matrix $A\in\mathbb{C}^{N\times N}$ has a black-box description if there exists two unitary oracles $\soracle_A$ and $\soracle_F$ such that:
\begin{itemize}
    \item  if the query input to oracle $\soracle_A$ is a row index $i\in [N]$, a column index $j\in [N]$ and $z\in\{0,1\}^b$, the oracle returns the matrix element $A_{ij}=\matrixel{i}{A}{j}$ represented as a $b$-bit string
    \begin{equation}
    \soracle_A\ket{i}\ket{j}\ket{z}=\ket{i}\ket{j}\ket{z\oplus A_{ij}}.
    \end{equation}
    \item if the query input to oracle $\soracle_F$ is a row index $i\in [N]$ and an index $l\in [d]$, the oracle returns the column index $f(i,l)$ of the $l$th non-zero matrix entry in the $i$-th row
    \begin{equation}
        \soracle_F\ket{i}\ket{l}=\ket{i}\ket{f(i,l)}.
    \end{equation}
\end{itemize}    
\end{definition}
Many of the algorithms for QLSP such as HHL in \cref{subsec:hhl} or the algorithm in \cref{sec:lcu_inversion} give their complexity in terms of this type of input access, while later works often simply assume a block-encoding of the matrix $A$. The distinction between input models is important when considering questions about the query complexity of the algorithms. For instance, the discrete adiabatic algorithm \cite{costa_optimal_2021} presented in \cref{discrete-adiabatic} has a query complexity given by $\spr \kappa \log(1/\varepsilon)$, with respect to block-encoding access. It remains unclear whether this is jointly optimal for $\spr, \kappa$ and $\varepsilon$. We know of other methods which allow to obtain a $\sqrt{\spr}$ dependence on the sparsity \cite{Low2019hamsim} by block-encoding a sparse matrix with a query complexity depending on $\sqrt{\spr}$, yet it is not known whether this type of encoding can be used in algorithms with optimal dependence in $\kappa$ and $\varepsilon$ such as the discrete adiabatic algorithm~\cite{costa_optimal_2021}.
We provide a more detailed discussion of optimal scaling of algorithms solving the QLSP in \cref{sec:optimal}.

Another way to access the matrix $A$ is to encode it into a block of a larger unitary. This type of access is known as block-encoding and is given formally as follows. 
\begin{definition}[Block-Encoding]\label{Definition:Block_encoding}
Let $A$ be an $n$-qubit matrix, $\alpha, \varepsilon \in \mathbb{R}_+$ and $a \in \mathbb{N}$. We say that the $(n+a)$-qubit unitary $U$ is an $(\alpha,a,\varepsilon)$-block-encoding of $A$ if
$
\|A-\alpha(\bra{0^a}\otimes I_n)U(\ket{0^a}\otimes I_n)\| \leq \varepsilon.
$
\end{definition}

Gilyen et al. \cite{gilyen2019quantum} provide constructions for block-encodings for sparse matrices, assuming access to the relevant oracles. We state a simplified version (omitting some technicalities) below and refer to Ref.~\cite{gilyen2019quantum} for the full details.

\begin{proposition}[Block-encoding of sparse-access matrices --- Lemma 48, \cite{gilyen2019quantum}]\label{Proposition:block_encoding_sparse_matrices}
Let $A \in \mathbb{C}^{2^n \times 2^n}$ be an $s$-sparse matrix. Suppose we have access to the following $(n+1)$-qubit sparse-access oracles:
\begin{align*}
    &O_r: \ket{i}\ket{k} \to \ket{i}\ket{r_{ik}} \qquad \forall i \in [2^n]-1, k \in [s]\\
    &O_c: \ket{l}\ket{j} \to \ket{c_{lj}}\ket{j} \qquad \forall l \in [s], j \in [2^n]-1
\end{align*}
where $r_{ik}$ is the index for the $k$th nonzero entry of the $i$th row of $A$ and $c_{lj}$ is the index for the $l$th nonzero entry of the $j$th column of $A$. Additional assume we also have access to a third oracle
\begin{align*}
    O_A: \ket{i}\ket{j}\ket{0^b} \to \ket{i}\ket{j}\ket{A_{ij}} \qquad \forall i,j \in [2^n]-1
\end{align*}
where $A_{ij}$ is a $b$-bit description of the $ij$-matrix element of $A$. Then there is a $(s,n+3,\varepsilon)$-BE of $A$, whose implementation makes $O(1)$ queries to $O_r,O_c$ and $O_A$.
\end{proposition}

\section{Quantum Linear Systems Algorithms} \label{sec:improv}
This section reviews fault-tolerant quantum linear system algorithms with demonstrable speedups for the QLSP. It proposes a taxonomy and outlines the main solvers developed so far for the QLSP. Research into the QLSP within the near-term quantum computing paradigm is discussed briefly in \cref{sec:nearterm}.

A taxonomy, or even a branching of some sort, of QLSP solvers based on certain criteria is not straightforward.
To some extent, the most ``obvious'' solvers simply invert matrix $A$ in the context of QLSP. This is the key idea behind HHL. Since then, 
we observe that in the past recent years there has been a heavy reliance in analyzing the adiabatic theorem for solving the QLSP. Therefore, the section on ``direct inversion'' takes its name retrospectively, as it precedes the discussion of more recent advancements in adiabatic quantum computation covered in the following sections. Then within the works that rely on analyzing the adiabatic theorem, we find methods such as trial state preparation or polynomial filtering. One exception is the discrete adiabatic method in Ref.~\cite{costa_optimal_2021}, which implements a discrete version of adiabatic evolution. Additionally, augmentation and kernel reflection in Ref.~\cite{dalzell2024shortcutoptimalquantumlinear} does not rely on inversion by adiabatic evolution, but uses an extra variable in the problem definition. This is all to say that there are subtleties in any of the categorizing we considered.

We propose the following taxonomy.
\begin{itemize}
    \item \textbf{Direct inversion.} This refers to methods that straightforwardly  invert the matrix $A$ in a spectral sense. More specifically, they devise algorithms that apply $A^{-1}$ (or an approximation to it) to a state that encodes $\bv$. In \cref{subsec:direct}, we highlight works based on direct inversion, namely the HHL algorithm \cite{harrow_quantum_2009}, LCU implementation of inverse function in Fourier and Chebyshev bases \cite{childs_quantum_2017} and inversion based on QSVT \cite{gilyen2019quantum}.
    \item \textbf{Inversion by adiabatic evolution.} This includes solvers that use AQC or AQC-inspired methods to invert the matrix $A$. Namely, they aim to encode the inversion process into an adiabatic evolution. In \cref{subsec:adiabaticevolution}, we highlight the adiabatic randomization method \cite{Subasi2019adiabatic} and the time-optimal adiabatic method \cite{an_quantum_2020}.
    \item \textbf{Trial state preparation and filtering.} The idea here is to efficiently prepare an ansatz state (trial state) which is in some sense as close as possible to the solution vector and afterwards use eigenstate filtering to project, rotate, or reflect towards the solution vector. In \cref{subsec:filteringandtrialstate} we highlight eigenstate filtering and quantum Zeno method \cite{Lin2020optimalpolynomial}, the discrete adiabatic method \cite{costa_optimal_2021}, and the augmentation and kernel reflection method \cite{dalzell2024shortcutoptimalquantumlinear}.
\end{itemize}

\begin{figure}
    \centering
    \includegraphics[width=1\linewidth]{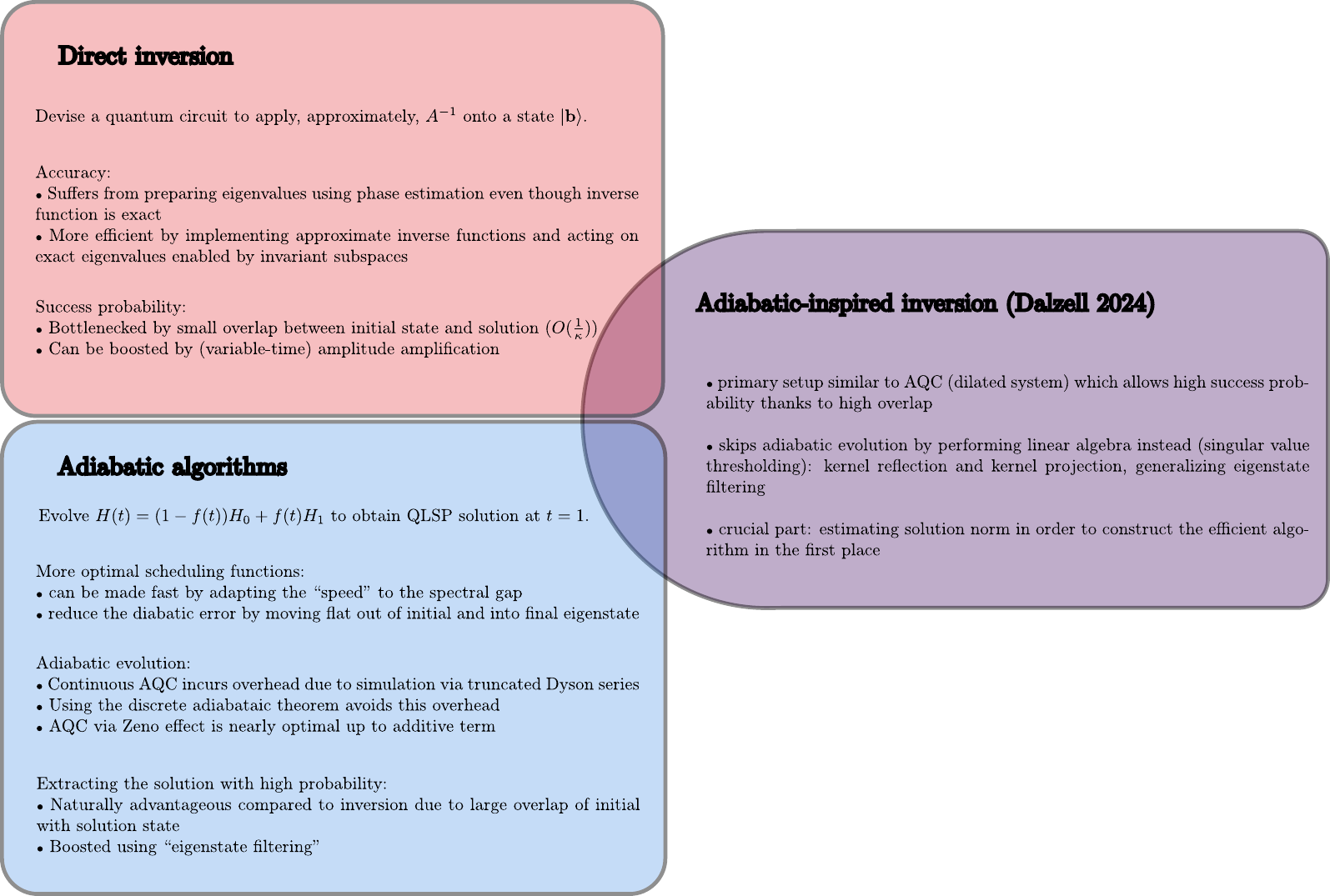}
    \caption{Quantum algorithms for linear systems: main contributions to their complexity and techniques.}
    \label{fig:figure-two}
\end{figure}

In the following parts of this section, we provide a more in-depth discussion of the works referenced above that push the state-of-the-art of quantum linear solvers, sorted into the broad categories we propose.
We relate the methods and summarize main techniques and contributing factors to their complexities in \cref{fig:figure-two}.

Additionally, for an overview on main characteristics of the quantum linear solvers, we refer the reader to \cref{tab:tableQLSP} which compares the runtime and query complexity of the algorithms we discuss in this section. In the same context, we point to iterative classical algorithms that solve the LSP, such as the conjugate gradient method \cite{shewchuk_introduction_1994}. It is also to be noted that the quantum algorithms we consider use different query access models and therefore care must be taken when comparing their query complexities.

\begin{table}[htbp]
\centering
\adjustbox{max width=\textwidth}{
\renewcommand{\arraystretch}{3.2}
\begin{tabular}{p{4.5cm}@{\hskip 4mm} p{4.5cm}@{\hskip 4mm} l @{\hskip 4mm} l} 
\toprule
\textbf{Reference} & \textbf{Characteristics} & \textbf{Input Model} & \textbf{Runtime/Query Complexity} \\
\midrule
Iterative methods (CG) \cite{shewchuk_introduction_1994} & Solving LSP & Classical &$O(N \spr \kappa \log(1/\varepsilon))$ \\
\rowcolor{blue!5}
HHL \cite{harrow_quantum_2009} & The first quantum linear system solver; utilizes quantum phase estimation & Sparse matrix oracle & $O(\log(N)\spr^2 \kappa^2 / \varepsilon)$ \\
Variable-time amplitude amplification \cite{ambainis_variable_2010} & Improved query complexity with $\kappa$ & Sparse matrix oracle & $O(\log(N)\spr^2 \kappa / \varepsilon)$ \\
\rowcolor{blue!5}
LCU implementation of inverse function in Fourier and Chebyshev basis \cite{childs_quantum_2017} & Improved time complexity & Sparse matrix oracle  & $O(\log(N)\spr\kappa \text{polylog}(\spr \kappa / \varepsilon))$ \\
QSVT \cite{gilyen2019quantum} & Based on block-encoding framework & Block-encoding & $O(\kappa^2 \log(\kappa/\varepsilon))$ \\
\rowcolor{blue!5}
Phase Randomization method \cite{Subasi2019adiabatic} & Adiabatic randomization & Sparse matrix oracle & $O(\log(N)\spr\kappa \log \kappa /\varepsilon)$ \\
Time-optimal adiabatic method \cite{an_quantum_2020} & Defines a schedule function & Sparse matrix oracle & $O(\log(N)\spr\kappa \mathrm{poly}(\log(\spr\kappa/\varepsilon)))$ \\
\rowcolor{blue!5}
Zeno eigenstate filtering method \cite{Lin2020optimalpolynomial} & Optimal polynomial filtering & Block-encoding & $O \left(\spr \kappa (\log \left( 1/\varepsilon \right)+(\log\log \kappa)^2) \right)$ \\
Discrete adiabatic method \cite{costa_optimal_2021} & Optimal scaling of $\kappa$ and $\varepsilon$ & Block-encoding & $O(\spr\kappa \log(1/\varepsilon))$ \\
\rowcolor{blue!5}
Augmentation and kernel reflection \cite{dalzell2024shortcutoptimalquantumlinear} & Augments QLSP with an extra variable removing trial state dependency & Block-encoding & $O(\spr\kappa \log(1/\varepsilon))$ \\
\bottomrule
\end{tabular}
}
\caption{\label{tab:tableQLSP} \textbf{Overview of the algorithmic improvements in solving the QLSP.} Here we list the main characteristics of the proposed QLSP solvers and their asymptotic runtime/query complexity. Note that for those algorithms assuming block-encodings as inputs, there is no dependence on the matrix size $N$ in their query complexities. Instead, $N$ appears in the complexity of constructing the block-encodings themselves.}
\end{table}

\subsection{Direct inversion}\label{subsec:direct}
In this section, we discuss three algorithms that we sort into the ``direct inversion'' category and directly implement an application of $A^{-1}$ on $\ket{\bv}$. The presented algorithms mostly differ in the way they construct a circuit implementation of the action of the inverse matrix. The first is the HHL algorithm~\cite{harrow_quantum_2009} that uses phase estimation and controlled rotation of the thus retrieved eigenvalues to apply $A^{-1}$. Further, various algorithmic improvements upon HHL have been introduced in \cite{childs_quantum_2017}, where an LCU implementation of the inverse function expressed in Fourier and Chebyshev bases allows to apply an inverse matrix. Finally, we elaborate on how the quantum singular value transformation~\cite{gilyen2019quantum} that enables the implementation of polynomial functions on the singular values of an operator, can be used for matrix inversion.

\subsubsection{The Harrow-Hassidim-Lloyd Algorithm}\label{subsec:hhl}
The HHL algorithm was the first and is to date the most famous algorithm to solve QLSP.
While the HHL algorithm was shown to have some advantage against classical algorithms, its limitations in practical applicability were soon challenged~\cite{aaronson_read_2015}. Subsequently, improvements of the algorithm have been achieved on the dependence of the query complexity on sparsity $s$, error $\varepsilon$ and condition number $\kappa$. Below, we provide a high-level idea as well as implementation details along with a complexity analysis.

\textit{High-level idea} -- The high-level outline of HHL is as follows. As per assumptions from \cref{prob:QLSP}, we have that $A$ is Hermitian, PSD and normalized. That means we can write its spectral decomposition as $A=\sum_{j=0}^{N-1} \lambda_j \ketbra{\psi_j}{\psi_j}$ with $\frac{1}{\kappa(A)} \leq \lambda_j \leq 1$. Further, write the expansion of $\ket{b}$ in terms of the eigenbasis of $A$ as $\ket{\mathbf{b}}=\sum_{j=0}^{N-1} \beta_j \ket{\psi_j}$. We wish to obtain $\ket{\mathbf x} \propto A^{-1}\ket{\mathbf b} = \sum_{j=0}^{N-1} \frac{\beta_j}{\lambda_j} \ket{\psi_j}$. Suppose $\ket{\mathbf b}=\ket{\psi_j}$ is an eigenvector of $A$. To get $\ket{\mathbf x}$ in this case we have to somehow extract the eigenvalue $\lambda_j$ associated to $\ket{\psi_j}$, invert it, and append it onto $\ket{\psi_j}$. The general case of inverting $\ket{\mathbf b}=\sum_j \beta_j \ket{\psi_j}$ then follows easily if these tasks could be performed in a way that preserves linearity. As we shall see below, this is indeed the case: the eigenvalue extraction is carried out via quantum phase estimation (QPE), and the appending of $\lambda_j$ to their respective eigenvectors $\ket{\psi_j}$ is performed with the assistance of controlled unitaries and appropriate postselection.\\

\begin{figure}[H]
    \centering
    \begin{quantikz}[column sep=0.3cm]
    \lstick{$\ket{0^t}$} & \qwbundle{t} && \gategroup[2,steps=4,style={dashed,rounded corners}]{$\mathsf{QPE}$} & \gate{H^{\otimes t}} & \ctrl{1} & \gate{\mathsf{QFT}^{-1}} && \gategroup[3,steps=3,style={dashed,rounded corners}]{$\mathsf{ctrl}\text{-}\mathsf{rot}$} & \ctrl{2} &&& \gategroup[2,steps=4,style={dashed,rounded corners}]{$\mathsf{QPE}^{-1}$} & \gate{\mathsf{QFT}} & \ctrl{1} & \gate{H^{\otimes t}} && \rstick{$\ket{0^t}$}\\
    \lstick{$\ket{b}$} & \qwbundle{} &&&& \gate{U} &&&&&&&&& \gate{U^{-1}} &&& \rstick{$\ket{x}$}\\
    \lstick{Ancilla $\ket{0}$} &&&&&&&& \gate{CA} & \gate{R} & \gate{CA^{-1}} &&&&&& \meter{} & \rstick{$\ket{1}$}
    \end{quantikz}
    \caption{\textbf{Circuit implementing HHL (without amplitude amplification).} After applying the unitary ${\mathsf{QPE}^{-1} \circ \mathsf{ctrl}\text{-}\mathsf{rot} \circ \mathsf{QPE} \circ U_b}$ on $\ket{0^t}\ket{0^{\log N}}\ket{0}$, measure the ancilla qubit and postselect on $\ket{1}$. This results in the state $\ket{x}$ on the second qubit register. For brevity we have omitted further ancillary qubits required to help implement the classical arithmetic unitary $CA$.}
    \label{figure:HHL_circuit}
\end{figure}
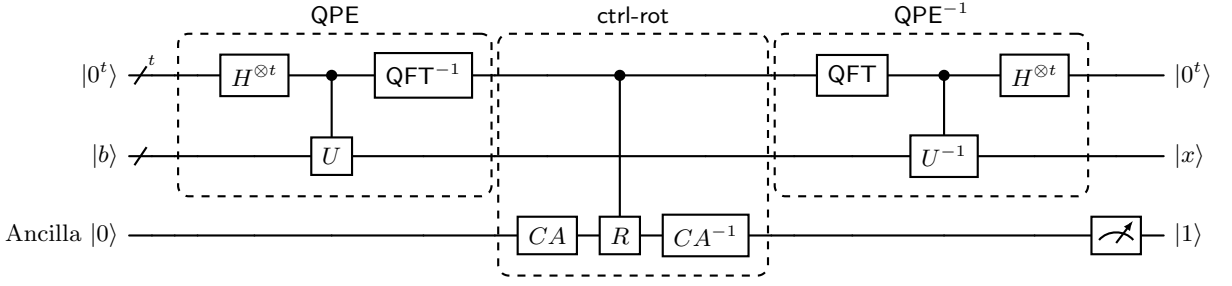

\textit{Details} -- The main steps of HHL are summarized in \cref{Algorithm:HHL} and further expounded below. The corresponding circuit is illustrated in \cref{figure:HHL_circuit}.
\begin{algorithm}[H]
\caption{Harrow-Hassidim-Lloyd (HHL)}
\label{Algorithm:HHL}
\begin{algorithmic}[1]
    \STATE \textbf{Input:}\\ 
    Oracle access to $A=\sum_{j=0}^{N-1} \lambda_j \ketbra{\psi_j}{\psi_j}$ where $1/\kappa(A) = \lambda_{\min} \leq \lambda_{\max}  = 1$ and $A$ is $s$-sparse;\\
    Efficient oracle preparing $\ket{\mathbf b}$;\\
    Output error tolerance $\varepsilon>0$;\\
    $t=\log T$ ancillary qubits initialized to $\ket{0^t}$, where $T=\widetilde{O}(\kappa s^2 \log N/\varepsilon)$.
    \STATE Prepare $\ket{\mathbf b}$.
    \STATE Apply $\mathsf{QPE}$ on the first and second qubit registers.
    \STATE Apply controlled-$\mathsf{rot}$ on the ancilla qubit, conditioned on the $t$-qubit register.
    \STATE Apply $\mathsf{QPE}^{-1}$ on the first and second qubit registers.
    \STATE Repeat steps 2-5 ${O}(\kappa)$ times. 
    \STATE Measure the ancilla qubit with respect to the computational basis. Postselect on the state $\ket{1}$.
    \STATE \textbf{Output:} $\ket{\tilde{\mathbf x}}$ such that $\|\ket{\tilde{\mathbf x}} - \ket{\mathbf x}\| < \varepsilon$, where $\ket{\mathbf x} = A^{-1}\ket{\mathbf b}/\|A^{-1}\ket{\mathbf b}\|$.
\end{algorithmic}
\end{algorithm}

Now we discuss \cref{Algorithm:HHL} in detail starting from Step 2. To highlight the salient features of the algorithm we make the further simplifying assumptions that $\ket{b}$ can be prepared without error, and Hamiltonian simulation $e^{iA\tau}$ can also be perfectly executed.
\begin{description}
    \item[Step 2] Prepare $\ket{\mathbf b} = \sum_{j=0}^{N-1} \beta_j \ket{\psi_j}$ using the efficient oracle: $U_b\ket{0^{\log N}} = \ket{\mathbf b}$. Here, ``efficient'' means $U_b$ is of size polylogarithmic in the system size $N$.
    
    \item[Step 3] With $U=e^{2\pi i A}$, $\mathsf{ctrl}\text{-}U = \sum_{j=0}^{2^t-1} \ketbra{j}{j} \otimes (e^{2\pi i A})^j$. Applying $\mathsf{QPE}$ (see \cref{sec:QuantumAlgorithmPrimitives}) gives $\mathsf{QPE}\ket{0^t}\ket{\mathbf b} = \sum_{j=0}^{N-1} \beta_j \ket{2^t\lambda_j}\ket{\psi_j}$.
    
    \item[Step 4] On the first and third registers, define the controlled unitary $\mathsf{ctrl}\text{-}R = \sum_{j=0}^{2^t-1} \ketbra{j}{j} \otimes (e^{-i\sigma_y})^j$. Ideally, we want
    \begin{align*}
        \mathsf{ctrl}\text{-}R\left(\sum_{j=0}^{N-1} \beta_j \ket{2^t\lambda_j}\ket{\psi_j}\ket{0}\right) \overset{?}{=} \sum_{j=0}^{N-1} \beta_j \ket{2^t\lambda_j} \ket{\psi_j}\left( \sqrt{1-\frac{1}{\lambda_j^2 \kappa^2}}\ket{0} + \frac{1}{\lambda_j \kappa}\ket{1} \right),
    \end{align*}
    so that the eigenvalues listed on the first register are appended onto the ancilla qubit. Working out the details, one realizes $\mathsf{ctrl}\text{-}R$ alone is insufficient; this part is often glossed over in many expositions on HHL. There are multiple ways to do this. The less efficient way relying on implementing arithmetics is to introduce an auxiliary parameter $\theta_k$ to store the information of $\lambda_k$ as $\theta_k = \arcsin(\frac{1}{\lambda_k\kappa})$. Then one first transforms $\ket{2^t \lambda_k}$ into $\ket{\theta_k}$ before applying $\mathsf{ctrl}\text{-}R$. The mapping $\ket{2^t \lambda_k} \mapsto \ket{\theta_k}$ can be implemented using \textit{classical arithmetic circuits}. These require $O(\text{poly}(t))$ elementary gates and ancilla qubits if $\theta_k$ is represented up to $t$-bit precision like $\lambda_k$. More details on this can be found in~\cite{lin2022lecture}. 
    A more efficient approach compared to using arithmetic circuits is to use inequality testing as described in \cite{sanders2012blackbox}, reducing the gate complexity by a considerable constant factor compared to arithmetics. 
    
    Thus, we define the controlled rotation step as $\mathsf{ctrl}\text{-}\mathsf{rot} = CA^{-1}(\mathsf{ctrl}\text{-}R)CA$, where we again omit tensor products with identities for brevity. This gives
    \begin{align*}
        \mathsf{ctrl}\text{-}\mathsf{rot}\left(\sum_{j=0}^{N-1} \beta_j \ket{2^t\lambda_j}\ket{\psi_j}\ket{0}\right) = \sum_{j=0}^{N-1} \beta_j \ket{2^t\lambda_j} \ket{\psi_j}\left( \sqrt{1-\frac{1}{\lambda_j^2 \kappa^2}}\ket{0} + \frac{1}{\lambda_j \kappa}\ket{1} \right)
    \end{align*}
    as desired.
    
    \item[Step 5] Uncomputing with $\mathsf{QPE}^{-1}$, we get the state $\sum_{j=0}^{N-1} \beta_j \ket{0^t}\ket{\psi_j}\left( \sqrt{1-\frac{1}{\lambda_j^2 \kappa^2}}\ket{0} + \frac{1}{\lambda_j \kappa}\ket{1}\right)$. After this, we discard the first register.

    \item[Step 6] Amplitude amplification (see \cref{sec:QuantumAlgorithmPrimitives}): with Steps $2-5$ constituting our unitary $U$ in \cref{AA} above, we implement the circuit shown in \cref{AACircuit}. This entails rerunning Steps $2-5$ $O(\kappa)$ times.
    
    \item[Step 7] Measure the ancilla qubit in the computational basis. Postselect on $\ket{1}$, then discard the ancilla qubit. This results in the state
    \begin{align*}
        \ket{\mathbf x} \propto \sum_{j=0}^{N-1} \frac{\beta_j}{\lambda_j} \ket{\psi_j} = A^{-1}\ket{\mathbf b}.
    \end{align*}
    
\end{description}

\textit{Analysis} -- Next, we briefly discuss the complexity of the HHL algorithm.  
\begin{enumerate}[\itshape a.]
\item The efficiency of preparing $\ket{\mathbf b}$ is crucial here, otherwise the resources required for state preparation could suppress any quantum speed-up gained. Let $T_b$ denote the gate complexity required to implement $\ket{\mathbf b}$. Assuming that preparing $\ket{b}$ is efficient, i.e., $T_b = O(\text{polylog}(N))$, the dominant resource expenditure then comes from the QPE subroutine. Henceforth for clarity we mostly omit $T_b$ from our complexity count.
\item It was shown in Ref.~\cite{berry2007efficient} that to simulate $e^{iA\tau}$ for $s$-sparse $A$, the (query) complexity required is $T= {O}(\tau s^2 (\frac{\tau}{\varepsilon})^{1/2k} \log N)$. `Query' here refers to the calling of oracles accessing the entries of $A$, see \cref{def:sparse_oracle} above. It is shown in \cite{harrow_quantum_2009} that in order to have $\|\ket{\tilde{\mathbf{x}}} - \ket{\mathbf x}\| < \varepsilon$ it is required that $\tau = O(\kappa/\varepsilon)$. Thus, in \cref{Algorithm:HHL} above we have $T={O}(\kappa s^2 \frac{\log N}{\varepsilon^{1+1/2k}})$, where we can say that $\frac{1}{\varepsilon^{1+1/2k}}\sim\frac{1}{\varepsilon^{1+o(1)}}$, as increasing $k$ in \cite{berry2007efficient}, the error due to Hamiltonian simulation is suppressed. The gate complexity of the arithmetic circuits goes as $O(\text{polylog}(T))$. This is dominated by $T$, i.e., the query complexity, and can thus be suppressed. 
\item At this stage, the probability of obtaining $\ket{1}$ upon measuring the ancilla qubit is $p(1)=\sum_{j=0}^{N-1} \frac{\beta_j^2}{\lambda_j^2\kappa^2} = \Omega\left(1/\kappa^2\right)$. That is, we are expected to run HHL $O(\kappa^2)$ times to obtain $\ket{1}$. Using amplitude amplification, which entails repeating steps $2-5$ $O(\kappa)$ times, we boost the success probability $p(1)$ to near $1$. Therefore, the overall complexity of HHL is ${O}(\kappa T_b + \kappa^2 s^2 \log N/\varepsilon^{1+o(1)})$, or simply ${O}(\kappa^2 s^2 \log N/\varepsilon^{1+o(1)})$ if we suppress the $T_b$ term.
This is amplitude amplification with an unknown amplitude, performed it repeatedly with varying numbers of steps until it succeeds. It is also possible to use fixed-point amplitude amplification with a logarithmic overhead \cite{Yoder2014fixed}.
\end{enumerate}

Since our intention is to present the most essential features of HHL, our presentation thereof is a ``bare-bones'' version. For a more detailed analysis taking into account implementation issues such as numerical stability and other matters, and techniques to handle them, we refer the reader to the original work \cite{harrow_quantum_2009} and the primer in Ref.~\cite{dervovic_quantum_2018}. In \Cref{subsec:rereading} we discuss the caveats of the algorithm as presented in Ref.~\cite{aaronson_read_2015}, and how they have held up with the latest improvements.

\subsubsection{LCU implementations of inverse function} \label{sec:lcu_inversion}
In the HHL algorithm~\cite{harrow_quantum_2009}, the dependence of the query complexity on the error $\varepsilon$ was dominated by the use of phase estimation. Concretely, the dependence on the approximation error of $O(1/\varepsilon)$ comes from the need to perform phase estimation which requires $\Theta(1/\varepsilon)$ uses of the unitary operation $e^{-iAt}$ to estimate the eigenvalues.

The work by Childs et al. \cite{childs_quantum_2017} presented in this section improves upon the HHL algorithm by circumventing the phase estimation algorithm and directly applying $A^{-1}$ on $\ket{\mathbf b}$.  This is carried out by implementing the matrix inverse as a linear combination of unitaries (LCU), as presented in the Hamiltonian simulation algorithm in \cite{berry_optimal_2015}. These changes are shown to give an exponential improvement in the error  dependence $\varepsilon$. For the most part, this can be described to the rapidly converging approximation of the inverse, while LCU allows to implement a given approximation exactly. 
Note that Ref.~\cite{childs_quantum_2017} did not make (direct) use of quantum signal processing~\cite{low2017optimal,martyn2021grand} or quantum singular value transformation techniques~\cite{gilyen2019quantum}.
Specifically, they present two approaches to implement the inverse. In the first, they use an integral identity to rewrite $1/x$, followed by a Fourier transformation so that the argument appears as a Hamiltonian Simulation. The appearing integrals are discretized, which on one hand introduces discretization error, on the other hand allows representation as a LCU.  
The second approach approximates the inverse function in a basis of Chebyshev polynomials and then show that this can be implemented by a quantum walk. We note that this is closely related to inversion by QSVT, cf.~\Cref{sec:qsvt_inversion}.

The algorithm presented in Ref.~\cite{childs_quantum_2017} uses the sparse matrix oracle access described in \cref{def:sparse_oracle}. Following the notation in that paper, $\mathcal{P}_A$ corresponds to both $\soracle_A$ and $\soracle_F$ in \cref{def:sparse_oracle}. They further assume access to an oracle $\mathcal{P}_B$ that prepares the rigged Hilbert space (RHS) state $\ket{\bv}$ in time $O(\mathrm{poly(\log N)})$.

\paragraph {Linear Combination of Unitaries}
In order to apply $A^{-1}$, it is decomposed as a sum of unitaries. Specific constructions are provided below. It proves useful to to use the non-unitary LCU lemma \cite[Lemma 7]{childs_quantum_2017} where  an operator is written as a linear combination of not necessarily unitary operators (as is the case for unitary LCU). Still, these operators that make up the linear combination themselves are a (sub)block of a unitary on a larger space. This means that for $M=\sum_i \alpha_i T_i$ where all $\alpha_i>0$ and $T_i$ need not be unitary, for any state $\ket{\psi}$ it holds that 
\begin{equation}\label{eq:nonunitary-lcu-lemma}
    U_i \ket{0^t}\ket{\psi} = \ket{0^t}T_i\ket{\psi} + \ket{\perp_i}.
\end{equation}
Here, $t\in\mathbb{N}$ is the number of ancillae and $(\op{0^t}\otimes I )\ket{\perp_i} = 0$ for all $i$.~\cite[Lemma 7]{childs_quantum_2017}
The difference to unitary LCU in \cref{eq:nonunitary-lcu-lemma} is that the ``garbage states'' $\ket{\perp_i}$ are not necessarily the same across all terms in the linear combination. 
As is the case for unitary LCU, this can be implemented by two subroutines $V: \ket{0^m} \to \frac{1}{\sqrt\alpha} \sum_i \sqrt{\alpha_i}\ket{i}$ and $U = \sum_i \op{i}\otimes U_i$, oftentimes called PREP and SEL, so that $V^\dagger U V \ket{\bv} \propto \frac{1}{A}\ket{\bv}$. 

Given an algorithm $\mathcal{P}_B$ that prepares the right-hand side $\ket{\bv}$, application of the non-unitary LCU in combination with amplitude amplification yields a state that can be retrieved with constant probability of success with using $O(\frac{\alpha}{\norm{M\ket{\psi}}})$ calls to $\mathcal{P}_B$, $V$, and $U$. 
Then, as per \cite[Corollary 10]{childs_quantum_2017}, we have the following:
\begin{corollary}[Corollary 10 in Ref.~\cite{childs_quantum_2017}]\label{cor:LCU-f-approx}
Let $A$ be a Hermitian operator with eigenvalues in $D\subseteq \mathbb{R}$. Suppose $f:D\to\mathbb{R}$ fulfills $\abs{f(x)}>1$ for all $x\in D$ and is $\varepsilon$-close to $\sum_i \alpha_i T_i$ in $D$ for some $\varepsilon \in (0,1/2)$, $\alpha_i > 0$ and functions $T_i: D\to \mathbb{C}$. Let $\{U_i\}$ be a set of unitaries such that 
\begin{equation}
         U_i \ket{0^t}\ket{\phi} = \ket{0^t} T_i(A) \ket{\phi} + \ket{\Phi^\perp_i} 
\end{equation}
for all states $\ket{\phi}$, where $t$ is non negative integer and $(\ketbra{0^t}{0^t} \otimes I) \ket{\Phi^\perp_i}=0$. Given an algorithm $\mathcal{P}_B$ to create state $\ket{b}$, there is a quantum algorithm that prepares a state $4\varepsilon$-close to $f(A)\ket{b}/\norm{f(A)\ket{b}}$, succeeding with constant probability that makes an expected $O(\alpha/\norm{f(A)\ket{b}})=O(\alpha)$ uses of $\mathcal{P}_B$, $U$ and $V$ where $U = \sum_i \ketbra{i}{i} \otimes U_i$, $V\ket{0^m} = \frac{1}{\sqrt{\alpha}} \sum_i \sqrt{\alpha_i} \ket{i}$,  $\alpha = \sum_i \alpha_i$.
\end{corollary}
We make the following remark: The function $f(x)=1/x$ to be approximated is considered over the domain $D_\kappa=[-1,-\kappa^{-1}]\cup[\kappa^{-1},1]$. Considering matrices as outlined in \cref{prob:QLSP} with $\norm{A}=1,\norm{A^{-1}}=\kappa$, this is the relevant spectral range of $A$ and satisfies the condition \cref{cor:LCU-f-approx} $f(x)>1$ for all $x$ in the domain of $f$. 

\paragraph{Fourier approach}\label{sec:Fourier}
The starting point for this approach is identifying the inverse function through an integral identity -- namely, for any odd function $f(y)$ over the real numbers so that $\int_{\mathbb{R}_+} dyf(y) =1$, integrating $f(xy)$ over the same domain is equal to $\frac{1}{x}$ for $x\neq 0$. Then, \cite{childs_quantum_2017} choose $f(y)=ye^{-y^2/2}$ and additionally use a Fourier transform representation in the variable $z$. 
Then, we obtain
\begin{equation}\label{eq:fourier-representation}
    \frac{1}{x}= \frac{i}{\sqrt{2\pi}}\int_{0}^\infty dy \int_{-\infty}^\infty dz \;z e^{-z^2/2} e^{-ixyz}  \approx_{\varepsilon} h(x) = \frac{i}{\sqrt{2\pi}} \sum_{j=0}^{J-1} \Delta_y \sum_{k=-K}^K \Delta_z z_k e^{-z_k^2/2} e^{-ixy_jz_k} .
\end{equation}
The reason for using the Fourier transform is that this leads to terms $e^{-ixyz}$, where $x$ comes from the system matrix $A$, hence discretizing the integral will lead to a linear combination of unitaries that are Hamiltonian simulation steps. Note that this procedure is related to an algorithm called Linear Combination of Hamiltonian Simulations~\cite{an2023linear,an2023quantum}, that has been proposed to solve differential equations, where now, the task is to find an integral identity for time propagation rather than inversion.

To implement an algorithm based on this decomposition, we need to impose a cutoff on the integrals and discretize, as shown in \cref{eq:fourier-representation}. Appropriate choices of $\Delta_y,\Delta_z$ and $J,K$ then lead to an $\varepsilon$-approximation of $1/x$ by $h(x)$ on $D_\kappa$. The details for the cutoff and discretization are given in Lemma 11 and Lemma 12 in Ref.~\cite{childs_quantum_2017}. The query complexity will depend on the chosen cutoff and the size of the discretization, which in turn can be chosen depending on the desired target error $\varepsilon$ and the condition number $\kappa$.
With suitable discretization $\Delta_y,\Delta_z$ and cutoffs $J,K$, it is possible to prove the following result:
\begin{theorem}[Theorem 3 in Ref.~\cite{childs_quantum_2017}]\label{thm:fourier-childs}
The QLSP can be solved with  $O(\kappa \sqrt{\log(\kappa/\varepsilon)})$ uses of a Hamiltonian simulation algorithm that approximates $e^{-iAt}$ for $t=O(\kappa \log(\kappa/\varepsilon))$ with precision $O(\varepsilon/\kappa \sqrt{\log(\kappa/\varepsilon)})$. Using the best algorithms for Hamiltonian Simulation (at the time of publication of \cite{childs_quantum_2017}), this makes $O(\spr\kappa^2 \log^{2.5}(\kappa/\varepsilon))$ queries to $\mathcal{P}_A$, makes $O(\kappa \sqrt{\log(\kappa/\varepsilon)})$ uses of $\mathcal{P}_B$ and has gate complexity $O(\spr\kappa^2 \log^{2.5}(\kappa/\varepsilon)(\log N + \log^{2.5}(\kappa/\varepsilon)))$.
\end{theorem}
\begin{remark}
    Note that the complexity of the theorem above can be improved as in Ref.~\cite{sanders2012blackbox} by using techniques that avoid arithmetic and calculation of trigonometric functions.
\end{remark}

\paragraph{Chebyshev approach}\label{sec:Chebyshev}
An alternative to the above integral identity is expressing $1/x$ directly in a basis of Chebyshev polynomials.  
The Chebyshev polynomials of the first kind are defined by the recurrence relation $\mathcal{T}_0(x)=1$, $\mathcal{T}_1(x)=x$ and $\mathcal{T}_{n+1}=2x\mathcal{T}_n(x)-\mathcal{T}_{n-1}(x)$; they are defined over $x\in[-1,1]$ but can be applied more generally by appropriate transformations. These polynomials form a complete, orthogonal basis. In this section they will be used to approximate $1/x$ on the domain $D_\kappa$. 
Note that \cite{childs_quantum_2017} does not directly decompose $1/x$ with Chebyshev polynomials but uses $\frac{1}{x}(1-(1-x^2)^\beta)$ instead due to the discontinuity near $x\to 0$. Note that using QSVT directly, as we discuss in the next section, one can approximate $1/x$ more directly in a Chebyshev basis via the bounded approximation theorem. 

The decomposition given in Ref.~\cite{childs_quantum_2017} is given in~\cite[Lemma 14]{childs_quantum_2017}, namely 
\begin{equation}\label{eq:chebyshev-decomp}
    \textstyle g(x) = 4 \sum_{j=0}^{j_0} (-1)^j \left[\frac{\sum_{i=j+1}^\beta \binom{2\beta}{\beta+i}}{2^{2\beta}} \right] \mathcal{T}_{2j+1}(x)
\end{equation}
where $j_0=\lceil\sqrt{\beta\log(4\beta/\varepsilon)}\,\rceil$ and $b=\lceil\kappa^2\log(\kappa/\varepsilon)\rceil$. Then, $g(x)$ is $2\varepsilon$-close to $1/x$ on $D_\kappa$.
Using the notation of \cref{cor:LCU-f-approx}, we need to find operators $U_i$ such that 
\begin{equation}
         U_i \ket{0^t}\ket{\phi} = \ket{0^t} T_i(A) \ket{\phi} + \ket{\Phi^\perp_i},
\end{equation}
where in this case $T_i(A)=\mathcal{T}_i (A)$.
This will lead to a different algorithmic structure as in the ``Fourier approach'', as a Chebyshev basis compared to Fourier does not lead to unitaries via Hamiltonian simulation. 
To implement such a $U_i$, the authors use a method based on quantum walks. Given a $d$-sparse $N\times N$ Hamiltonian $A$, the quantum walk is defined on a set of states $\{\ket{\psi_j} \in \mathbb{C}^{2N} \otimes \mathbb{C}^{2N} \}_{j=1}^{N}$. Each of the states in this set is defined as 
\begin{equation}
    \ket{\psi_j} = \ket{j} \otimes \frac{1}{\sqrt{d}}\sum_{k\in [N]: A_{jk} \neq 0} \sqrt{A_{jk}^\ast} \ket{k} + \sqrt{1-\abs{A_{jk}}}\ket{k+N}.
\end{equation}
The quantum walk operator $W:=S(2TT^{\dag}-I) $ is defined in term of the isometry $T=\sum_{j\in [N]}\ketbra{\psi_j}{j}$  and the swap operator $S$, which acts as $S\ket{j,k}=\ket{k,j}$ for all $j,k\in[2N]$.  Such a walk operator can be implemented with a constant number of queries to $\mathcal{P}_A$ \cite{berry_blackbox_2012,berry_optimal_2015}.
Given an eigenvector $\ket{\lambda}$ of $H$ with eigenvalue $\lambda \in (-1,1)$, then the operator $W$ can be written as a block in the space $\mathrm{span}\{T\ket{\lambda},ST\ket{\lambda} \}$, 
\begin{equation}
    \begin{pmatrix}
\lambda & -\sqrt{1-\lambda^2} \\
\sqrt{1-\lambda^2} & \lambda 
\end{pmatrix},
\end{equation}
where the first row/column corresponds to $T\ket{\lambda}$ and the second row column to the orthogonal state. In this block form it is possible to show that in the previously mentioned subspace we have 
\begin{equation}
    W^n =     \begin{pmatrix}
\mathcal{T}_n(\lambda) & -\sqrt{1-\lambda^2} \mathcal{U}_{n-1} (\lambda) \\
\sqrt{1-\lambda^2}\mathcal{U}_{n-1} (\lambda) & \mathcal{T}_n (\lambda)
\end{pmatrix}
\end{equation}
which can be shown by a simple induction. 
The same argument argument about the invariant 2D subspace induced by $T\ket{\lambda}$ as in QSVT~\cite{gilyen2019quantum} now allows to notice that this also holds for any function of $H$ rather than only a single eigenstate.
Any state $\ket{\psi}\in \mathbb{C}^{N}$ can be written as a linear combination of the eigenvectors $\ket{\lambda}$, which implies
\begin{equation}
    W^n T \ket{\psi} = \mathcal{T}_n (H) T \ket{\psi} +  \ket{\perp_{\psi}} 
\end{equation}
where $\ket{\perp_\psi}$ is orthogonal to $\mathrm{span}\{T\ket{j} : j\in [N]\}$, which must be true in general since the state obtained is orthogonal to $T\ket{\lambda}$.

By implementing the unitary $\ket{0^{m}}\ket{\psi}\to T\ket{\psi}$, the following operation can be done: Act with this unitary on $\ket{0^{m}}\ket{\psi}$, then act with $W^n$ and undo the first operation. The overall transformation is $\ket{0^m}\ket{\psi} \to \ket{0^m}\mathcal{T}_n(H)\ket{\psi} + \ket{\Psi^\perp}$ where $\Pi \ket{\Psi^\perp}=0$ with $\Pi = \ketbra{0^m}{0^m} \otimes I$. $W$ and $T$ are implemented with $O(1)$ calls to $\mathcal{P}_A$ as seen in Ref.~\cite{berry_optimal_2015}, which implies that implementing the whole operator takes $O(n)$ queries. In this way we are able now to implement operators $U_i$ of \cref{cor:LCU-f-approx}. The operator $V$ is defined by the coefficients in the decomposition of $1/x$ as in~\cref{eq:chebyshev-decomp}. The query complexity and gate complexity results for the Chebyshev method are summarized in the following theorem.
 \begin{theorem}[Theorems 4 in Ref.~\cite{childs_quantum_2017}]\label{thm:chebyshev-childs}
 The QLSP can be solved using $O(\spr\kappa^2 \log^2 (\spr\kappa/\varepsilon))$ queries to $\mathcal{P}_A$ and $O(\kappa \log (d\kappa/\varepsilon))$ uses of $\mathcal{P}_B$ with gate complexity $O(\spr \kappa^2 \log^2 (\spr\kappa/\varepsilon) (\log N + \log^{2.5} (\spr\kappa/\varepsilon)))$
 \end{theorem}
\begin{remark}
    This result can be recovered using QSVT when considering the same approximating polynomial, namely $f(x) = \frac{1}{x}(1-(1-x^2)^\beta)$, $\beta =\lceil \kappa^2\log(\frac{\kappa}{\varepsilon})\rceil$.
    For details, \cite[Lemma 9]{gilyen2019quantum} and \cite[Theorem 41]{gilyen2019quantum}.
\end{remark}

\paragraph{Improvements by Variable Time Amplitude Amplification}
So far, the Fourier approach and the Chebyshev approach have given algorithms with a quadratic dependence (up to logarithmic factors) on the condition number $\kappa$. In Ref.~\cite{childs_quantum_2017}, the authors improve this to a linear dependence using the so-called variable-time amplitude amplification technique (VTAA). 

The quadratic dependence on the condition number in \cref{thm:chebyshev-childs} and \cref{thm:fourier-childs} comes from two aspects. The first one comes from the fact that \cref{cor:LCU-f-approx} uses $O(\alpha)$ oracle calls from using amplitude amplification so that $A^{-1}$ is correctly applied, where $\alpha=O(\kappa)$. This is because the subnormalization of the LCU, which comes from the numerical discretization, i.e., is required to represent the inverse function up to the target precision. The second contribution comes from the gate cost of implementing the unitary $U$ for the LCU in \cref{cor:LCU-f-approx}. In both the Fourier (see after \cref{thm:fourier-childs}) and Chebyshev case (see after \cref{thm:chebyshev-childs}) this dependence is linear in $\kappa$. Let us be a bit more precise here.

In the Fourier case we, using the Hamiltonian simulation algorithm from \cite{berry_optimal_2015} for a simulation time $t$ and error $\varepsilon$ requires $O(\spr t\log(t/\varepsilon))$ queries to a $\spr$-sparse Hamiltonian. This gives a number of query calls which is nearly linear in $\kappa$. In the case of the Chebyshev approach, implementing $U$ had a cost of $O(j_0)$ where $j_0$ is the highest order in the Chebyshev polynomials which is nearly linear as well (see after \cref{thm:chebyshev-childs}). 

To improve the performance on $\kappa$, the authors in Ref.~\cite{childs_quantum_2017} assume that $\ket{\bv}$ is contained in a subspace of $A$ so that all eigenvalues values in this subspace are close to 1 in magnitude. Then the problem is easier as the effective condition number $\kappa'$ is smaller than the original one. If the eigenvalues are close to $1/\kappa$, then the complexity for doing amplitude amplification diminishes. This can be exploited by the algorithm in Ref.~\cite{ambainis_variable_2010} which performs the VTAA. We do not present this algorithm in detail, but we do point out that it uses a low-precision version of the phase estimation primitive, which allows to both keep the dependence on $\kappa$ linear and also achieve an exponential reduction in the error.
We further point out that an new version through a tunable VTAA has been developed in \cite{low2024quantumlinearalgorithmoptimal}  and is applied to the QLSP. This will be discussed further in \cref{subsec:optimal-scaling}.

\subsubsection{Matrix inversion based on QSVT}\label{sec:qsvt_inversion}
In this section, we discuss QLSP through the lens quantum singular value transformation. In particular, this section also gives more general results as the last section, as it can work with more general block-encodings than block-encoding through LCU and further can represent more general approximations to the inverse.

We assume the same properties for $A$ as stated in \cref{prob:QLSP}, i.e., $A$ is Hermitian and $\norm{A}=1$, where $\norm{A^{-1}} = \kappa^{-1}$ and $A$ is $\spr$-sparse. The generalization to arbitrary $A$ so that the algorithm implements the Moore-Penrose pseudoinverse is straightforward.
In this QSVT variant of QLSP, we assume a block-encoding input model of $A$ using a unitary $U_A$. The goal is to produce an approximation to $A^{-1}$ through a sequence of operations which involves querying $U_A$. 
This is done by interleaving application of parametrized exponentiated reflections around the spaces spanned by the singular vectors with $U_A$ (e.g., see \cite[Theorem 17]{gilyen2019quantum}). Note that if the reflections are parametrized with all-zeroes, this corresponds to a Chebyshev basis -- and recovers the quantum walk in the Chebyshev section above. 

QSVT is able to produce a block-encoding of $p(A)$, where $p$ is a polynomial of a certain degree and definite parity (arbitrary polynomials of a certain degree can be implemented using an even/odd decomposition, e.g. see \cite[Theorem 5]{gilyen2019quantum}). 
Then, to tackle a more general function $f(\cdot)$, the first step is to find a polynomial that approximates it well enough over the interval of interest, namely, the spectrum of $A$ scaled to the range $[-1,1]$. Then, this interval is the same as the  domain $D_\kappa = [-1,-\kappa^{-1}]\cup[\kappa^{-1}, 1]$ as in \cref{sec:lcu_inversion}. 

Observe that QSVT thus is able to approximate a matrix inverse by any polynomial of degree $m$ that gives an $\varepsilon$ approximation to $1/x$ over $D_\kappa$ and further,  is bounded on the interval $[-\kappa^{-1}, \kappa^{-1}] = [-1,1]\setminus D_\kappa$. Using the bounded approximation result in  \cite[Corollary 66]{gilyen2019quantum}, one can show that a general approximation to the inverse has degree $m=O(\frac{1}{\delta}\log\frac{1}{\varepsilon})$, with $0<\varepsilon\le \delta \le \frac{1}{2}$~\cite[Corollary 69]{gilyen2019quantum}. This observation is also discussed in \cite{tang2024cs}. 

We now look at specific examples for the approximating polynomial in the literature. The choice  $\frac{1}{x}(1-x^2)^\beta$,  $\beta=\lceil \kappa\log(\frac{\kappa}{\varepsilon})\rceil $ with a Chebyshev expansion was done in \cite{childs_quantum_2017} and observed again in \cite[Theorem 41]{gilyen2019quantum}. Furthermore, \cite{ying2022stable} employs a polynomial approximation to the function $\frac{1-e^{-(5\kappa x) ^2}}{x}$. 

We are essentially done --- it remains to apply $U_{A^{-1}}$ to $\ket{\bv}\ket{000\dots}$ and then project onto the proper subspace. Finally, the query complexity of preparing $U_{A^{-1}}$ up to error $\varepsilon$ is given by $m = O(\kappa \log \frac{\kappa}{\varepsilon})$, namely through the degree of a polynomial approximation. 
Note that while the sparsity $s$ and matrix size $N$ does not appear in the query complexity, they feature in the construction of the block-encoding of $A$ itself, see \cref{Proposition:block_encoding_sparse_matrices}.
To conclude, to apply $U_{A^{-1}}$, we require $O(\kappa)$ applications to reproduce this with high probability. This gives a final query complexity of $O(\kappa^2 \log(\frac{\kappa}{\varepsilon}))$.

\subsection{Inversion by adiabatic evolution}\label{subsec:adiabaticevolution}
The next section discusses algorithms that invert linear systems based on adiabatic evolution. Here, the system matrix is typically embedded into a larger-dimensional parametrized Hamiltonian. Then, the latter is evolved according to the adiabatic principle, given an initial state that is easy to prepare, and the final state approximates the sought after solution state. A proper initial state generally improves the success probability by a factor of $\kappa$ in the more advanced techniques compared to most direct inversion approaches, as it can be chosen to have constant overlap with the solution state. The gap of the adiabatic evolution, as discussed below, can be related to the condition number of the linear system. 

We start this section by a high-level description of adiabatic evolution for the sake of inverting a linear system. In adiabatic algorithms, an adiabatic parameter is varied continuously and slow enough to stay in the ground state of a Hamiltonian as the system evolves. Similar to other applications of AQC, the main idea is to write a time-dependent Hamiltonian resulting from a linear interpolation between two time-independent Hamiltonians as follows
\begin{equation}
\label{eq:Ham_prob}
    H(t) = (1-f(t))H_0 + f(t)H_1,
\end{equation}
where the function $f(t)$ is a scheduling function, which is a continuous map from $[0,1]$ to $[0,1]$ so that $f(0)=0$ and $f(1)=1$.
For the quantum linear systems problem, there exists some eigenstate of $H_1$ that encodes the solution of the linear systems problem. The basic idea of the AQC method is to have an eigenstate of $H_0$ that is easy to prepare in order to move from the eigenstate of $H_0$ to the eigenstate $H_1$. This can be done by staying in the same eigenstate of \cref{eq:Ham_prob} from $t=0$ to $t=1$, since $H(0)=H_0$ and $H(1)=H_1$.

\subsubsection{Adiabatic randomization method}\label{sec:adiabatic-rand}
The first method to introduce adiabatic inspired methods for solving the QLSP problem was given in Ref.~\cite{Subasi2019adiabatic}. This work proposes two algorithms, the best one achieving a dependence of $O(\kappa \log \kappa/\varepsilon)$.
The technique used is referred to as the phase randomization method, which consists in performing an adiabatic evolution with discrete changes in the adiabatic parameter. We often refer to it as simply randomization method. 
The two algorithms correspond to different choices of the Hamiltonian used in the adiabatic evolution.
The time complexity of this method is linear in $1/\varepsilon$, although this dependence can be improved to poly-logarithmic in $1/\varepsilon$. This improvement requires repeated use of phase estimation which would incur a high cost in required ancillary qubits.

To solve \cref{prob:QLSP}, construct a Hamiltonian $H(t)$ dependent on a parameter $t\in [0,1]$ for which the solution state $\ket{\mathbf x}=A^{-1}\ket{\mathbf b}/\norm{A^{-1}\ket{\mathbf b}}_2$ from \cref{prob:QLSP} is a ground state when $t=1$. More specifically, define the operator $A(t)= (1-t) X \otimes I + t Z \otimes A$ where $X$ and $Z$ are Pauli operators. Note that the extra qubit ensures that $A(t)$ is invertible for all $t\in [0,1]$. Then we have $A(1)\ket{+}\otimes \ket{\mathbf x}\propto\ket{+}\otimes \ket{\mathbf b}$ and therefore defining $P^\perp_{b}=I-\ketbra{+\mathbf b}{+\mathbf b}$ gives $P^\perp_{b}A(1)\ket{+}\otimes \ket{\mathbf x}=0$.
This motivates the following definition for the Hamiltonian:
\begin{equation}\label{eq:RandMethod_Ham}
    H(t)=A(t)P^{\perp}_{b} A(t),
\end{equation}
which satisfies $H(t) A(t)^{-1}\ket{+ \mathbf b}=0$. The phase randomization method, first proposed in Ref.~\cite{Boixo2009eigenpath}, is a method which allows to traverse eigenstate paths of a Hamiltonian using a sequence of evolutions by a random time. This method is based on the quantum Zeno effect. If one has a family of states parametrized by $t$, $\{\ket{\psi(t)}\}$, then the final state $\ket{\psi(1)}$ can be prepared from the initial state $\ket{\psi(0)}$ by choosing a discretization $0=t_0<t_1<\ldots<t_q=1$ which is fine-grained enough so that $\ket{\psi(t_j)}$ is close to $\ket{\psi(t_{j+1})}$. Then, starting from $\ket{\psi(0)}$, one can successively project onto the state next in the discretized sequence and with high probability go from $\ket{\psi(t_j)}$ to $\ket{\psi(t_{j}=1)}$.

More precisely, the quantum operation $M_{t_j}(\rho)=\Pi_j \rho \Pi_j + \mathcal{E}((1-\Pi_j)\rho(1-\Pi_j)$ is applied between successive states $\ketbra{\psi(t_{j-1})}$ and $\ketbra{\psi(t_j)}$, where $\Pi_j=\ketbra{\psi(t_j)}{\psi(t_j)}$ and $\mathcal{E}$ is some arbitrary quantum operation which depends on the problem. Therefore at $t_j$ we get  $\ketbra{\psi(t_j)} = M_{t_j}(\ketbra{\psi(t_{j-1})})$, with $M_{t_j}$ being quantum projective measurement operations. 

\begin{lemma}[Zeno effect \cite{Boixo2009eigenpath}]\label{lem:Zeno}
Consider a collection of states along a continuous path $\{\ket{\psi(l)}\}_{l\in[0,L]}$ and assume that for fixed $a$ and any $\delta\in\mathbb{R}$,
\begin{equation}
    \abs{\braket{\psi(l)| \psi(l+\delta)}}^2 \geq 1-a^2\delta^2.
\end{equation}
Then, starting from $\ket{\psi(0)}$  the state $\ket{\psi(L)}$ can be prepared with fidelity $p>0$ with $\lceil L^2 a^2/(1-p)\rceil$ intermediate projective measurement operations.
\end{lemma}
Importantly, one does not need to keep track of the intermediate measurement results. For the QLSP, we would like the final state $\ket{\psi(t=1)}$ to be the ground state of the Hamiltonian in \cref{eq:RandMethod_Ham}.
The intermediate projective measurements required by \cref{lem:Zeno} can be implemented by evolutions under the adiabatic Hamiltonian for random times (for details, see \cite{Boixo2009eigenpath}). 

When choosing the adiabatic path, the natural parametrization is chosen such that the rate of change of the eigenstate is bounded by a constant. The details can be found in the supplementary material of \cite{Subasi2019adiabatic}. The method sketched above provides an algorithm which requires a time complexity of $O(\kappa^2 \log(\kappa)/\varepsilon)$. To improve it to $O(\kappa \log(\kappa)/\varepsilon)$, another family of Hamiltonians is chosen such that the gap is greater, using the gap amplification technique from \cite{Somma2013gapamplification}.

\subsubsection{Time-optimal adiabatic method}\label{sec:time-opt-adiabatic}
Reference~\cite{an_quantum_2020} proposes another quantum linear system algorithm based on adiabatic quantum computation with a time-optimal scheduling function. 
The algorithm closes the gap between the randomization method and the adiabatic quantum computation by showing that a direct implementation of near-adiabatic dynamics without phase randomization suffices for quantum linear system problems. 
The overall query complexity of the algorithm is $O(\kappa \mathrm{poly}(\log(\kappa/\varepsilon)))$, which is similar to that of \cite{childs_quantum_2017} yet without relying on the VTAA subroutine. 

The algorithm first reduces the quantum linear system problem to an eigenstate preparation problem, following the randomization method \cite{Subasi2019adiabatic}. 
Specifically, the algorithm considers a parametrized Hamiltonian $H(t)=(1-f(t))H_0+f(t)H_1$ for $t \in [0,1]$ where the zero-energy eigenstate of $H_1$ encodes the solution of the linear system problem.
We recall from \cref{subsec:adiabaticevolution} that the scalar function $f(t)$ is called the scheduling function.
The corresponding eigenstate of $H_0$ can be simply constructed from $\ket{\mathbf b}$. More specifically, define the Hamiltonians $H_0=\begin{pmatrix}
        0 & Q_{\mathbf{b}}\\
        Q_{\mathbf{b}} & 0
    \end{pmatrix}$ and $H_1=\begin{pmatrix}
        0 & AQ_{\mathbf{b}}\\
        Q_{\mathbf{b}}A & 0
    \end{pmatrix}$, where $Q_{\mathbf{b}}=I-\op{\mathbf{b}}$ and $A$ is defined by the QLSP problem. If $A\ket{\mathbf{x}}=\ket{\mathbf{b}}$ then the the solution to the linear problem $\ket{\mathbf{x}}$ is encoded in the null space of $H_1$. 
To find the desired eigenstate of $H_1$, the algorithm then uses the adiabatic approach which considers the time-dependent Hamiltonian simulation problem 
\begin{equation}
    i \frac{d}{dt}\ket{\psi(t)} = H(t/T) \ket{\psi(t)}, \quad 0 \leq t \leq T.   
\end{equation}
The adiabatic theorem guarantees that this dynamics drives the initial eigenstate $\ket{\psi(0)}$ of $H_0$ to a good approximation of the target eigenstate of $H_1$, as long as $T$ is sufficiently large. 
Since the target eigenstate of $H_1$ actually encodes the solution of the linear system problem, solving this time-dependent Hamiltonian simulation problem for sufficiently large $T$ will solve the quantum linear system problem. 
The algorithm in~\cite{an_quantum_2020} uses the truncated Dyson series method for Hamiltonian simulation \cite{Low2018interaction} to achieve near-optimal time and precision dependence. 

A key parameter affecting the overall query complexity of the algorithm is the adiabatic evolution time $T$, as Hamiltonian simulation algorithms typically have at least linear dependence on $T$ \cite{berry2007efficient,Atia2017,Haah2018}.
If we choose the scheduling function $f(t) = t$, i.e., linear interpolation between $H_0$ and $H_1$, then the adiabatic theorem \cite{jansen2007bounds} implies $T = O(\kappa^3/\varepsilon)$, which is sub-optimal in both $\kappa$ and $\varepsilon$. 
To reduce the scaling of $T$, Ref.~\cite{an_quantum_2020} takes the strategy of designing the scheduling function carefully and finds two choices. 

The first choice, called AQC($p$) scheduling function, named after Adiabatic Quantum Computing dependent on a fixed parameter $p$, tunes the speed of $f(t)$ proportional to the size of the spectral gap of $H(t)$, so the AQC($p$) scheduling function slows down when the time-dependent gap of $H(t)$ is small to control the diabatic error and speeds up when the gap is large to shorten the evolution time. 
The scheduling function $f(t)$ is chosen to satisfy the differential equation 
\begin{equation}
    \dot{f}(t) = c_p \big(\Delta(f(t))\big)^p, \quad f(0) = 0, \quad 1 < p < 2. 
\end{equation}
Here, $\Delta(t)$ is the spectral gap and $c_p$ is a normalization factor such that $f(1) = 1$. 
This differential equation can be explicitly solved as 
\begin{equation}
\label{eq:sched_QLSP}
    f(t) = \frac{\kappa}{\kappa - 1}\left[1-\left(1+t(\kappa^{p-1}-1)\right)^{\frac{1}{1-p}}\right]. 
\end{equation}
With AQC($p$) as scheduling function, the evolution time $T$ scales according to $O(\kappa/\varepsilon)$~\cite{an_quantum_2020}. This is optimal in the condition number $\kappa$. 

To further improve the dependence on the error $\varepsilon$, \cite{an_quantum_2020} proposes a second choice called the AQC(exp) scheduling function by imposing the boundary cancellation condition. This condition says that all the derivatives of $H(f(t))$ vanish at the boundaries $t=0$ and $t=1$.  
The AQC(exp) scheduling function is given by 
\begin{equation}
    f(t) = c_e^{-1} \int_0^t \exp\left(-\frac{1}{t'(1-t')}\right) d t',
\end{equation}
where $c_e = \int_0^1\exp\left(-1/(t'(1-t'))\right) d t'$ is a normalization constant ensuring $f(1) = 1$. 
The AQC(exp) scheduling inherits the adaptive speed property of the AQC($p$) scheduling, but also becomes flat at $s = 0$ and $s = 1$ to exponentially reduce the diabatic error, i.e., $H^{(k)}(0)=H^{(1)}(1)=0$ for all $k\geq 1$, where $H^{k}(t)$ is the $k$-th derivative of $H$ at $t$. 
As a result, the evolution time $T$ is able to follow $O(\kappa \mathrm{~poly}(\log(\kappa/\varepsilon)))$. 
By implementing the Hamiltonian simulation problem using the truncated Dyson series method~\cite{Low2018interaction}, the resulting quantum algorithm with the AQC(exp) scheduling achieves query complexity $O(\kappa \mathrm{~poly}(\log(\kappa/\varepsilon)))$. This is near-optimal in both $\kappa$ and $\varepsilon$. 

\subsection{Trial state preparation and Filtering}\label{subsec:filteringandtrialstate}
The following algorithms achieve (near-)optimal scaling in both the condition number and the solution error. While greatly inspired by adiabatic techniques, we deemed it appropriate to allocate an own section for them. These algorithms are based on the idea of preparing a trial state which can be efficiently transformed into the solution vector of the QLSP. The main idea is to prepare the trial state close to the solution vector (for example with overlap $\Omega(1)$) and then filtering the state to give the solution. Filtering is a technique that allows to project onto a subspace spanned by a subset of the eigenvectors of a matrix, i.e., an implementation of a spectral projection. Some of the methods such as that based on the quantum Zeno effect in \cref{sec:zeno-filtering} and the method in \cref{sec:kernelreflection}  consist in taking a trial state and evolving it as following a path as in the adiabatic method. We include these methods in this section as they also require the initial preparation of a trial state and then implementing a version of eigenstate filtering.
  
\subsubsection{Eigenstate filtering and quantum Zeno effect}\label{sec:zeno-filtering}
The first work to introduce the trial state and eigenstate filtering technique for the QLSP provided two algorithms based on this method \cite{Lin2020optimalpolynomial}.
These algorithms further improved the query complexity,  achieving near optimal complexity for a $\spr$-sparse matrix, $\kappa$ and $\varepsilon$ in $O \left(\spr \kappa \log \left( \frac{1}{\varepsilon} \right) \right)$.
Both algorithms are based on the eigenstate filtering technique which consists of approximating a projector $P_{\lambda}$ onto some eigenspace associated to an eigenvalue $\lambda$. The approximation is carried out using quantum signal processing (QSP) by choosing an adequate polynomial constructed from Chebyshev polynomials.

The first method consists in using the time-optimal adiabatic quantum computation (AQC) method from \cref{sec:time-opt-adiabatic} to prepare the trial state and then using the eigenstate filtering to project the trial state onto the solution of the QLSP. The AQC can be implemented with the truncated Dyson series method \cite{Low2018interaction}. The trial state can be prepared to constant precision, therefore if AQC($p$) is used, the dependence of the runtime on the error can be disregarded. Then the eigenstate filtering can be applied which can be shown to give the solution with success probability $\Omega(1)$.
The second algorithm is based on the quantum Zeno effect (QZE) as used in \cite{Boixo2009eigenpath} and summarized in \cref{sec:adiabatic-rand}. In this case, rather than running the time-dependent evolution obtained from the adiabatic method, a series of projections are implemented giving as a result an evolution along the adiabatic path.
As mentioned above, the query complexity of this algorithm for both methods is near-optimal for a $\spr$-sparse matrix in $\kappa$ and $\varepsilon$, namely $\tilde{O} \left(\spr \kappa \log \left( \frac{1}{\varepsilon} \right) \right)$. In what follows we give a brief explanation of the AQC based algorithm and a more extended discussion of the QZE based algorithm.

Just as in Ref.~\cite{an_quantum_2020}, a Hamiltonian needs to be specified which encodes the solution to the QLSP. Let $H(t)$ be defined as in \cref{sec:time-opt-adiabatic}, which we write as $H=(1-f(t))H_0 + f(t) H_1$. The lower bound on the gap of $H(f(t))$ is given by $\Delta^*(f(t))=1-f(t)+\frac{f(t)}{\kappa}$. We can then run AQC($p$) with constant precision as mentioned before, giving an algorithm with runtime $O(\kappa)$. Finally the eigenstate filtering procedure can be applied, which we detail below.
Then the eigenstate filtering operator $R_{\ell}(\frac{H_1}{\spr};\frac{1}{\spr \kappa})$ is applied where
\begin{align}
    R_\ell (x;\Delta)=\frac{T_\ell\left(-1+2\frac{x^2-\Delta^2}{1-\Delta^2}\right)}{T_\ell \left(-1+2\frac{-\Delta^2}{1-\Delta^2}\right)}
\end{align}
with $T_\ell$ the $\ell$-th Chebyshev polynomial of the first kind. This polynomial has several properties (see Lemma 2 in \cite{Lin2020optimalpolynomial}) which allows to approximate a projector $P_\lambda$ by applying $R_\ell$ to $H-\lambda I$. After the operator $R_\ell (\frac{H_1}{\spr};\frac{1}{\spr\kappa})$ is applied, the solution $\ket{\mathbf{x}}$ is obtained with $\Omega(1)$ success probability. 

We will now show that the QZE-based algorithm also provides a simple digital implementation of the adiabatic evolution, which can yield the nearly optimal query complexity $\tilde{O}(\kappa \log(1/\varepsilon))$~\cite{Lin2020optimalpolynomial}.
 
We choose the following scheduling function
\begin{equation}
\label{eq:scheduling_qze}
f(t)=\frac{1-\kappa^{-t}}{1-\kappa^{-1}}.
\end{equation} 
Consider, as usual in this review, a Hermitian positive definite matrix $A$, and let $\ket{x(f)}$ be a normalized vector such that 
\begin{equation}
\label{eq:def_xf}
((1-f)I+fA)\ket{x(f)}\propto \ket{b}.
\end{equation}
We define a Hamiltonian $H(f)$ along the path as
\begin{equation}
    {H}(f) = \left(\begin{array}{cc}
        0 & ((1-f)I + fA)Q_b \\
        Q_b((1-f)I + fA) & 0
    \end{array}\right) {.}
\end{equation}
Then the null space of $H(f)$ is spanned by $\ket{\psi(f)}=\ket{0}\ket{x(f)}$ and $\ket{1}\ket{b}$. By adding the additional constraint
\begin{equation}
\label{eq:geometric_phase}
\braket{x(f)|\partial_f x(f)}=0,
\end{equation}
the eigenpath $\{\ket{x(f)}\}$ becomes uniquely defined with the initial condition $\ket{x(0)}=\ket{b}$. 

Define the eigenpath length $L(a,b)$ between $0<a<b<1$ as
\begin{equation}
L(a,b) = \int_a^b \|\partial_f \ket{x(f)}\| \, \mathrm{d}f,
\end{equation}
and it is upper bounded by
\begin{equation}
\label{eq:bound_segment_eigenpath_length}
L(a,b)\leq \int_a^b \frac{2}{\Delta^*(f)}\mathrm{d} f= 
\frac{2}{1-1/\kappa}\log\left(\frac{1-(1-1/\kappa)a}{1-(1-1/\kappa)b}\right)=:L_{*}(a,b).
\end{equation}
In particular, the upper bound for the entire path is given by
\begin{equation}
L(0,1)\le L_*(0,1)=\frac{2\log(\kappa)}{1-1/\kappa}.
\end{equation}
We then have\begin{equation}
|\braket{x(f_j)|x(f_{j-1})}| \geq 1-\frac{1}{2}\|\ket{x(f_{j-1})}-\ket{x(f_j)}\|^2\geq 1-\frac12 L_*(f_{j-1},f_j)\ge 1-\frac{2\log^2(\kappa)}{M^2(1-1/\kappa)^2}.
\end{equation}

To bound the success probability, for simplicity, we assume all block-encodings are implemented exactly. Then if we choose $M\geq \frac{4\log^2(\kappa)}{(1-1/\kappa)^2}$, the success probability satisfies
\begin{equation}
 \prod_{j=1}^{M}\|P_{f_j}\ket{\psi(f_{j-1})}\|^2 = \prod_{j=1}^{M}|\braket{x(f_j)|x(f_{j-1})}|^2\geq \left(1-\frac{2\log^2(\kappa)}{M^2(1-1/\kappa)^2}\right)^{2M} \geq \left( 1-\frac{2\log^2(\kappa)}{M(1-1/\kappa)^2} \right)^2\geq \frac{1}{4}.
\end{equation} 
A more careful analysis shows that if the projection $P_{f_j}$ can only be approximately implemented to precision $\epsilon_i$, then it is sufficient to choose $\varepsilon_1=\ldots=\varepsilon_{M-1}=O(M^{-2})$, and $\varepsilon_M=\varepsilon$, so that the success probability is $\Omega(1)$, and the trace distance between the final state and $\ket{0}\ket{x}$ is $O(\varepsilon)$.

The spectral projector can be implemented using an even approximation to the rectangular function~\cite{gilyen2019quantum}. The number of queries to $A$ needed for implementing $U_{P_j}\in BE_{1,a}(P_j,\varepsilon_j)$ is $O(\Delta_*(f_j)^{-1} \log \varepsilon_j^{-1})$. Along the path, the number of queries for the first $M-1$ steps of the projection is of order 
\begin{equation}\label{eqn:calc_qze_path1}
\begin{aligned}
\log\left(\frac{1}{\varepsilon'}\right)\sum_{j=1}^{M-1}\frac{1}{1-f(s_j)+f(s_j)/\kappa} &\leq \log\left(\frac{1}{\varepsilon'}\right)M\int_0^1 \frac{1}{1-f(s)+f(s)/\kappa}\mathrm{d} s \\
&=\log\left(\frac{1}{\varepsilon'}\right)M\int_0^1 \kappa^{s}\mathrm{d} s\\
&\le \log\left(\frac{1}{\varepsilon'}\right)M \frac{\kappa}{\log \kappa}.
\end{aligned}
\end{equation}  
Plug in $M=\Theta(\log^2 \kappa), \varepsilon'=O(M^{-2})=O(\log^{-4}\kappa)$, we find that the number of queries to $A$ is $O(\kappa\log\kappa \log\log \kappa)$.
The last step of the projection should be implemented to precision $\varepsilon$, and the number of queries to $A$ is $O(\kappa \log(1/\varepsilon))$. For practical purposes, $\log\log \kappa$ can be treated as a constant (e.g., $\log\log 10^{12}\approx 3.3$). So neglecting $\log\log \kappa$ factors, the total query complexity of the algorithm is thus $O(\kappa \log(\kappa/\varepsilon))$. 

The query complexity of the algorithm above can be slightly improved to remove the $\log \kappa$ factor. At each step of the algorithm,  $\|P_{f_j}\ket{\psi(f_{j-1})}\|$ can be slightly smaller than $1$. Therefore $M$ needs to be chosen to be $O(\log^2(\kappa))$ to ensure that the final success probability is $\Omega(1)$. However, by choosing $M=\Theta(\log(\kappa))$, it is already sufficient to guarantee
\begin{equation}
\|P_{f_j}\ket{\psi(f_{j-1})}\| = |\braket{x(f_j)|x(f_{j-1})}|\geq 1-\frac{2\log^2(\kappa)}{M^2(1-1/\kappa)^2}=\Omega(1).
\end{equation}
So we may use the fixed point amplitude amplification in Refs.~\cite{grover2005fixed, Yoder2014fixed}
to prepare a state $\ket{\widetilde{\psi}(f_{j})}$ so that 
\begin{equation}
|\braket{\widetilde{\psi}(f_j)|0,x(f_{j-1})}|\ge 1-\frac{1}{M}.
\end{equation}
This process uses the block-encoding of $P_{f_j}$ for $O(\log(M))=O(\log\log \kappa)$ times. The overall success probability after $M$ steps is lower bounded by $(1-M^{-1})^M\approx e^{-1}$. 

Repeating the calculation in \cref{eqn:calc_qze_path1}, we find that the number of queries for the first $M-1$ steps of the projection is proportional to 
\begin{equation}
\begin{aligned}
(\log M)\left(\log\varepsilon'^{-1}\right)\sum_{j=1}^{M-1}\frac{1}{1-f(s_j)+f(s_j)/\kappa} &\leq (\log M)\left(\log\varepsilon'^{-1}\right)M\int_0^1 \frac{1}{1-f(s)+f(s)/\kappa}\mathrm{d} s \\
&\le (\log M)\left(\log\varepsilon'^{-1}\right) \frac{M\kappa}{\log \kappa}.
\end{aligned}
\end{equation}  
Plug in $M=\Theta(\log \kappa), \varepsilon'=O(M^{-2})=O(\log^{-2}\kappa)$, we find that the number of queries to $A$ is $O(\kappa (\log\log \kappa)^2)$. So neglecting $\log\log \kappa$ factors, the total query complexity of this improve algorithm is thus $O(\kappa \log(1/\varepsilon))$. 

\subsubsection{Discrete adiabatic method}\label{discrete-adiabatic}
The algorithm using the discrete adiabatic method proposed in Ref.~\cite{costa_optimal_2021} achieves optimal scaling $O(\kappa \log(1/\varepsilon))$ in terms of the query complexity in $\varepsilon$ and $\kappa$. The adiabatic methods for solving quantum linear systems, such as that in \cref{sec:time-opt-adiabatic}, are based on continuous adiabatic theorems.

The adiabatic theorem indicates the optimal rate for updating the Hamiltonian from $t=0$ in order to minimize the error between the actual, $\ket{\widetilde{\psi}(t)}$, and the ideal, $\ket{\psi(t)}$, eigenstate of $H(t)$. Given a scheduling function as introduced in \cref{subsec:adiabaticevolution}, the adiabatic theorem then explores properties of the Hamiltonian. Important studied properties are the gap $\delta(f(t))$ between the eigenvalue of the desired eigenstate and the rest of the spectrum of $H(t)$, and its first and second derivative with respect to the parameter $t$, i.e., $H^{(k)}(t)\coloneqq d^k H(f(t))/dt^k$ for $k=1,2$. More concretely, from Ref.~\cite{jansen2007bounds}, the ideal evolution can be analyzed by building the ideal adiabatic Hamiltonian $H_A(t)$, defined as
\begin{equation}
\label{eq:Adiab_Ham}
H_A(t) = H(t) + \frac{i}{T}[\dot{P}(t), P(t)],    
\end{equation}
where $T$ is the parameter called the runtime of AQC and
\begin{equation}
P(t) = \frac{1}{2\pi i} \oint _{\Gamma (t)}\left(H(t) - z\right)^{-1}dz,
\end{equation}
is the resolvent operator of $H(t)$ that returns the projection over the desired eigenstates by choosing a suitable contour $\Gamma (t)$. The evolution of the ideal Hamiltonian $H_A(t)$ defines the adiabatic unitary $U_A$ that describes the ideal evolution. We can quantify how much the ideal eigenstate deviates from the actual state by computing the following difference,
\begin{equation}
\label{eq:Diffe_bound}
\eta(t) = \|U(t)P_0U^{\dagger}(t) - U_A(t)P_0U_A^{\dagger}(t)\|,   
\end{equation}
where $P_0 = \ket{\psi(0)}\bra{\psi(0)}$. By exploring properties of the resolvent and the adiabatic operators, namely $H_A$ and $U_A$, and using several inequalities and approximations, Theorem 3 of Ref.~\cite{jansen2007bounds} provides an upper bound for $\eta(t)$,
\begin{align}
 \label{eq:cont_adiabatic}
 \eta(t) &\leq \frac{1}{T} \frac{m(t)\|H^{(1)}(0)\|}{\delta^2(f(0))} +  \frac{m(t)\|H^{(1)}(t)\|}{\delta^2(f(t))}\nonumber\\
 &\quad+  \frac{1}{T}\int_0^t dt'\; \left(\frac{m(t')\|H^{(2)}(t')\|}{\delta^2(t')} + 7m(t')\sqrt{m(t')}\frac{\|H^{(1)}(t')\|}{\delta^3(f(t'))}\right),    
\end{align}
which depends on the scheduling function as in \cref{eq:Ham_prob}. The Hamiltonian $H(t)$ restricted
to $P(t)$ consists of $m(t)$ eigenvalues separated by the Hamiltonian gap $\delta(t)$. 

When applying the quantum adiabatic theorem \cref{eq:cont_adiabatic} from Ref.~\cite{jansen2007bounds} to the QLSP, the idea is that $H(t)$ is constructed by the interpolation Hamiltonians that embed the QLSP, $A\ket{x}=\ket{b}$:
\begin{equation}
H_1 =  \begin{pmatrix}
0 & AQ_b \\
Q_bA & 0
\end{pmatrix}.  
\end{equation}
Here, $Q_b=I -\ket{b}\bra{b}$. By substituting the Hamiltonian for the quantum linear system problem into the adiabatic theorem \cref{eq:cont_adiabatic}, a straightforward expression for the gap of $H(t)$ can be derived in terms of the condition number $\kappa$ of $A$ and the scheduling function
\begin{equation}
\delta(f(t)) \geq 1- f(t) +f(t)/\kappa.   
\end{equation}
E.g., this is stated in Ref.~\cite{an_quantum_2020}. In the context of the QLSP, we target an approximation error $\varepsilon$ such that $\eta(1)\leq \varepsilon$.
The adiabatic theorem states that the total evolution time is linear in the condition number and inversely proportional to the target error of the solution, namely $T=O(\kappa/\varepsilon)$. However, there is a logarithmic overhead arising from approximating the time evolution of the time-dependent Hamiltonian using the truncated Dyson series, resulting in a dependence of $O(\kappa \log(\kappa))$.

We first recap complexities of solving the QLSP based on AQC using the continuous adiabatic theorem. In Ref.~\cite{Lin2020optimalpolynomial}, AQC is applied together with eigenstate filtering, which yields a nearly optimal dependence in $\kappa$ and an optimal dependence on the target error of the solution. The fundamental concept of using eigenstate filtering to solve the QLSP involves initially executing the AQC while aiming for a constant precision in the solution error. This approach results in a runtime of $O(\kappa)$ and a logarithmic overhead in $\kappa$ in the query complexity due to emulating Hamiltonian simulation via the truncated Dyson series. Next, the eigenstate filtering algorithm~\cite{Lin2020optimalpolynomial}  achieves a query complexity of $\tilde O(\kappa\log(1/\varepsilon))$, which is optimal in the solution error, and the total complexity is near-linear in $\kappa$. 
In what following, we demonstrate how optimal dependence in the combination of the parameters $\kappa$ and $\varepsilon$ for solving the QLSP can be achieved by considering the AQC based on the discrete adiabatic theorem in conjunction with eigenstate filtering. 

In the discrete version of the adiabatic theorem, the complexity analysis studies the properties of the unitary operator that is applied to move  in discrete steps from to the initial state the final state.
This is in contrast to the continuous adiabatic theorem, where properties of the Hamiltonian are used to infer how long it takes to ``move'' from the initial to the final state.
More formally, the model of the adiabatic evolution is based on a sequence of $T$ \textit{walk operators} $\{W_T(n/T): n \in \mathbb{N}, 0\leq n \leq T-1\}$.
That is, if the system is initially prepared in a state $\ket{\psi_0}$, then the sequence of unitary transformations $W_T(n/T)$ effectuates that
$\ket{\psi_0} \mapsto \ket{\psi_1} \mapsto \cdots$.
To model this evolution with $t := n/T$, we can write
\begin{equation}
U_T(t) = \prod_{n = 0}^{tT-1} W_T\left(n/T\right);
\end{equation}
with $U_T(0) = I$, this means that $\ket{\psi_n} = U_T(t)\ket{\psi_0}$. 
Now, relevant properties of the overall ``adiabatic walk'' are the strict upper bounds of the multistep differences of the walk operator $W_T$,  
\begin{equation}\label{eq:ineq_Dk}
\left\|D^{(k)} W_T(t)\right\| \leq \frac{c_k(t)}{T^{k}}.
\end{equation}
For $k=1$, we have $D^{(1)}W_T(t) =
W_T \left( t + \frac{1}{T} \right) - W_T (t)$ and for $k>1$,
\begin{equation}
 D^{(k)}W_T(t)=D^{(k-1)}W_T(t+\frac{1}{T})-D^{(k-1)}W_T(t).   
\end{equation}
In \cref{eq:ineq_Dk}, the assumption is that $W_T(t)$ is a smooth operator such that $c(t)$ can be chosen independently of $T$. Another important quantity is the gap $\Delta(t)$. Since the discrete adiabatic theorem deals with unitary operators, we know that the relevant eigenvalues to consider lie on the complex unit circle. The projector $P$ defines a set of eigenvalues $\sigma_P$, and we call the eigenvalues associated to the complementary projector $Q$ $\sigma_Q$. The gap $\Delta$ is defined as the minimum arc distance between the eigenvalues in $\sigma_P$ and $\sigma_Q$.

For the continuous adiabatic theorem in \cref{eq:cont_adiabatic} we refer to Theorem 3 in Ref.~\cite{costa_optimal_2021}. The discrete version goes as follows,
\begin{align}
\label{eq:disc_Ad}
\quad \|U_T(t) - U_T^{A}(t)\| 
& \leq \frac{12\hat{c}_1(0)}{T\check{\Delta}(0)^2}+  \frac{12\hat{c}_1(t)}{T\check{\Delta}(t)^2} + \frac{6\hat{c}_1(t)}{T\check{\Delta}(t)} + 305\sum_{n=1}^{tT-1}\frac{\hat{c}_1(n/T)^2}{T^2\check{\Delta}(n/T)^3} \nonumber \\
& \quad  + 44\sum_{n=0}^{tT-1} \frac{ \hat{c}_1(n/T)^2}{ T^2 \check{\Delta}(n/T)^2} +  32\sum_{n=1}^{tT-1}\frac{\hat{c}_2(n/T)}{T^2\check{\Delta}(n/T)^2} 
\end{align}
In the expression above, $U^A_T$ is the ideal adiabatic unitary, where we do not explicitly state the dependency on the adiabatic Hamiltonian, as done in the continuous version \cref{eq:Adiab_Ham}.
The quantity $\hat{c}_1(t)$ is defined as the maximum value among the neighbouring time steps $t$, i.e., the maximum in $\{c_1(t-1/T), c_1(t), c_1(1+1/T)\}$. Similarly, the notation $\check{\Delta}(t)$ denotes the minimum between neighbouring time steps.
Comparing the discrete \cref{eq:disc_Ad} with the continuous adiabatic theorem \cref{eq:cont_adiabatic}, we can identify direct continuous-discrete correspondances in the expressions. That is, the function $c_k(t)$ is a discrete representation of the derivatives in the continuous version, i.e., $H^{(k)}(t)$. The sum expressions in the last three terms in \cref{eq:disc_Ad} serve as the discrete analog of the integration part in \cref{eq:disc_Ad}. Additionally, we observe that the same ratio order related to the gap in the continuous formulation is present in the discrete version as well.

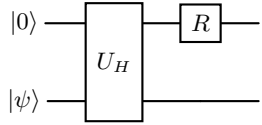
\begin{figure}[H]
\centering
\begin{quantikz}
\ket{0}\;&\gate[2]{U_H}&\gate{R}&  \\
\ket{\psi}\;&& &
\end{quantikz}
\caption{Circuit representation for the qubitized quantum walk.}
\label{figure:QW_circuit}
\end{figure}

Ref.~\cite{costa_optimal_2021} utilizes the qubitized walk $W$ to implement the unitary operators required for the discrete adiabatic theorem. This operator, first analyzed in Ref.~\cite{szegedy2004quantum}, has since been the subject of additional studies focused on block-encoding~\cite{berry2018improved}. We provide a high level depiction of $W$ in \cref{figure:QW_circuit}. It differs from the block-encoding operator of $H\in \mathbb{C}^{n\times n}$, $U_H$, by incorporating a reflection about the ancilla qubit (or qubits) given by $R=2\ket{0}\bra{0}\otimes I_n -I$, contrary to projective measurement in the block-encoding.

The spectral analysis of this operator within the relevant two-dimensional subspace  of the block-encoding indicates that it behaves similarly to
\begin{equation}
    W  = e^{\pm i\arccos{(H)}}.
\end{equation}
This emphasizes why the qubitized walk is a good unitary choice for solving the QLSP. By considering the Hamiltonian used to solve the QLSP with the continuous AQC, given in \cref{eq:Ham_prob}, we can construct the quantum walk operator $W_T(t)$. 
This approach enables us to map all eigenvalues and the spectral gap of $H(t)$ onto $W_T(t)$ and allows the application of the discrete adiabatic theorem. In this framework, we perform the analysis at the level of the unitary operator $W_T(t)$ and its gap. This greatly simplifies the evolution as it removes the need for `compiling' the evolution via the truncated Dyson series and consequently eliminates the logarithmic complexity overhead in the adiabatic component. Similarly to the continuous approaches, the algorithm concludes with the eigenstate filtering algorithm. In the discrete case, this is simpler as it is not necessary to search for angles to perform QSVT; then, the number of walk steps are $O(\kappa \log(1/\varepsilon))$.

\subsubsection{Augmentation and kernel reflection}\label{sec:kernelreflection}
As outlined previously, the strategy followed in \cref{sec:zeno-filtering} and \cref{sec:adiabatic-rand} consists in, first, using an adiabatic procedure to prepare some trial state which has a  constant overlap with the solution of the QLSP, and second, using the eigenstate filtering method to project the trial state onto the solution.
Following these lines, Ref.~\cite{dalzell2024shortcutoptimalquantumlinear} proposes a modified strategy which bypasses the need to go through the rather involved analysis of the adiabatic methods.

The starting point for this method is  augmenting the linear system with an extra, scalar variable $t$. This extra variable corresponds to an estimate of the norm of the solution $\norm{\mathbf{x}}$. If $t$ estimates the norm up to a multiplicative factor, then it can be shown that an easy-to-prepare initial state can be transformed into the correct solution using QSVT with an optimal query complexity of $O(\kappa \log(1/\varepsilon))$. In general, one would not expect to know the norm of the solution before solving the problem. Therefore an algorithm scaling linearly with $\kappa$ to estimate this norm is provided in Ref.~\cite{dalzell2024shortcutoptimalquantumlinear}.
The augmentation of the system is done as follows. The original system is by a matrix $A\in\mathbb{C}^{N\times N}$ with $N=2^n$. Then, one defines a qubit system with $s=\lceil \log_2 (1+N)  \rceil$ qubits. This increases the dimension of the linear system by $1$ and defining a new matrix with the increased dimension which depends on $A$.

By increasing the dimensionality of the system, there is a new orthogonal vector to the original basis which can be used as an ansatz. In particular, this vector will be orthogonal to the solution of the original linear problem $A\mathbf{x}=\mathbf{b}$.  By using the so-called kernel reflection method that Ref.~\cite{dalzell2024shortcutoptimalquantumlinear} introduces based on QSVT, the state can be rotated to the solution.

We continue by describing the algorithm in more detail. Let $\ket{\mathbf{e}_0},\cdots, \ket{\mathbf{e}_N}$ be an orthonormal basis of $\mathbb{C}^{N+1}$. 
The matrix $A$ is given as input through a $(1,a)$-Block-encoding $U_A$. The matrix $A$ is augmented to $A_t \in \mathbb{C}^{(N+1)\times (N+1)}$ defined as 
$$A_t = A+t^{-1}\ketbra{\mathbf{e}_N}{\mathbf{e}_N}, \quad t\in [1,\kappa] .$$ 
As mentioned before, the role of the variable $t$ is to be an estimation for $\norm{\mathbf x}$. As will be seen later when $t=\norm{\mathbf{x}}$, the operation to obtain the solution will be very simple and will require a constant number of repetitions.

We now can define a new QLSP of the form $A_t \mathbf{x}_t=\mathbf{b}'$ where $\mathbf{b}'=\frac{1}{\sqrt{2}}(\mathbf{b}+\ket{\mathbf{e}_N})$ and $\mathbf{x}_t=\frac{1}{\sqrt{2}}(\mathbf{x}+t\ket{\mathbf{e}_N})$. This new equation holds true by construction so that $\mathbf{x}$ is orthogonal to $\ket{\mathbf{e}_N}$: 
Let $\theta_t = \arctan{\frac{\norm{\mathbf{x}}}{t}}$, then we can write
\begin{align}
\ket{\mathbf{e}_N}=\cos(\theta_t)\ket{\mathbf{x}_t}+\sin(\theta_t)\ket{\mathbf{y}_t}
\end{align}
where $\mathbf{y}_t$ is a vector orthogonal to $\mathbf{x}_t$ and in the plane spanned by $\mathbf{x}$ and $\mathbf{e}_N$. 
Note that if $t=\norm{\mathbf{x}}$, then the angle between $\ket{\mathbf{e}_N}$ and $\ket{\mathbf{x}_t}$ is $\frac{\pi}{4}$, i.e., reflecting $\ket{\mathbf{e}_N}$ around $\ket{\mathbf{x}_t}$ would bring the vector to $\ket{\mathbf{x}}$.

To reflect $\ket{\mathbf{e}_N}$ around $\ket{\mathbf{x}_t}$, a technique called kernel reflection is used. The motivation behind this technique is to increase the overlap of the state with $\ket{\mathbf{x}}$.
The general idea for kernel reflection is as follows. Let $G$ be some operator and $\ket{w}$ a vector in the kernel of $G$. Starting from a state $\alpha\ket{w}+\beta\ket{w_{\perp}}$, kernel reflection maps this to 
$ \alpha\ket{w}+\beta\ket{w_{\perp}} \to    \alpha\ket{w}-\beta(1-\delta_1)\ket{w_{\perp}}+\beta\delta_2\ket{w'_{\perp}}.$
The new amplitudes, dependent on $\delta_1$ and $\delta_2$, ensure that the overlap with $\ket{w}$ increases and $\ket{w'_\perp}$ is orthogonal to the kernel and to $\ket{w}$. For more details we refer to Appendix~B~in Ref.~\cite{dalzell2024shortcutoptimalquantumlinear}). Such a reflection can be implemented using QSVT based on a block-encoding of $A$.

We consider the operator $G_t=Q_{\mathbf{b}'}A_t$ where $Q_{\mathbf{b}'}=I-\ketbra{\mathbf{b}'}{\mathbf{b}'}$. Note that $\ket{\mathbf{x}_t}$ is in the kernel of $G$ and therefore applying the kernel reflection on $\ket{\mathbf{e}_n}$ increases the overlap with $\ket{\mathbf{x}}$. Finally, the resulting state is projected onto the span of $\{\ket{\mathbf{e}_j}\}_{j=0}^{N-1}$. It can be shown that if the norm of $\mathbf{x}$ is known up to some constant factor ($t\in [c^{-1}\norm{\mathbf{x}},c \norm{\mathbf{x}}]$ with $c$ constant), then the query complexity to perform this algorithm is $O(\kappa \log(1/\varepsilon))$. Though, access to an estimate for the norm of the solution is non-trivial. The next paragraph discusses the algorithm in Ref.~\cite{dalzell2024shortcutoptimalquantumlinear} to determine such an estimate.

\vspace*{.6em}
\paragraph{Estimation of the solution norm}
As per the assumptions, we are promised that $1 \leq \norm{\mathbf{x}}\leq \kappa$. Thus, one possible approach is an exhaustive search for an additive $\ln(2)$-approximation to $\ln(\norm{\mathbf{x}})$. To perform this search, all the values from $\mathcal{T}=\{0,\,\ln(2),\,2\ln(2),\ldots,\,\lceil \log_2(\kappa)\rceil \ln(2)\}$ are tested sequentially using the algorithm based on the kernel reflection explained previously, until a solution has been found. It can be shown that with high probability the returned candidate solution will be a successful approximation. The query complexity for such a search is $O(\kappa \log(\kappa) \log\log(\kappa))$. 

To improve this dependence on the condition number, an alternative is   binary search rather than exhaustive search. To do so, we need a method to determine when a candidate value for $t$ is too large or too small. This can be done by modifying the algorithm described previously and replacing the kernel reflection by a so-called kernel projection. The kernel projection procedure gives a similar transformation as the kernel reflection. Given a state  $\alpha\ket{w}+\beta\ket{w_{\perp}}$, the kernel projection returns the state $\alpha\ket{w}+\beta\delta_1\ket{w_{\perp}}+\beta\delta_2\ket{w'_{\perp}}.$
This has success probability $\abs{\alpha}^2 + \abs{\nu}^2 (\delta_1^2+\delta_2^2)$. As in kernel reflection, the $\delta_1$ and $\delta_2$ are such that the overlap with the kernel is increased. A crucial difference with respect to kernel reflection is that when implementing kernel projection for operator $G_t$, the success probability increases monotonically with $t/\norm{\mathbf{x}}$, achieving a value of $1/2$ when $t=\norm{\mathbf{x}}$ and is going to $1$ as $t/\norm{\mathbf{x}}\to \infty$. We remark that for simplicity we have omitted a few parameters appearing in the success probability.
 
Since the kernel projection gives a method to detect closeness of $t$ to $\norm{\mathbf{x}}$, we can use it for a binary search procedure. Starting from the set $\mathcal{T}$ as above, the median of this set is computed and rounded to the closes element in $\mathcal{T}$. If the value estimated is greater than $1/2$, the lower half of the candidate set is eliminated and the search continues with the rest and similarly for the case when the estimation is smaller than $1/2$. By this procedure, at least a third of the candidate set is eliminated and therefore by repeating this procedure with the rest of the candidates gives that $O(\log \abs{\mathcal{T}})$ estimations are required. The cost of running kernel projections is linear in $\kappa$ and each kernel projection is run $O(\log \log \log \kappa)$ times, giving a query complexity of $O(\kappa \log \log(\kappa) \log \log \log(\kappa))$. Note that this already gives an algorithm close to optimal.

\vspace*{.6em}
\paragraph{Algorithm with optimal scaling}
To obtain an algorithm with optimal scaling in $\kappa$ and $\varepsilon$, an adiabatic-inspired variation of the above algorithm is considered. The super-linear cost in $\kappa$ for finding an appropriate $t$ comes from the size of the search space, namely $\abs{\mathcal{T}}=O(\log(\kappa))$. If the search space is constant instead,  it can be shown that the algorithm for solving QLSP can be linear in $\kappa$ as it would only depend on the cost of implementing the filtering procedure.
A parametrized family of matrices is defined with increasing condition number such that the norm does not change by more than a constant factor from one member of the family of linear systems to the next. This allows to consider a search space $\mathcal{T}$ of constant size.

Consider a parameter $\sigma\in [\kappa^{-1},1]$ and $f(\sigma)=\sqrt{\frac{\sigma^2 \kappa^2 -1}{\kappa^2-1}}$ which is monotonically increasing and such that $f(\kappa^{-1})=0$ and $f(1)=1$.
Then, the aforementioned parametrized family of matrices is chosen to be $\bar{A}_\sigma$ so that $A$ gives a family of linear systems $\bar{A}_\sigma \overline{\mathbf{x}}_\sigma = \overline{\mathbf{b}}$ where $\overline{\mathbf{b}}$ is a vector of length $2^s$ that corresponds to the vector $\mathbf{b}$ padded with $2^{s-n}$ zeros.
Omitting details about the padding, the parametrized family of matrices $A_\sigma$ is defined as a $(s+1)$-qubit operator
$$\bar{A}_\sigma= \sqrt{1-f(\sigma)^2}\ketbra{0}{0}\otimes A + f(\sigma)\ketbra{0}{1}\otimes I_n.$$

The ``adiabatic-like evolution'' is started at  $\sigma=1$. Then, as $\sigma$ is decreased towards $\kappa^{-1}$, more information about the matrix $A$ is introduced. At the same time, as $\sigma$ decreases, the condition number of $\bar{A}_\sigma$ increases. In fact, it is shown that the condition number for a given $\sigma$ is bounded by $\sigma^{-1}$.
Furthermore, the solution $\overline{\mathbf{x}}_\sigma$ satisfies that $\norm{\overline{\mathbf{x}}_1}=1$, $\overline{\mathbf{x}}_{1/\kappa}=\norm{\mathbf{x}}$ and crucially, for $\kappa^{-1}\leq \sigma \leq \sigma' \leq 1$, we have that $1\leq\frac{\norm{\overline{\mathbf{x}}_\sigma}}{\norm{\overline{\mathbf{x}}_{\sigma'}}}\leq \sigma'/\sigma$. This means by following this evolution, if we approximate the norm of $\overline{\mathbf{x}}_\sigma$ as $\sigma$ decreases, then an approximation to the norm of $\norm{\mathbf{x}}$ can be obtained. By the third property, the ratio between the norms is bounded by the quotient between parameters, this will allow to sequentially approximate the norm and obtain a constant factor approximation for the norm of the solution. 

Specifically, the proposed algorithm sequentially approximates the sequence of norms $\norm{\overline{\mathbf{x}}_{2^{-j}}}$ for $j=0,1,\cdots, \log_2(\kappa)$. A sequence of estimations for the norm $t_0,\cdots, t_{\log(\kappa)}$ is obtained, such that each estimate is a multiplicative approximation of the corresponding norm. Note that due to the properties mentioned before -- namely, that the ration between two norm estimates lies between 1 and the inverse ratio of their associated parameters --  we have that 
$$1\leq \frac{\norm{\overline{\mathbf{x}}_{2^{-j}}}}{\norm{\overline{\mathbf{x}}_{2^{-(j-1)}}}}\leq 2.$$
Therefore, if $\overline{\mathbf{x}}_{2^{-(j-1)}} \in [t_{j-1}/2,2 t_{j-1}]$ then $\overline{\mathbf{x}}_{2^{-j}} \in [t_{j-1}/2,4 t_{j-1}]$ -- based on an initial estimate, we can deduce the following ones. This means that the search for an estimate of $\overline{\mathbf{x}}_{2^{-j}}$ is sufficient within a constant-sized interval, hence we can proceed with a binary search with search space $\abs{\mathcal{T}}=O(1)$. When the kernel projection is run for some $\sigma=2^{-j}$, the condition number is bounded by $2^j$ and therefore the query complexity will be given by $O(2^j)$. To achieve the final, optimal scaling, a step to amplify the probability of success is required which is done using the so-called ``log log trick'' \cite{kothari2023estimation}. This amplification then allows to increase the query complexity for a fixed $j$ to $O(2^j (1+\log_2(\kappa)-j))$. Finally, the total query complexity for the estimation of the norm is given by $\sum_{j=1}^{\log_2(\kappa)}O(2^j (1+\log_2(\kappa)-j))\leq O(\kappa)$.

\section{Rereading the Fine Print}\label{subsec:rereading}
In the critique commentary entitled ``Read the Fine Print''~\cite{aaronson_read_2015} on the assumptions and performance of the HHL algorithm, Scott Aaronson distills a set of five following main caveats. As it has been nearly ten years since this work was published, we revisit the noted caveats with respect to the latest quantum linear systems algorithms and take stock of the current state of the art. 
In particular, we want to base our discussion on the ``checklist of caveats'' put forward in Ref.~\cite{aaronson_read_2015}:
\begin{enumerate}
    \item Initial state preparation is required for all algorithms that solve QLSP. For the procedure to maintain a quantum speedup, preparing the initial state needs to be efficient. State preparation of arbitrary vectors in general is exponentially hard. Thus, in the worst case, for a $2^n$-dimensional vector, this scales exponentially in $n$ as well. This is a severe limitation if there is no structure in the RHS $\mathbf{b}$ that can be exploited and will likely remain a constraint for the foreseeable future.
    \item Loading the data into quantum random access memory (QRAM) \cite{giovannetti2008quantum} and accessing it efficiently. Presently, QRAM is one of the most notorious aspects of quantum computation. A non-trivial assumption of the HHL is the presence of QRAM, and that the vector $\mathbf{b}$ is loaded into QRAM efficiently. Nevertheless, post-HHL solvers do not rely on utilizing QRAM in their solution. Therefore this caveat is no longer a concern in newer algorithms.
    \item Efficient Hamiltonian simulation of the matrix $A$. The simulation of sparse, Hermitian matrices $H$ with favorable norm growth, i.e., $\norm{H}=\text{poly}(n)$ is generally known to be efficient. This means it has polynomial dependency in relevant input quantities \cite{aharonov2003adiabaticquantumstategeneration, berry2007efficient, childs_simulating_2011}. Hamiltonian simulation techniques have also improved since 2015 via methods such as quantum signal processing \cite{low2017optimal} and linear combination of unitaries \cite{low2016methodology}. However, the main concern that efficiency is tied to sparsity remains unchanged remains even with the advancement of block-encoding based techniques.
    \item Scaling in the condition number $\kappa$. Compared to some classical iterative methods like Conjugate Gradients, the lower bound for quantum algorithms to solve QLSP is $\kappa$ as opposed to $\sqrt\kappa$ (cp. discussion in \cref{sec:optimal} and Ref.~\cite{costa2024discreteadiabaticquantumlinear}). Since 2015, the quantum query lower bound here has been shown and establishes a separation in complexity in this quantity towards classical computing. However, the quantum query complexity and time complexity in the convergence theorem of CG are not directly comparable. Furthermore, it is likely that if a quantum speedup is maintained through input state preparation and measurement, the worse scaling in the condition number might be compensated. For instance, it is known that the condition number of discretized Laplacians over a rectangular grid, as often appearing in the solution of PDEs, grows quadratically with $N$ for $N$ gridpoints \cite{bagherimehrab2023fast}. 
    Ideas on preconditioning have also been explored in various contexts, entailing efficient transformations of the QLSP \cite{tong_fast_2021,clader2013preconditioned,bagherimehrab2023fast} (cp. a discussion in the context of differential equations \cref{subsub:de-stationary}, and \cref{sec:optimal}).
    \item Lastly, and also very crucial, maintenance of quantum speedups for linear systems algorithms is impossible if we are interested in reading out the entire solution vector --- with length $2^n$, state tomography will scale on this order. Therefore, applications are limited to the case where a lower-dimensional quantity is of interest, such as the expectation values of observables of interest. Efficient algorithms have been developed to address these problems, see \cite{knill2007optimal,alase_tight_2022,huggins2022nearlyoptimal}.
\end{enumerate}

\section{Remarks on optimal scaling and constant factors}\label{sec:optimal}

\subsection{Optimal scaling}\label{subsec:optimal-scaling}
When it comes to approaching problems algorithmically, an interesting question is what is the minimum complexity, be it query or gate complexity, that is needed to solve an as large as possible class of problem instances --- in other words, finding a lower bound on the complexity. In this section we address the question of lower bounds for algorithms solving the QLSP.
The original work on the HHL algorithm gives a lower bound for the matrix inversion problem  \cite{harrow_quantum_2009} in form of a linear lower bound for $\kappa$. Concretely, it was shown that the problem of simulating a poly-depth quantum circuit and measuring an output qubit is reducible to the problem of producing the state $A^{-1}\ket{\mathbf{b}}$ and measuring an output qubit --- for example measuring a 1-qubit observable. As a result of this reduction, it is shown that if there is an algorithm for the QLSP that runs in time $O(\kappa^{1-\delta} \poly\log(N))$ with $\delta>0$, then $\mathsf{BQP}=\mathsf{PSPACE}$ \footnote{Although we do not go into the details of complexity theory, for completeness we briefly comment on complexity classes. $\mathsf{BQP}$ and $\mathsf{PSPACE}$ correspond to classes of decision problems, i.e., subsets of the set of all bitstrings $\{0,1\}^*$. Given such set $L$, an algorithm must decide if some particular bitstring $z$ is in $L$ or not. The class $\mathsf{BQP}$ corresponds to problems that can be decided in polynomial time by quantum algorithms and $\mathsf{PSPACE}$ corresponds to problems decidable in polynomial space. A more detailed exposition on these definitions and background on theoretical computer science can be found in Ref.~\cite{Arora2009}}, which is widely believed to be false. 

By a small modification of the reduction in Ref.~\cite{harrow_quantum_2009}, a joint lower bound $\Omega(\kappa \log(1/\varepsilon))$ can be given for the condition number and precision. A proof of this is given in Appendix A of Ref.~\cite{costa2024discreteadiabaticquantumlinear}. This lower bound is obtained by combining lower bounds on computing the parity of a bitstring and constructing a quantum circuit which computes said parity. 

As shown in \cref{tab:tableQLSP}, the current algorithms most efficient with respect to $\kappa$ and $\varepsilon$ dependence in $\kappa$ and $\varepsilon$ have query complexity $O(\spr\kappa \log(1/\varepsilon))$.
Unpublished results \cite{harrow_inpreparation_2021} suggest
a lower bound of $O(\sqrt{\spr}\kappa\log(1/\varepsilon))$ for solving the QLSP. Methods with optimal scaling in $\kappa$ and $1/\varepsilon$ such as the discrete adiabatic or the augmented system method in \cref{tab:tableQLSP} are not known to achieve this lower bound due to their dependence on $\spr$. One possible way to achieve the lower bound could be to block-encode the matrix with complexity $\sqrt{\spr}$, but there is no known method that can do this.

In Ref.~\cite{Low2019hamsim}, a Hamiltonian simulation algorithm is given with a $\sqrt{\spr}$ scaling for the number of queries. This technique is then used to give a quantum algorithm solving the QLSP problem with $(\kappa\sqrt{\spr})^{1+o(1)}/\varepsilon^{o(1)}$ queries. To implement the algorithm, a recursive version of the interaction Hamiltonian approach is utilized \cite{Low2018interaction}.
Given a Hamiltonian which is decomposed in $m$ terms $H=\sum_{i=1}^m H_j$, with each term described by a block-encoding, The evolution $e^{-i(H_1+\cdots+ H_k)}$ can be simulated with $e^{-i(H_1+\cdots+H_{k-1})}$ and the block-encoding of $H_k$ through the interaction picture introduced in \cite{Low2018interaction}.
Directly applying this procedure with the discrete adiabatic or augmented system method would unfortunately reintroduce factors which would worsen the dependence on other parameters in the complexity.
This factors are introduced due to the worse dependence on parameters such as $\kappa$ which would no longer be linear.
Therefore, the problem of whether this lower bound can be achieved is open and seems to require new techniques, different from the ones reviewed in this survey. Such techniques developed to improve on the sparsity might also have an impact in the design of algorithms for other problems.

Recently, the optimal query complexity in terms of the success amplitude $\sqrt{p}=\frac{\norm{A^{-1}\ket{b}}}{\norm{A^{-1}}}$ together with $\kappa$ and $\varepsilon$ has been studied in Ref.~\cite{low2024quantumlinearalgorithmoptimal}. As pointed out in this work, techniques like the discrete adiabatic theorem and augmenting the linear system  do not provide optimal query complexity for both oracles $\mathcal{P}_A$ and $\mathcal{P}_{B}$ as defined in \cref{sec:lcu_inversion}. An algorithm is given which achieves optimal query complexity $\Theta(\frac{1}{\sqrt{p}})$ for $\mathcal{P}_B$ and nearly-optimal query complexity $O(\kappa \log(1/p)(\log \log(1/p)+\log(1/\epsilon)))$ for $\mathcal{P}_A$. The key technique to obtain this algorithm is a modified version of VTAA denoted as Tunable VTAA. This tunable version allows to tune the number of repetitions in the amplitude amplification step. 

Aside from general linear systems, one can consider lower bounds for the QLSP with a restricted class of matrices. In Ref.~\cite{Orsucci2021solvingclassesof}, a lower bound for restricted families of positive-definite matrices is given. As mentioned before, the general case in a worst-case analysis has a $\Omega(\kappa)$ lower bound. For classical algorithms, scaling in terms of $\sqrt{\kappa}$ can be achieved when the matrices are positive definite. Then, a natural question is whether quantum algorithms can also achieve this improvement for the QLSP. The short answer is that this is not true for generic positive-definite matrices \cite{Orsucci2021solvingclassesof}. Nonetheless, some results are given which allow a $\sqrt{\kappa}$ dependence for an even more restricted family of positive-definite matrices. This leaves open the question of whether there are other families where such speed-ups are possible, either on the worst- or average-case.

Lower bounds have also been given in the setting of parallel quantum computing \cite{wang_tight_2024}. In this setting, the complexity metric used is that of the quantum query-depth, defined as the minimal depth (in terms of query calls) required in a circuit. Put in another way, those query calls that can be performed in parallel (at the same depth) are only counted once. In Ref.~\cite{wang_tight_2024} it is shown that for both the sparse-matrix oracle and block-encoding, the query-depth is $\Omega(\kappa)$.

Beyond the core algorithmic improvements, significant efforts have been directed towards enhancing the performance of QLSP solvers through techniques such as preconditioning. Preconditioning aims to transform the original linear system into an equivalent system with a more favorable condition number $\kappa$, thereby improving the efficiency of the solver~\cite{tong_fast_2021}, as long as the new problem does not incur higher costs than the original. Typically, preconditioners are selected based on specific instances to optimize their performance, rather than being designed for general applicability. For instance, the preconditioners discussed in Ref.~\cite{clader2013preconditioned}, the wavelet preconditioner in Ref.~\cite{bagherimehrab2023fast}, and the approach outlined in Ref.~\cite{tong_fast_2021}, illustrate this tailored selection process. Preconditioning in context with PDEs (i.e., \cite{bagherimehrab2023fast}) is discussed in \Cref{subsec:de}. More recently, Ref.~\cite{low2024quantumlinearalgorithmoptimal} introduces block-preconditioning as a strategy to enhance query complexity and reduce initial state preparation costs, where$\ket{\bv}$ describes the preconditioner through the scaling matrix $S=s\op{\bv} + (I-\op{\bv})$ with $0<s<1$, so that the action of the inverse matrix on the subspace spanned by the initial state is amplified when inverting the preconditioned system $SA$.

\subsection{Constant factors}\label{subsec:optimal-constant-factors}
Simulation cost in asymptotic scaling does not paint the full picture as these expressions are equivalent up to multiplication by a constant that is independent of relevant order parameters. A discussion of constant factors is thus important in the following sense: Given two algorithms, where one has a much lower constant factor yet only slightly worse asymptotic scaling, the one with the lower constant factor may be more favorable within a significant portion of a regime of system parameters.
This will become increasingly relevant as these algorithms are implemented in practice.
Hence, this subsection focuses on the dependence on constant factors which has been the focus of some research after the publication of the discrete adiabatic algorithm \cite{costa_optimal_2021}.

While the discrete adiabatic method already achieves the optimal scaling, in Ref. \cite{jennings_efficientquantumlinearsolver_2023} the authors give an upper bound on the number of queries for a modified version of the adiabatic randomization method which improves over the bounds given in Ref.~\cite{costa_optimal_2021} for the discrete adiabatic. The relevant parameters in the costing are $\epsilon, \kappa$ and the rescaling constant used in the block encoding $\alpha$. For $\alpha=1$ and $\epsilon=10^{-10}$ the query complexity upper bound for the randomized method is $4.8$ to $8.8$ times more efficient than the discrete adiabatic when $\kappa\in [10^{2},10^6]$ and it outperforms the discrete adiabatic up to values of $\kappa=10^{32}$. Note that this improvement is despite the fact that the randomized method has a worse asymptotic scaling than the discrete adiabatic.

Comparing analytical upper bounds is a useful proxy to the performance of algorithms but it must be kept in mind that these bounds could be much looser than the actual performance of the algorithm. In Ref. \cite{costa2024discreteadiabaticquantumlinear}, a numerical approach is taken to compare the discrete adiabatic and randomization methods. The upshot of this numerical study is that the constant factor for the discrete adiabatic method is actually $1500$ times better than the loose upper bound would suggest, and in fact, the discrete adiabatic method is $20$ times more efficient on average than the randomization method. The work tested Hermitian positive definite as well as general non-Hermitian matrices  $A\in \mathbb{R}^{16\times 16}$  with $\kappa\in\{10,20,30,40,50\}$. Whether there are other regimes where the randomized or some other method is preferable, is not clear. This work makes the point that when actually implementing quantum linear system solvers, classical numerics will play an important role when determining which algorithm to employ.

Ref.~\cite{dalzell2024shortcutoptimalquantumlinear} establishes upper bounds that improve upon earlier estimates by more than an order of magnitude. Whether this algorithm has an advantage over others under numerical tests is open. In fact, a broad comparison of the different methods remains an open interesting question which may be relevant for certain regimes. As quantum linear solvers become feasible, numerics may play a role in determining which algorithm to implement. How the algorithm is implemented in practice may greatly affect these comparisons; for instance, while classical and discrete adiabatic approaches present a more complicated method for assessing algorithm complexity, their implementation remains quite straightforward. Once the constant factor is estimated, we just have to run the same circuit repeatedly, where, in particular, for the discrete adiabatic, the quantum circuit for the Hermitian and non-Hermitian matrices are provided in Ref.~\cite{costa2024discreteadiabaticquantumlinear}. Moreover, the eigenstate filtering routine in the discrete adiabatic seems simpler than those in other works. One may expect that this would introduce some savings in practical implementations.

\section{Near-Term and Early Fault-Tolerant Solvers}\label{sec:nearterm}
On a parallel note, it is worth highlighting that there are efforts to design QLSP solvers in near-term and early fault-tolerant models of quantum computation \cite{preskill_quantum_2018, bharti_noisyintermediatealgorithms_2022, katabarwa_early_2023, liang_modeling_2024, ni_lowdepth_2023}, which do not assume the fully fault-tolerant model. There exists several such studies as well as comparative studies of these solves \cite{huang_near_2021, BravoPrieto_variational_2023, xu_variational_2021, perelshtein_solving_2022, chen2024enablinglargescalehighprecisionfluid, pellowjarman_nearterm_2023, omalley_neartermquantumalgorithmsolving_2024, ghisoni2024shadowquantumlinearsolver}. Given the heuristic nature of these algorithms, a rigorous complexity analysis is not feasible. In such cases, numerical simulations are used to gauge the runtime behavior of different parameters. Below, we highlight two main classes of solutions that do not rely on the fault-tolerant sub-routines we discussed throughout this manuscript so far. These are a variational and an early-fault tolerant approach.

\paragraph{Variational quantum linear solver.} On one hand, there exist the fully variational approaches such as in Refs.~\cite{BravoPrieto_variational_2023, xu_variational_2021, perelshtein_solving_2022, chen2024enablinglargescalehighprecisionfluid}. \citet{BravoPrieto_variational_2023} introduce the variational quantum linear solver (known as VQLS), a hybrid quantum-classical architecture which involves a classical optimizer to minimize a cost function, which are constructed in a way so that the overlap of the parametrized quantum state with the space that is orthogonal to the span of the solution vector is minimized  --- the construction of the Hamiltonian corresponding to the cost function is inspired by the generator of the dynamics in the AQC approach proposed in Ref.~\cite{Subasi2019adiabatic}. The cost function is measured via a quantum device and parameters are  updated iteratively until convergence. The solution presented in this work is tested on quantum hardware on a specific problem size of $1024 \times 1024$. Barren plateaus are a common occurrence in hardware-efficient parametrized quantum circuits. These are regions where the gradient of the cost function is nearly zero, making it infeasible to develop the optimization process towards a local minima. While  barren plateaus have been demonstrated for VQLS, there are techniques such as local cost functions and clever circuit design to alleviate their occurrence. 

\paragraph{Classical combination of quantum states.} On the other hand, a non-variational proposal that goes more along the lines of early fault-tolerant quantum computation is introduced in Ref.~\cite{huang_near_2021}. This work introduces the approach called classical combination of  quantum states (CQS), which, as the name suggests, expresses the solution of the linear system as a linear combination of quantum states. While the quantum states in principle could be variational states, the core approach classically only optimizes the coefficients associated with this linear combination. Similar to the variational solvers, CQS assumes that the linear system is defined by a linear combination of unitary matrices, and each matrix is associated with an efficiently executable quantum circuit. Using these circuits, CQS creates an ansatz tree of quantum states by applying these circuits sequentially. CQS is proposed in response to the optimization plateaus present in variational architectures~\cite{larocca2024reviewbarrenplateausvariational}, as it focuses on optimizing the combination parameters only and considers the two-norm and Tikhonov regression settings.  The complexity of the problem is moved into the construction of a potentially very large Ansatz tree, and there are situations where CQS shows benefits over VQLS, while also not requiring coherent superpositions involving many ancilla qubits as the FTQC approaches.

\textit{Further details on CQS} -- Let $L$ be a loss function $L(\theta) = \|A x(c) - \ket{b}\|^2$ where $A$ is a Hermitian matrix given as a linear combination of unitaries, $x(c)$ is the solution parametrized by a set of linear combination parameters $c$, and $\ket b$ is the right-hand side of the linear system.
Just as in variational algorithms in general, the objective is to minimize $L$ by optimizing the quantum state parametrized by classical variables. In terms of the optimization landscape, CQS optimization becomes convex and hence avoids barren plateaus mentioned above. The study investigates a loss function that is the $L_2$-distance to the right-hand side and a Tikhonov-regularized version. The Tikhonov-regularized version allows to achieve a circuit depth that does not depend on the condition number, as the optimization is reduced to a strongly convex optimization. 
The CQS is inspired by Krylov subspace methods and also Coupled Cluster ansatz techniques in quantum chemistry in the following sense. Assume that $A$ is given by a LCU, $A = \sum_j \alpha_j U_j$. A Krylov subspace of order $r$ induced by a matrix $A$ and a vector $b$ is given as $\{b, Ab, A^2b, \ldots, A^{r-1}b\}$. CQS builds a set of states $\{\ket{\psi_j}\}$ where each $\ket{\psi_j}$ is generated by the unitaries defining the decomposition of the matrix $A$ as $U_{j_1} U_{j_2} \cdots$. The use of these non-orthogonal states  is inspired by the linear combination of atomic orbital (or LCAO) approach in quantum chemistry and is related to variational algorithms for quantum chemistry the assembly in that work being performed adaptively, as described in Ref.~\cite{grimsley_adaptive_2019}. More details follow below. 

\textit{Method} -- The CQS training can be decomposed in the following main steps.
\begin{itemize}
    \item \textbf{Step 1:} Define a set of quantum states $\ket{\psi_j}$ by  corresponding sequences of the circuits that define $A$. 
    \item \textbf{Step 2:} Combine the generated quantum states classically as in
$x_{\text{combined}} = \sum_{i=1}^{N} c_i \ket{\psi(\theta_i)}$, where $c_i\in\mathbb{C}$ are classical coefficients. These coefficients are then determined by a classical optimization (quadratic program) that depends on measurements of overlaps and matrix elements of various combinations of states.
    \item \textbf{Step 3:} Use an ansatz tree to navigate to a different combinations of states. Typically, the ansatz tree is extended by adding states coming from a higher order in the Krylov space hierarchy. While different heuristics exists for selecting the next states, the gradient expansion heuristics selects the next states to be added  by the criterion of having the largest gradient with respect to the loss function.
    \item Next, Steps 1-3 are repeated until convergence.
\end{itemize}
For more details on the training procedure, we refer to the original work.

\textit{Discussion} -- The optimization process entails applying a heuristic method to explore the tree node by node. Simulations carried on large system sizes up to $2^{300} \times 2^{300}$ achieve similar performance to existing algorithms, using fewer quantum gates. However, it is worth noting that the ansatz tree grows exponentially if selecting all possible combinations in the tree, thus creating scalability challenges in optimizing larger trees. 
This structure is closely related to Coupled Cluster wavefunctions in quantum chemistry and the adaptive construction of Ref.~\cite{grimsley_adaptive_2019}. There, it is essential to note that due to the exponential increase in size, coupled cluster techniques truncate at a depth of two to three. ADAPT only constructs a single wavefunction that is adaptively grown, thus it also does not run into the issue of exponentially increasing basis set size. The necessary depth for CQS is $O(\kappa \log(\kappa/\varepsilon))$ for an unregularized loss function and $O(\log(1/\varepsilon))$ for Tikhonov regularization. Hence, for the unregularized case, in order to be computationally viable, this approach is restricted to the case of $\kappa\in O(1)$. Furthermore, even for Tikhonov regularization, achieving a high precision might be challenging depending on the specific system. 

CQS has been tested on a real quantum device using three qubits in Ref.~\cite{pellowjarman_nearterm_2023} with favorable performance for a chosen task.

\section{Applications of quantum linear systems solvers} \label{sec:applications}
Now we discuss some of the main fields of research which apply solutions of the QLSP to problems, namely in differential equations in \cref{subsec:de} and quantum machine learning in \cref{subsec:qml}. \ref{sec:greensfunction} presents a computation of Green's functions in quantum many-body systems \cite{tong_fast_2021}. Beyond the applications we contextualize and highlight in greater depth below, one can mention applications in quantum eigenvalue processing \cite{low2024quantumeigenvalueprocessing}, quantum interior point estimation \cite{Augustino_2023}, and applications in calculating electromagnetic scattering \cite{clader2013preconditioned}.

\subsection{Differential Equations}\label{subsec:de}
Many phenomena in science and engineering disciplines and beyond, like finance, are described by differential equations. Solving these equations for practical applications rarely can be done analytically. Various discretization techniques such as finite difference, finite volume, finite element methods, discontinuous Galerkin, wavelet discretizations, etc., are used to determine a finite-dimensional approximation that can be processed by computers~\cite{butcher2016classicalodesbook,ames2014classicalpdesbook1,evans2012classicalpdesbook2,cockburn2012classicaldgbook,dahmen1997classicalmultiscalebook}. Desired quantities to extract from the resulting finite-dimensional setup then either relate to the spectrum of the encoded operators or to the solution of discretized system. Examples for the former are modal analysis of engineering structures or the ground state problem in quantum chemistry. The latter deems to extract quantities out of a solution (vector). Linear systems problems frequently appear in this solution step when solving discretized differential equations, which naturally lead to consider the role of quantum linear solvers in this section.
Clearly, the hope for quantum computers to be useful in this area is through speedups, like the linear systems speedup, and the ability to encode large amounts of data. In what follows, we will discuss quantum algorithms to solve differential equations via the QLSP and discuss their potential, taking into account the caveats of the QLSP in practice. 

\begin{remark}
 In our considerations below, we assume sufficient regularity on the solution, well-posedness and a proper, consistent discretization scheme that is used for the differential equations --- conditions that need to be met regardless of quantum or classical solutions. For any subtleties regarding this, we refer to extensive work in numerical analysis.
\end{remark}

To begin with, we will formally state the general problems associated with differential equations. We will discuss how these problems are typically tackled and then how quantum algorithms for linear systems can come into play. 
Evolution equations are oftentimes tackled in a different computational manner than stationary problems, which is why we will also make this distinction here. 

\subsubsection{Stationary problems}\label{subsub:de-stationary}
\paragraph{Setup}
We first consider problems that do not explicitly depend on time, and are built by equating the action of a differential operator with a right-hand side. This often comes from stationary points of evolution equations in the sense that $u_\infty$ is a fixed point of the dynamics, i.e., $\partial_t u_\infty =  \mathcal{L}u_\infty + b =0$. Then, setting $u=u_\infty$, a linear stationary PDE problem may look like $\mathcal{L}u=-b$.  
\
\begin{problem}[Quantum Differential Equation Problem (stationary)]\label{problem:qde-stat}
    Let $\mathcal L: A\to B$ be a linear, elliptic differential operator, $u\in A$ be a function so that $\mathcal L u = f$ and $L\in\mathbb{R}^{n}\to\mathbb{R}^n$ be a finite-dimensional representation through a suitable discretization scheme, which admits solving the linear system $L \bar u = \bar f$, $\bar u \in \mathbb{R}^n$. Then, for some $\varepsilon>0$, we seek to find a quantum state $\ket{\bar u}$ so that  
    \begin{equation}
        \norm{\ket{u} - \ket{\bar u}} \le \varepsilon.
    \end{equation}
\end{problem}
\begin{remark}
    This error $\varepsilon$ consists both of discretization error $\varepsilon_{\text{disc}}$ and error in approximately solving the resulting linear system $\varepsilon_{\text{LS}}$.
    For the sake of this review, we may ignore the discretization error and refer to the abundant literature in numerical analysis of differential equations.
    We assume it is properly designed and well-behaved in the sense that $\varepsilon_{\text{disc}}O{\varepsilon_{\text{LS}}}$. 
\end{remark}

As we can see from \cref{problem:qde-stat}, a discretization of a stationary PDE problem immediately produces a linear systems problem. Thus, treating the discretized differential operator $L$ to be like $A$ in \cref{prob:QLSP} and the right-handside $b$ like $f$, we directly have a quantum linear systems problem. Note that in \cref{prob:QLSP}, we assumed that $A\succeq 0$ and normalized in the sense that $\norm{A}=1$. The normalization can be realized for any differential operators with bounded spectrum --- while this is not necessarily true in the infinite-dimensional case for $\mathcal{L}$, it then is for a finite-dimensional $L$. Operators like the Laplacian also comply the positive-definiteness requirement. 

\paragraph{Potential and pitfalls}
References \cite{cao2013quantum,montanaro_quantum_2016,childs_highprecision_2020} discuss the quantum implementation of elliptic PDEs --- while \cite{cao2013quantum} provides a complete algorithm, it uses rather ``old'' techniques and relies on HHL. A more modern and also complete approach is the one in Ref.~\cite{childs_highprecision_2020}, using finite-difference approximations and a spectral method that has been previously used in the quantum solution for ODEs in Ref.~\cite{childs2020quantum}.
Solving DEs brings more or less the same pitfalls as solving linear systems in general, as discussed in \cref{subsec:rereading}. That is, problems of interest are when the quantum linear solver can exploit exponential speedup with respect to space. Then, it is important that state preparation of the RHS (source vector) can be done efficiently. Furthermore, final quantities of interest have to be restricted to quantities like expectation values of observables that describe a physical quantity of interest. A simple example here could be the average heat flux across a part of the domain when discretizing the stationary heat equation. Though, we note that in particular the heat equation is not a candidate for an exponential quantum speedup as it can be reduced to a search problem that inhibits a lower bound (square-root speedup)~\cite{linden2020quantumvsclassicalalgorithms}. 

Another thing to note is that it is well-known that the condition number of elliptic operators, such as the Laplacian, grows with the dimension. To that end, preconditioning techniques, which are also very common in classical numerical solutions to PDEs~\cite{mardal2011classicalpreconditioningbook}, can be of great help to accelerate the computation. 
For Poisson equation with a potential function, Ref.~\cite{tong_fast_2021} applies a direct inverse of the discrete Laplacian as the preconditioner and gets rid of the dimension dependence in the query complexity. 
Ref.~\cite{bagherimehrab2023fast} devised a wavelet-based preconditioner for elliptic operators that achieves a condition number that is bounded by a constant and shows that the application of the preconditioner does not introduce a significant additional cost. Even earlier on, Ref.~\cite{clader2013preconditioned} devised a preconditioner for a linear system stemming from discretized electromagnetic scattering and the first discussion of using the HHL algorithm for general finite element methods pointed out the importance of preconditioners as well \cite{montanaro_quantum_2016}.

\vspace*{3em}
\subsubsection{Evolution equations}
We next discuss DEs undergoing explicit time-evolution, where we restrict the discussion to ODEs. The occurring linear operator $L$ may stem from the spatial discretization of a PDE problem. 
\paragraph{Setup}
\begin{problem}[Quantum Differential Equation Solver (evolution)]
    Let $t\in [0,T]$ and $L\in\mathbb{C}^{2^n\times 2^n}$. Then, we consider evolution ODEs of the form $\partial_t  u(t) = {L}  u(t) +  b$, with solution vector $u: [0;T]\to \mathbb{C}^{2^n}$. $L, b$ may or may not be time-dependent. We seek  to prepare a quantum state $\ket{\bar{u}(T)} \propto \bar{u}(T)$ for some final time $T$ so that $\norm{\ket{u(T)} - \ket{\bar{u}(T)}}_{\ell_2}\le \varepsilon$.  
    A sufficient condition for stability of the dynamics that is chosen by many quantum implementations is $ L + L^{\dagger} \preceq 0$. 
\end{problem}
For evolution equations, there is a higher variety in the proposed quantum solutions.
The arguably most straightforward way to solve a evolution equation classically is the ``forward Euler'' scheme, which proceeds along a series of time-steps with a first-order difference formula of the time derivative. As outlined in Ref.~\cite{fang2023time}, a naive implementation of this is quantumly prohibitive due to exponentially diminishing success probability with the number of time steps. To see this, consider  $\ket{u(t_j)} = (I+\Delta t L)\ket{u(t_{j-1})}$. Then, for $I+\Delta t L$ non-unitary, the subnormalization factor is lower-bounded by the spectral norm, $\alpha\ge 1 + \Delta t \norm{L}$. So, the success probability to find the solution at time $T>0$ after $n_t$ time-steps goes exponentially small as $\left(\frac{1}{(1+T/n_t \norm{L} )^{n_t}} \frac{\norm{\ket{u(T)}}}{\norm{\ket{u(0)}}}\right)^2$. 
To counter this, \cite{fang2023time} made use of uniform singular value amplification and achieved an efficient time-stepping based quantum algorithm which has near-optimal performance in the number of calls to the state preparation oracle, however, quadratic complexity in the simulation time.   

Alternatively, there are approaches that map the differential equation to a Hamiltonian simulation problem, including methods for specific types of differential equations with energy conservation~\cite{costa2019quantum,babbush2023exponential} as well as methods for generic dissipative systems \cite{an2023linear,an2023quantum,jin2022quantumsimulationpartialdifferential}. While are very efficient as they can resort to highly optimized quantum simulation methods, they cannot be applied to dissipative systems in an obvious manner. 

Beyond that, one of the most prevalent approaches to solve evolution equations quantumly is making use of quantum linear system solvers. The idea for this goes back to Feynman's clock register construction, also oftentimes called history state~\cite{feynman1982simulating,feynman1986quantum}.
Here, the evolution is stored as a superposition over time-steps \cite{berry_highorder_2014}. Applying this to differential equations means that one can write a linear system, where each row is composed by a finite-dimensional discretization of the dynamics between time-steps.  
The history state for $\bar{u}(t)\in\mathbb{C}^{2^n}$ in an amplitude encoding, is defined as follows, where we discretize time for the sake of simplicity in $n_t$ equidistant steps of $\Delta t = \frac{T}{n_t}$:
\begin{equation}
    \frac{1}{\norm{u_{\text{hist}}}}\sum_{\tau=0}^{n_t-1} \sum_{j=0}^{2^n-1} \ket{\bar{u}_j(\tau \Delta t)} \ket{\tau},
\end{equation}
where $\norm{u_\text{hist}} = \norm{\sum_\tau \sum_j \bar{u}_j(\tau\Delta t)\ket{x}\ket{\tau}}$ is the norm of the history state. 
We can immediately notice that this way, the success probability to measure the final time decreases with the number of time steps, which is not desirable. This was already observed by Feynman in his construction and mitigated by a simple trick, and also carried over in Ref.~\cite{berry_highorder_2014} and subsequent works. One can hold the solution constant for a few steps to improve upon the success probability at final time. We will discuss  the resulting linear system in the following.

Let us assume that at time-step $\tau$, the local propagator is defined by $\bar u((\tau+1 )\Delta t) = (I+V_\tau)\bar u(\tau\Delta t)$.  
The associated linear system in the history state can be written as follows:
\begin{equation}\label{eq:ode-big-system}
\mathcal{A}\overline{\mathbf{u}} = 
    \begin{bmatrix}
       I &      &      &    &           &    &    &    &    &   \\
    -V_1 & I    &      &    &           &    &    &    &    &   \\
         & -V_2 & I    &    &           &    &    &    &    &   \\
         &      &      & \ddots& \ddots &    &    &    &    &   \\
         &      &      &    & -V_{n_t}  & I  &    &    &    &   \\
         &      &      &    &           &-I  &  I &    &    &   \\
         &      &      &    &           &    & -I &  I &    &   \\
         &      &      &    &           &    &    & \ddots &  \ddots &   \\
         &      &      &    &           &    &    &    & -I & I
    \end{bmatrix}
    \begin{bmatrix}
        \bar{u}(0) \\ 
        \bar{u}(\Delta t) \\ 
        \bar{u}(2 \Delta t) \\ 
        \vdots\\
        \bar{u}(n_t \Delta t) \\
        \bar{u}(n_t \Delta t) \\
        \bar{u}(n_t \Delta t)  \\
         \vdots  \\
        \bar{u}(n_t \Delta t)  
    \end{bmatrix}
    =
    \begin{bmatrix}
        \bar b(0) \\ 
        \bar{b}(\Delta t) \\ 
        \bar{b}(2 \Delta t) \\ 
        \vdots\\
        \bar{b}(n_t \Delta t) \\ 
        0\\
        0\\
        \vdots \\
        0
    \end{bmatrix}
\end{equation}
As an example, for a Forward Euler scheme on $\partial u = L u$, we have $V_\tau = \Delta t L$ for all $\tau \in [n_t-1]_0$.
In the literature, a variety of approaches are considered --- from multi-step schemes \cite{berry_highorder_2014}, pseudospectral methods \cite{childs2020quantum}, to recently increasingly more methods based on truncated Taylor- \cite{berry2017quantum,krovi2023improved} and truncated Dyson-series~\cite{berry2022quantum}. The latter also allows to tackle systems with $L,b$ time-dependent. 

\paragraph{Discussion}
We next discuss the impact on complexity of ODE systems encoded like \cref{eq:ode-big-system}.
Generally, the condition number of $\mathcal{A}$ depends on the number of steps $n_t$, including the ones that hold the solution constant, discussed e.g. in Ref.~\cite{childs2020quantum,berry2022quantum}. Additionally, depending on the method, it may depend on the conditioning of the matrix that diagonalizes $L$ --- this is intrinsic to approaches such as \cite{berry_highorder_2014,childs2020quantum}, not to the ones that implement truncated series expansions. 
In the complexity of the ODE solver in Ref.~\cite{berry2022quantum}, which use the optimal scaling linear solver from \cite{costa_optimal_2021}, the condition number appears as a factor of $\alpha_{\mathcal{A}} T$, namely the time scaled by the subnormalization of the encoded evolution matrix, in our notation here that is $L$. This comes from the fact that time-steps are chosen so that the subnormalization of the history state matrix $\mathcal{A}$ is bounded by a constant, hence a single time step will be inverse proportional to the subnormalization of $L$.
Very recently, Ref.~\cite{jennings2024costsolvinglineardifferential,an2024fastforwardingquantumalgorithmslinear} have observed that the condition number of $\mathcal{A}$ can be sub-linear in $n_t$ if the differential equations satisfy certain stronger stability conditions, resulting in fast-forwarded quantum algorithms for these particular differential equations with sub-linear scaling in the evolution time $T$. 

Apart from this, limitations regarding state preparation and solution extraction hold equivalently to the stationary case.

Recently, several algorithms for nonlinear ODEs have been proposed. Early work on this \cite{leyton_quantum_2008} encoded the ODE system as a QLSP as well, though their reported complexity is exponential in evolution time and the degree of nonlinearity. Recent attempts mostly rely on Carleman linearization~ \cite{carleman1932application} \cite{liu2021efficient,liu2023efficient,krovi2023improved,costa2023further} or homotopy perturbation methods~\cite{xue2021quantum} to map a nonlinear ODE to a large system of linear ODEs. For Carleman linearization, given a stability condition that requires a rather high degree of dissipation (this is also similar for the homotopy perturbation method), it can be shown that the linearization error is quite well behaved and decays exponentially in the truncation number of the linearization \cite{forets2017explicit}.
Overall, tackling nonlinearities adds a multiplicative cost linear in the truncation of the linearization, and further impacts the success probability on top of the history state due to enlarging the system; this can be mitigated through a rescaling of the solution~\cite{liu2021efficient,krovi2023improved,costa2023further}. 
The likely largest limitation for nonlinear equations is that, as already mentioned, the ratio  strength of nonlinearity versus dissipation needs to be quite small. For instance, while there is great interest from the fluid dynamics community for accurate solutions, with a large grid where the favourable space complexity of quantum computers be of advantage, this community is also interested particularly in simulating turbulence, where this stability constant is not small. It is still left up to further research whether quantum computers can help here. A promising direction might be representation in a Lattice-Boltzmann picture, such as done in Ref.~\cite{li2023potential}.

\subsection{Quantum Machine Learning}\label{subsec:qml}
Machine learning techniques leverage solving systems of linear equations in the context of model training and optimization \cite{mitchell_machine_1997, goodfellow_deep_2016}. In the emerging field of quantum machine learning (QML), similar principles apply, but the algorithms leverage quantum properties and operations to potentially achieve computational advantages over classical approaches. By extension, many of the proposed QML algorithms incorporate solving linear systems of equations as a fundamental subroutine in both supervised and unsupervised learning methods \cite{schuld_machine_2021}. QML algorithms employ quantum states $|\psi\rangle$ and unitary operators $U$ to process data. Given a dataset $\{(x_i, y_i)\}_{i=1}^N$, QML uses quantum feature maps $\phi(x)$ to encode classical data into quantum states $|\phi(x)\rangle$ for potentially achieving computational complexity reductions in tasks such as classification, clustering and regression. Naturally, the QLSP approach, particularly through the HHL algorithm, is instrumental in various QML methods. Quantum linear systems of algorithms solve the linear systems arising in the optimization functions in the training of several quantum learning models, such as the quantum neural networks. In addition, as it has been pointed out in Ref.~\citep{adcock_advances_2015}, due to its close relationship to solving data-fitting procedures such as linear regression, HHL can be considered as the equivalent of gradient descent from classical machine learning. This implies its ability to generalize well over different environments as an optimization method. Consequently, the QML techniques based on HHL solvers, also inherit their limitations, which will be discussed further in subsequent sections. We note that summaries of quantum linear systems of equations in QML techniques have been carried out in Ref.~\cite{duan_survey_2020} and very briefly in Ref.~\cite{dervovic_quantum_2018}. The next part gives a brief overview and taxonomy of some of the most important quantized machine leaning techniques, under the lens of the QLSP solvers.

\subsubsection{Quantum Linear Solvers based Quantum Learning}
In supervised learning a model is trained on a labeled dataset $\{(x_i, y_i)\}_{i=1}^N$ to learn a mapping function $f: \mathcal{X} \rightarrow \mathcal{Y}$, where $\mathcal{X}$ is the input space and $\mathcal{Y}$ is the output space, such that it can predict the output $y$ for new input $x$ from the same distribution. Tasks such as classification and regression encompass most of the approaches to supervised learning. A notable technique commonly used for classification tasks is the support vector machine (SVM). SVM has been explored with a quantum lens to give rise to quantum support vector machine (QSVM) \cite{rebentrost_quantum_2014}. This algorithm is showcased below in terms of how it incorporates the HHL solver, how it compares with its classical counterpart, and its time complexity and limitations.

In the line of other supervised algorithms, regression presents a significant category of mathematical tasks for prediction. This statistical method models the relationship between a dependent variable $Y$ and one or more independent variables $X_1, X_2, \ldots, X_p$, with respect to estimating the coefficients $\beta_1, \beta_2, \ldots, \beta_p$ in the equation
\begin{equation}
Y = \beta_0 + \beta_1 X_1 + \beta_2 X_2 + \cdots + \beta_p X_p + \varepsilon,
\end{equation}
where $\beta_0$ is the intercept term and $\varepsilon$ represents the error term. A regression analysis makes predictions based on the changes in this relationship. Regression has been well studied in the quantum context \cite{wiebe_quantum_2012, liu_fast_2017, yu_improved_2019, schuld_prediction_2016, date_adiabatic_2021}. For instance, in Ref.~\cite{wiebe_quantum_2012} the runtime complexity of the proposed quantum linear regression based on the HHL solver is formulated as $O(\log(N)s^3 \kappa^6 / \varepsilon)$.
This improved complexity denotes exponential speedup in input size, but is sensitive to matrix sparsity, condition number, and precision requirements. While theoretically an achievement in comparison with the classical counterpart, practical applicability requires a well-conditioned and sparse matrix to maintain efficiency. In light of later solvers, such as Ref.~\cite{gilyen2019quantum}, the complexity of quantum linear regression in Ref.~\cite{wiebe_quantum_2012} can be updated if we naively assume a block encoded matrix instead of calling the matrix inversion procedure as carried out in the HHL solver.

Unsupervised learning entails the model trained on an unlabeled dataset $\{x_i\}_{i=1}^N$ to learn the underlying structure or distribution of the data, typically without specific output variables. In this paradigm, we take note of tasks such as clustering and dimensionality reduction, both of which have been extensively studied across the QML literature \cite{aimeur_2013_quantum, wiebe_quantum_2014, lloyd_quantum_2014}. A noteworthy example in this avenue is principal component analysis (or PCA) and the subsequent design of the quantum principal component analysis (QPCA) \cite{lloyd_quantum_2014}. PCA is used to reduce the dimensionality of a dataset while preserving variability. This involves calculating the covariance matrix, and performing eigenvalue decomposition to obtain to obtain eigenvalues and eigenvectors. Principal components are the top eigenvectors that capture the most important patterns in a dataset. QPCA as in Ref.~\cite{lloyd_quantum_2014} on the other hand, is designed to reveal the properties of an unknown quantum non-sparse low-rank density matrix. In principle, it does so similarly to the PCA by extracting the eigenvectors with the largest eigenvalues, using QPE. The time complexity assumption is that principal components for such density matrices can be prepared in $O(\log (d))$. This denotes an exponential speedup to the PCA. However, the practical efficiency of QPCA depends on the ability to efficiently prepare quantum states and perform quantum operations \cite{tang_quantum_2021}, which remains an open non-trivial research question. It is important to highlight that QPCA is mentioned here as a noteworthy example inspired by the HHL solution, rather than as a direct application.

Other notable algorithms that use QLSP solvers in their composition include recommendation systems \cite{kerenidis_quantum_2016} and generalizations of quantum hopfield neural networks \cite{rebentrost_quantumhop_2018}. Additionally, as mentioned in \cref{sec:improv}, Ref.~\cite{wossnig_quantum_2018} designs an algorithm that is more efficient in solving the QLSP for dense matrices, mentioning machine learning techniques as an example where this paradigm is useful. The highlighted techniques such as kernel methods and neural networks are some of the most prominent architectures in machine learning. Additionally, Ref.~\cite{an_quantum_2020} also extends solving the QLSP using the quantum approximate optimization algorithm (or better known as QAOA) \cite{farhi_quantum_2014} with optimal results.
To illustrate an example of how QLSP solvers are applied to learning techniques, we detail QSVM below.

\subsubsection{Example: Quantum Support Vector Machine}
Classically, an SVM aims to find the hyperplane that best separates different classes of data points in a high-dimensional space. The hyperplane is selected so that it maximizes the space between nearest data points from each class. The nearest data points are known as support vectors. Formally, given a dataset $(\mathbf{x}_1, y_1), (\mathbf{x}_2, y_2), \ldots, (\mathbf{x}_n, y_n)$ where $\mathbf{x}_i$ is the input data and $y_i$ is the label ($y_i \in \{-1, +1\}$), the classification function of binary form can be written as:
\begin{equation}
    f(\mathbf{x}) = \text{sign}\left(\sum_{i=1}^n \alpha_i y_i K(\mathbf{x}_i, \mathbf{x}) + b\right),
\end{equation}
where $K(\mathbf{x}_i, \mathbf{x})$ is the kernel function, $\alpha_i$ are the Lagrange multipliers, and $b$ is the bias. The follow-up optimization function can then be solved using quadratic programming methods. The complexity of solving this problem typically scales between $O(N^2)$ to $O(N^3)$, depending on the implementation and specifics of the optimization algorithm. 

\textit{Quantum Support Vector Machine}~\cite{rebentrost_quantum_2014} -- A QSVM is a quantum algorithm designed to classify data points using a quantum implementation of the SVM paradigm. Consider a training dataset $\{(\mathbf{x}_i, y_i)\}_{i=1}^N$ where $\mathbf{x}_i \in \mathbb{R}^d$ are feature vectors and $y_i \in \{-1, 1\}$ are binary labels. The goal of the QSVM is to find a hyperplane that maximizes the margin between the two classes in a feature space defined by a kernel function $K(\mathbf{x}_i, \mathbf{x}_j)$. Tangentially, the problem studied in the QSVM is known as least-squares SVM, which expresses the quadratic programming methods with equality constraints, thereby facilitating the use of the Moore-Penrose pseudoinverse, which is where HHL becomes relevant. Mathematically, this algorithm minimizes the following objective function:
\begin{equation}
    \min_{\mathbf{w}, b, \mathbf{e}} \frac{1}{2} \mathbf{w}^T \mathbf{w} + \frac{\gamma}{2} \sum_{i=1}^N e_i^2
\end{equation}
subject to the equality constraints $y_i (\mathbf{w}^T \phi(\mathbf{x}_i) + b) = 1 - e_i, \quad i = 1, \ldots, N$ where $\mathbf{w}$ is the weight vector, $b$ is the bias, \(\mathbf{e}\) is the error vector, $\gamma$ is a regularization parameter, $\phi(\mathbf{x}_i)$ is the feature map, and $y_i$ are the labels. In matrix form, the solution to this problem involves solving a system of linear equations given by:
\begin{equation}
\begin{bmatrix}
0 & \mathbf{y}^T \\
\mathbf{y} & K + \gamma^{-1} I
\end{bmatrix}
\begin{bmatrix}
b \\
\boldsymbol{\alpha}
\end{bmatrix}
=
\begin{bmatrix}
0 \\
\mathbf{1}
\end{bmatrix},
\end{equation}
where $K$ is the kernel matrix with entries $K_{ij} = \phi(\mathbf{x}_i)^T \phi(\mathbf{x}_j)$, $\boldsymbol{\alpha}$ is the vector of Lagrange multipliers, and $\mathbf{y}$ is the vector of labels. The pseudocode of least-squares QSVM has been sketched in \cref{alg:qsvm}.

\begin{center}
\begin{algorithm}[H]
\caption{Quantum Support Vector Machine (QSVM)\cite{rebentrost_quantum_2014}}\label{alg:qsvm}
\begin{algorithmic}[1]
    \STATE \textbf{Input:} Training dataset $\{(\mathbf{x}_i, y_i)\}_{i=1}^N$, Kernel matrix $K$, Regularization parameter $C$
    \STATE Construct the kernel matrix $K_{ij} = K(\mathbf{x}_i, \mathbf{x}_j)$
    \STATE Construct the matrix $H = K + \gamma^{-1} I$, where $\gamma$ is a regularization parameter
    \STATE Construct the vector $\mathbf{y} = [y_1, y_2, \ldots, y_N]^T$
    \STATE Invert the matrix $H$ to get $H^{-1}$
    \STATE Compute the coefficients $\alpha = H^{-1} \mathbf{y}$ using the least-squares solution
    \STATE Set $b = 0$
    \FOR{Each test data point $\mathbf{x}$}
        \STATE Compute the decision function $f(\mathbf{x}) = \sum_{i=1}^N \alpha_i K(\mathbf{x}_i, \mathbf{x})$
        \STATE Assign the label $sgn(f(\mathbf{x}))$ to the test data point $\mathbf{x}$
    \ENDFOR
\end{algorithmic}
\end{algorithm}

\end{center}
Much of the discussion in the previous sections centered on the runtime and query complexity analysis of QLSP solvers. As such, it is befitting to comment on the theoretical speedup of QSVM. Below we discuss the complexity measure.

\textit{Complexity measure for QSVM.} -- Consider a QSVM for classifying data points using a kernel matrix $K$ and a regularization parameter $\gamma$. The QSVM algorithm involves solving the linear system $(K + \gamma^{-1} I) \alpha = \mathbf{y}$ using the HHL subroutine. Let $N$ be the dimensionality of the feature space, $M$ the number of data points, $\kappa$ the condition number of the matrix $H$ and $\varepsilon$ precision parameter.
Let us be given access to a block-encoding of the matrix $K + \gamma^{-1} I$ with normalization $\alpha$. Then using the exponentially-improved linear systems solver of \cite{childs_quantum_2017,gilyen2019quantum} requires 
\begin{equation}
        \tilde O\left (\kappa \alpha\ {\rm poly}\log(NM)\log \left(\frac{1}{\varepsilon}\right)\right)\label{eq:complexity-qsvm-pre}
    \end{equation}
queries to the block-encoding to prepare a quantum state encoding the solution. 
In Ref.~\cite{rebentrost_quantum_2014}, the time complexity of preparing the quantum state is
    \begin{equation}
        \tilde O(\log(NM)\kappa^2 /\varepsilon^3)\label{eq:complexity-qsvm},
    \end{equation}
owing to the fact that the original HHL algorithm \cite{harrow_quantum_2009} and a sample-based  quantum simulation technique \cite{lloyd_quantum_2014} were used. Multiple copies of the quantum state have to be used to obtain the correct classification. This incurs typically an overhead of at least $\Omega (1/\varepsilon^2)$ for precision $\varepsilon$.

Following \cref{eq:complexity-qsvm}, in comparison with the SVMs where the training complexity scales between $O(N^2)$ to $O(N^3)$ depending on specifics of the optimisation function, QSVM provides a significant speedup. However, critical caveats exist. Below we highlight key assumptions and limitations.

\textit{Assumptions and limitations of QSVM.} -- QSVM places constraints on various variables, much of it inherited by the design of the HHL solver itself. A more in-depth analysis on the HHL was carried in \cref{subsec:hhl}. Regarding QSVM, the following can be highlighted.
\begin{itemize}
    \item Quantum encoding must be done efficiently. This limits two factors. First, the type and scale of data, and secondly, it requires an efficient encoding technique without a large encoding overhead. This is non-trivial, and an open research question.
    \item Constraints on condition number $\kappa$. A high $\kappa$ can lead to reduced quantum speedup. In fact, in general $\kappa$ would depend on the dimension $N$, eliminating any advantage compared to classical algorithms.
    \item Practical limitations. In line with fault-tolerant requirements, any hardware implementation will require a significant number of qubits, considerable gate depth and low error rates. These are all significant hurdles in the current state of quantum hardware development.
\end{itemize}

\textit{Discussion.} -- In closing this section, it is worth noting that a complexity analysis in comparison with the classical machine learning counterparts is much more subtle than it appears. We have seen how quantum versions require additional assumptions, all of which renders a fair complexity comparison questionable. In fact, there exists a whole body of research on dequantization of some of QML techniques \cite{tang_quantum_2019, tang_quantum_2021, tang_dequantizing_2022}. Simply stated, via dequantization, QML techniques can be interpreted in classical terms. The idea of dequantization is to design classical algorithms of QML versions that exhibit performance that is slower by only a polynomial factor compared to their quantum equivalents. And even though this means that quantum counterparts cannot give exponential speedups for classical data, there could still be a large polynomial gap for classical data \cite{tang_dequantizing_2022}. Furthermore, this comparison could account for the caveats and the practical aspects such as resource utilization, scalability, and robustness to noise. To further drive this point, the practical considerations are crucial for evaluating the real-world applicability of both classical and quantum learning algorithms.

\subsection{Green's function in fermionic systems} \label{sec:greensfunction}
Quantum linear system solvers can also find applications in computing quantum many-body systems, examples for which are based on green's functions \cite{tong_fast_2021} and correlation energies \cite{baskaran2023adapting}. Recently, an adapted measurement to improve upon the success probability due to large condition number when measuring correlation energies has been proposed as well~\cite{tsemo2024enhancing}.

For the remainder of this section, we briefly review the computation of Green's function as application and refer readers to Ref.~\cite{tong_fast_2021} for more details.
In particular, the routines to invert a matrix can be used to compute the one-particle Green's function in fermionic systems. While in other applications of linear systems, there is a difficulty in loading the classical data, this application does not suffer from such problems.  

Let us denote by 
$\{\hat{a}_i^{\dag}, \hat{a}_i\}_{i\in [N]}$ the creation and annihilation operators for a fermionic systems on $N$ sites. The Hamiltonian is a polynomial of these creation and annihilation operators. For instance, most of the Hamiltonians in quantum chemistry takes the form
\begin{equation}
\hat{H}=\sum_{i,j\in[N]} T_{ij} \hat{a}_{i}^{\dagger} \hat{a}_{j}+\sum_{i,j,k,l\in[N]} V_{ijkl} \hat{a}_i^{\dagger}\hat{a}_j^{\dagger}\hat{a}_l \hat{a}_k.
\end{equation}
Let $(E_0,\ket{\Psi_0})$ be the non-degenerate ground state eigenpair of $\hat{H}$. The Green's function contains the spectroscopic information due to excitations from the ground state. 
One type of single-particle Green's function is a matrix valued function $G(z)$ with entries
\begin{equation}
\label{eq:greens_function_def}
G_{i j}(z)=\Braket{\Psi_{0}|\hat{a}_{i}\left(z+E_0-\hat{H}\right)^{-1} \hat{a}_{j}^{\dagger}|\Psi_{0}}, \quad i,j\in [N].
\end{equation}
In other words,  $G$ can be viewed as a mapping $\mathbb{C} \mapsto \mathbb{C}^{N\times N}$.
Here the input $z=E-i\eta$ can  be interpreted as a complex energy shift, and $\eta>0$ is called the broadening parameter, and determines the resolution of Green's functions along the energy spectrum. Then  $G(z)$ is well-defined since $E_0 + z$ is not an eigenvalue of $\hat{H}$.  Also note that the dimension of the underlying Hilbert space for the problem is $2^{N}$. Hence the dimension of the matrix $G$ is much smaller compared to that of the Hamiltonian.

Now suppose we have an $(\alpha,m,\epsilon)$-block-encoding of the matrix inverse $\left(z+E_0-\hat{H}\right)^{-1}$ denoted by $U$ using a QLSP solver, then with a product of block encodings, we can construct an $(\alpha,m+2,\epsilon)$ block-encoding of $\hat{a}_i \left(z+E_0-\hat{H}\right)^{-1} \hat{a}_j^\dagger$.  Then the value of each entry $G_{ij}(z)$ can be estimated using the Hadamard test circuit. 

The problem can be further simplified if we are only interested in 
\begin{equation}
\Gamma(z)=\frac{1}{2 i }\left(G(z)-(G(z))^{\dag}\right),
\end{equation}
which is the anti-Hermitian part of $G(z)$.
This is often the most useful component of Green's functions as it encodes the spectral density.

\section{Final Remarks and Onto the Future} \label{sec:concl}
This work has surveyed the current state of algorithms and applications for the QLSP. The primary aim of this work has been to provide an overview of the key techniques used to develop these algorithms and the role these routines play in applications such as differential equation solvers and quantum machine learning, and in many-body physics.

In a nutshell, this work has analyzed a wide range of algorithms that solve the QLSP via (i) direct inversion, (ii) inversion by adiabatic evolution, and (iii) trial state preparation and filtering --- each method presenting advantages and challenges. We reviewed results on lower bounds for the optimal scaling and also recent work on reducing the constant factors as a critical aspect of improving these solvers. While the focus of this review was on provable fault-tolerant quantum algorithms for the QLSP, we have also considered near-term quantum solutions, which could provide benefits before full-scale quantum computing becomes feasible. The manuscript also identified how several applications of the QLSP solvers have already been explored, ranging from quantum machine learning to differential equations to many-body physics, namely in solving the Green's function in fermionic systems.

While in recent years great progress in improving the complexity of quantum linear system solvers has been achieved, several open questions remain. As discussed before, the question of achieving the scaling $\sqrt{\spr}\kappa \log(1/\varepsilon)$ is still open. 
Moreover, the quest to lower the constant factor in these algorithms remains an important problem to make the implementations of these algorithms more practical. 
Another interesting direction is that of finding relevant restricted families of instances where the QLSP could be solved more efficiently similar to what has been done in Ref.~\cite{Orsucci2021solvingclassesof} as we discuss in \cref{sec:optimal}.
Additionally, determining which algorithmic primitives should be preferred over others is just as crucial on an algorithmic level. On the hardware front, exploring ways for enhancing robustness to noise to achieve more efficient hardware implementations is an interdisciplinary task that intersects diverse areas of knowledge.

Looking ahead, there are several avenues where to seek improvement. Firstly, it is crucial to develop quantum algorithms with improved accuracy and robustness for solving a broader class of linear systems. Secondly, one potential avenue for advancement involves further applications of the quantum linear systems of equations. Here we focused on differential equations and QML, however the range of application of QLSP solvers is vast. Additionally, exploring the gap in the algorithmic design between near-term solutions and fault-tolerant algorithms could benefit QLSP solvers by identifying intermediate-term subroutines.

The QLSP also displays that it is crucial to think of quantum algorithms in a holistic way, from preparing the input state to the measurement step. The exponential speedup, compared to classical approaches, can only be sustained if for the specific problem instance, state preparation and asking the right questions through an observable are efficient. For the future, case studies of resource estimates for important problems can be of great impact.

As remarked in the introduction, it is important to keep in mind that solving the quantum linear systems problem is not the same as solving a linear system of equations. In the search for quantum solutions, it is necessary to reframe classical problems in quantum terms. However, this process may not preserve the core essence of the original problem, requiring additional hurdles, for instance, giving rise to the input and output problem in the QLSP. In pursuit of quantum advantage, proper benchmarks comparing classical and quantum algorithms are required, while not loosing track of the possible real-life practicality of these solvers.

In closing, given the time it has taken to polish and optimize the classical methods to solve linear systems, an area that remains actively researched to date, we expect this to be only the beginning of the collective efforts to make fundamental calculations more efficient using quantum subroutines --- the downstream impact of which will be evident in its applications.

\vspace{0.4cm}
{\em Acknowledgments:} We Alexander Dalzell, Chris Ferrie, Robin Kothari, Troy Lee and Rolando Somma for discussions and feedback. MESM acknowledges support from the U.S. Department of Defense through a QuICS Hartree Fellowship. LP, KK, and PR acknowledge support by the National Research Foundation, Singapore, and A*STAR under its CQT Bridging Grant and its Quantum Engineering Programme under grant NRF2021-QEP2-02-P05. A.A.-G. acknowledges the generous support from Dr. Anders G. Frøseth, Natural Resources Canada, and the Canada 150 Research Chairs Program. LL acknowledges support from the Challenge Institute for Quantum Computation (CIQC) funded by National Science Foundation (NSF) through grant number OMA-2016245.
\bibliography{bibliography}

\begin{thebibliography}{134}%
\makeatletter
\providecommand \@ifxundefined [1]{%
 \@ifx{#1\undefined}
}%
\providecommand \@ifnum [1]{%
 \ifnum #1\expandafter \@firstoftwo
 \else \expandafter \@secondoftwo
 \fi
}%
\providecommand \@ifx [1]{%
 \ifx #1\expandafter \@firstoftwo
 \else \expandafter \@secondoftwo
 \fi
}%
\providecommand \natexlab [1]{#1}%
\providecommand \enquote  [1]{``#1''}%
\providecommand \bibnamefont  [1]{#1}%
\providecommand \bibfnamefont [1]{#1}%
\providecommand \citenamefont [1]{#1}%
\providecommand \href@noop [0]{\@secondoftwo}%
\providecommand \href [0]{\begingroup \@sanitize@url \@href}%
\providecommand \@href[1]{\@@startlink{#1}\@@href}%
\providecommand \@@href[1]{\endgroup#1\@@endlink}%
\providecommand \@sanitize@url [0]{\catcode `\\12\catcode `\$12\catcode `\&12\catcode `\#12\catcode `\^12\catcode `\_12\catcode `\%12\relax}%
\providecommand \@@startlink[1]{}%
\providecommand \@@endlink[0]{}%
\providecommand \url  [0]{\begingroup\@sanitize@url \@url }%
\providecommand \@url [1]{\endgroup\@href {#1}{\urlprefix }}%
\providecommand \urlprefix  [0]{URL }%
\providecommand \Eprint [0]{\href }%
\providecommand \doibase [0]{http://dx.doi.org/}%
\providecommand \selectlanguage [0]{\@gobble}%
\providecommand \bibinfo  [0]{\@secondoftwo}%
\providecommand \bibfield  [0]{\@secondoftwo}%
\providecommand \translation [1]{[#1]}%
\providecommand \BibitemOpen [0]{}%
\providecommand \bibitemStop [0]{}%
\providecommand \bibitemNoStop [0]{.\EOS\space}%
\providecommand \EOS [0]{\spacefactor3000\relax}%
\providecommand \BibitemShut  [1]{\csname bibitem#1\endcsname}%
\let\auto@bib@innerbib\@empty
%</preamble>
\bibitem [{\citenamefont {Wendland}(2017)}]{wendland2017numerical}%
  \BibitemOpen
  \bibfield  {author} {\bibinfo {author} {\bibfnamefont {H.}~\bibnamefont {Wendland}},\ }\href@noop {} {\emph {\bibinfo {title} {Numerical linear algebra: An introduction}}},\ Vol.~\bibinfo {volume} {56}\ (\bibinfo  {publisher} {Cambridge University Press},\ \bibinfo {year} {2017})\BibitemShut {NoStop}%
\bibitem [{\citenamefont {Nielsen}\ and\ \citenamefont {Chuang}(2010)}]{nielsen_quantum_2010}%
  \BibitemOpen
  \bibfield  {author} {\bibinfo {author} {\bibfnamefont {M.~A.}\ \bibnamefont {Nielsen}}\ and\ \bibinfo {author} {\bibfnamefont {I.~L.}\ \bibnamefont {Chuang}},\ }\href@noop {} {\emph {\bibinfo {title} {Quantum computation and quantum information}}}\ (\bibinfo  {publisher} {Cambridge university press},\ \bibinfo {year} {2010})\BibitemShut {NoStop}%
\bibitem [{\citenamefont {Shor}(1997)}]{shor_polynomial_1997}%
  \BibitemOpen
  \bibfield  {author} {\bibinfo {author} {\bibfnamefont {P.~W.}\ \bibnamefont {Shor}},\ }\href {\doibase 10.1137/s0097539795293172} {\bibfield  {journal} {\bibinfo  {journal} {SIAM Journal on Computing}\ }\textbf {\bibinfo {volume} {26}},\ \bibinfo {pages} {1484–1509} (\bibinfo {year} {1997})}\BibitemShut {NoStop}%
\bibitem [{\citenamefont {Harrow}\ \emph {et~al.}(2009)\citenamefont {Harrow}, \citenamefont {Hassidim},\ and\ \citenamefont {Lloyd}}]{harrow_quantum_2009}%
  \BibitemOpen
  \bibfield  {author} {\bibinfo {author} {\bibfnamefont {A.~W.}\ \bibnamefont {Harrow}}, \bibinfo {author} {\bibfnamefont {A.}~\bibnamefont {Hassidim}}, \ and\ \bibinfo {author} {\bibfnamefont {S.}~\bibnamefont {Lloyd}},\ }\href {\doibase 10.1103/PhysRevLett.103.150502} {\bibfield  {journal} {\bibinfo  {journal} {Physical Review Letters}\ }\textbf {\bibinfo {volume} {103}},\ \bibinfo {pages} {150502} (\bibinfo {year} {2009})}\BibitemShut {NoStop}%
\bibitem [{\citenamefont {Montanaro}(2016)}]{montanaro_quantum_2016}%
  \BibitemOpen
  \bibfield  {author} {\bibinfo {author} {\bibfnamefont {A.}~\bibnamefont {Montanaro}},\ }\href {\doibase 10.1038/npjqi.2015.23} {\bibfield  {journal} {\bibinfo  {journal} {npj Quantum Information}\ }\textbf {\bibinfo {volume} {2}} (\bibinfo {year} {2016}),\ 10.1038/npjqi.2015.23}\BibitemShut {NoStop}%
\bibitem [{\citenamefont {Dalzell}\ \emph {et~al.}(2023)\citenamefont {Dalzell}, \citenamefont {McArdle}, \citenamefont {Berta}, \citenamefont {Bienias}, \citenamefont {Chen}, \citenamefont {Gily{\'e}n}, \citenamefont {Hann}, \citenamefont {Kastoryano}, \citenamefont {Khabiboulline}, \citenamefont {Kubica} \emph {et~al.}}]{dalzell2023quantum}%
  \BibitemOpen
  \bibfield  {author} {\bibinfo {author} {\bibfnamefont {A.~M.}\ \bibnamefont {Dalzell}}, \bibinfo {author} {\bibfnamefont {S.}~\bibnamefont {McArdle}}, \bibinfo {author} {\bibfnamefont {M.}~\bibnamefont {Berta}}, \bibinfo {author} {\bibfnamefont {P.}~\bibnamefont {Bienias}}, \bibinfo {author} {\bibfnamefont {C.-F.}\ \bibnamefont {Chen}}, \bibinfo {author} {\bibfnamefont {A.}~\bibnamefont {Gily{\'e}n}}, \bibinfo {author} {\bibfnamefont {C.~T.}\ \bibnamefont {Hann}}, \bibinfo {author} {\bibfnamefont {M.~J.}\ \bibnamefont {Kastoryano}}, \bibinfo {author} {\bibfnamefont {E.~T.}\ \bibnamefont {Khabiboulline}}, \bibinfo {author} {\bibfnamefont {A.}~\bibnamefont {Kubica}},  \emph {et~al.},\ }\href {https://arxiv.org/abs/2310.03011} {\bibfield  {journal} {\bibinfo  {journal} {arXiv preprint arXiv:2310.03011}\ } (\bibinfo {year} {2023})}\BibitemShut {NoStop}%
\bibitem [{\citenamefont {Aaronson}(2015)}]{aaronson_read_2015}%
  \BibitemOpen
  \bibfield  {author} {\bibinfo {author} {\bibfnamefont {S.}~\bibnamefont {Aaronson}},\ }\href {\doibase 10.1038/nphys3272} {\bibfield  {journal} {\bibinfo  {journal} {Nature Physics}\ }\textbf {\bibinfo {volume} {11}},\ \bibinfo {pages} {291} (\bibinfo {year} {2015})}\BibitemShut {NoStop}%
\bibitem [{\citenamefont {Childs}\ \emph {et~al.}(2017)\citenamefont {Childs}, \citenamefont {Kothari},\ and\ \citenamefont {Somma}}]{childs_quantum_2017}%
  \BibitemOpen
  \bibfield  {author} {\bibinfo {author} {\bibfnamefont {A.~M.}\ \bibnamefont {Childs}}, \bibinfo {author} {\bibfnamefont {R.}~\bibnamefont {Kothari}}, \ and\ \bibinfo {author} {\bibfnamefont {R.~D.}\ \bibnamefont {Somma}},\ }\href {\doibase 10.1137/16M1087072} {\bibfield  {journal} {\bibinfo  {journal} {SIAM Journal on Computing}\ }\textbf {\bibinfo {volume} {46}},\ \bibinfo {pages} {1920} (\bibinfo {year} {2017})}\BibitemShut {NoStop}%
\bibitem [{\citenamefont {Ambainis}(2010)}]{ambainis_variable_2010}%
  \BibitemOpen
  \bibfield  {author} {\bibinfo {author} {\bibfnamefont {A.}~\bibnamefont {Ambainis}},\ }\href@noop {} {\enquote {\bibinfo {title} {Variable time amplitude amplification and a faster quantum algorithm for solving systems of linear equations},}\ } (\bibinfo {year} {2010}),\ \Eprint {http://arxiv.org/abs/1010.4458} {arXiv:1010.4458 [quant-ph]} \BibitemShut {NoStop}%
\bibitem [{\citenamefont {Gily{\'e}n}\ \emph {et~al.}(2019)\citenamefont {Gily{\'e}n}, \citenamefont {Su}, \citenamefont {Low},\ and\ \citenamefont {Wiebe}}]{gilyen2019quantum}%
  \BibitemOpen
  \bibfield  {author} {\bibinfo {author} {\bibfnamefont {A.}~\bibnamefont {Gily{\'e}n}}, \bibinfo {author} {\bibfnamefont {Y.}~\bibnamefont {Su}}, \bibinfo {author} {\bibfnamefont {G.~H.}\ \bibnamefont {Low}}, \ and\ \bibinfo {author} {\bibfnamefont {N.}~\bibnamefont {Wiebe}},\ }in\ \href {https://dl.acm.org/doi/10.1145/3313276.3316366} {\emph {\bibinfo {booktitle} {Proceedings of the 51st Annual ACM SIGACT Symposium on Theory of Computing}}}\ (\bibinfo {year} {2019})\ pp.\ \bibinfo {pages} {193--204}\BibitemShut {NoStop}%
\bibitem [{\citenamefont {Subasi}\ \emph {et~al.}(2019)\citenamefont {Subasi}, \citenamefont {Somma},\ and\ \citenamefont {Orsucci}}]{Subasi2019adiabatic}%
  \BibitemOpen
  \bibfield  {author} {\bibinfo {author} {\bibfnamefont {Y.}~\bibnamefont {Subasi}}, \bibinfo {author} {\bibfnamefont {R.~D.}\ \bibnamefont {Somma}}, \ and\ \bibinfo {author} {\bibfnamefont {D.}~\bibnamefont {Orsucci}},\ }\href {\doibase 10.1103/PhysRevLett.122.060504} {\bibfield  {journal} {\bibinfo  {journal} {Physical Review Letters}\ }\textbf {\bibinfo {volume} {122}},\ \bibinfo {pages} {060504} (\bibinfo {year} {2019})}\BibitemShut {NoStop}%
\bibitem [{\citenamefont {An}\ and\ \citenamefont {Lin}(2022)}]{an_quantum_2020}%
  \BibitemOpen
  \bibfield  {author} {\bibinfo {author} {\bibfnamefont {D.}~\bibnamefont {An}}\ and\ \bibinfo {author} {\bibfnamefont {L.}~\bibnamefont {Lin}},\ }\href {\doibase 10.1145/3498331} {\bibfield  {journal} {\bibinfo  {journal} {ACM Transactions on Quantum Computing}\ }\textbf {\bibinfo {volume} {3}} (\bibinfo {year} {2022}),\ 10.1145/3498331}\BibitemShut {NoStop}%
\bibitem [{\citenamefont {Lin}\ and\ \citenamefont {Tong}(2020)}]{Lin2020optimalpolynomial}%
  \BibitemOpen
  \bibfield  {author} {\bibinfo {author} {\bibfnamefont {L.}~\bibnamefont {Lin}}\ and\ \bibinfo {author} {\bibfnamefont {Y.}~\bibnamefont {Tong}},\ }\href {\doibase 10.22331/q-2020-11-11-361} {\bibfield  {journal} {\bibinfo  {journal} {Quantum}\ }\textbf {\bibinfo {volume} {4}},\ \bibinfo {pages} {361} (\bibinfo {year} {2020})}\BibitemShut {NoStop}%
\bibitem [{\citenamefont {Costa}\ \emph {et~al.}(2021)\citenamefont {Costa}, \citenamefont {An}, \citenamefont {Sanders}, \citenamefont {Su}, \citenamefont {Babbush},\ and\ \citenamefont {Berry}}]{costa_optimal_2021}%
  \BibitemOpen
  \bibfield  {author} {\bibinfo {author} {\bibfnamefont {P.~C.~S.}\ \bibnamefont {Costa}}, \bibinfo {author} {\bibfnamefont {D.}~\bibnamefont {An}}, \bibinfo {author} {\bibfnamefont {Y.~R.}\ \bibnamefont {Sanders}}, \bibinfo {author} {\bibfnamefont {Y.}~\bibnamefont {Su}}, \bibinfo {author} {\bibfnamefont {R.}~\bibnamefont {Babbush}}, \ and\ \bibinfo {author} {\bibfnamefont {D.~W.}\ \bibnamefont {Berry}},\ }\href@noop {} {\enquote {\bibinfo {title} {Optimal scaling quantum linear systems solver via discrete adiabatic theorem},}\ } (\bibinfo {year} {2021}),\ \Eprint {http://arxiv.org/abs/2111.08152} {arXiv:2111.08152 [quant-ph]} \BibitemShut {NoStop}%
\bibitem [{\citenamefont {Dranov}\ \emph {et~al.}(1998)\citenamefont {Dranov}, \citenamefont {Kellendonk},\ and\ \citenamefont {Seiler}}]{dranov1998discrete}%
  \BibitemOpen
  \bibfield  {author} {\bibinfo {author} {\bibfnamefont {A.}~\bibnamefont {Dranov}}, \bibinfo {author} {\bibfnamefont {J.}~\bibnamefont {Kellendonk}}, \ and\ \bibinfo {author} {\bibfnamefont {R.}~\bibnamefont {Seiler}},\ }\href {https://pubs.aip.org/aip/jmp/article-abstract/39/3/1340/290167/Discrete-time-adiabatic-theorems-for-quantum?redirectedFrom=fulltext} {\bibfield  {journal} {\bibinfo  {journal} {Journal of Mathematical Physics}\ }\textbf {\bibinfo {volume} {39}},\ \bibinfo {pages} {1340} (\bibinfo {year} {1998})}\BibitemShut {NoStop}%
\bibitem [{\citenamefont {Dalzell}(2024)}]{dalzell2024shortcutoptimalquantumlinear}%
  \BibitemOpen
  \bibfield  {author} {\bibinfo {author} {\bibfnamefont {A.~M.}\ \bibnamefont {Dalzell}},\ }\href {https://arxiv.org/abs/2406.12086} {\enquote {\bibinfo {title} {A shortcut to an optimal quantum linear system solver},}\ } (\bibinfo {year} {2024}),\ \Eprint {http://arxiv.org/abs/2406.12086} {arXiv:2406.12086 [quant-ph]} \BibitemShut {NoStop}%
\bibitem [{\citenamefont {Cai}\ \emph {et~al.}(2013)\citenamefont {Cai}, \citenamefont {Weedbrook}, \citenamefont {Su}, \citenamefont {Chen}, \citenamefont {Gu}, \citenamefont {Zhu}, \citenamefont {Li}, \citenamefont {Liu}, \citenamefont {Lu},\ and\ \citenamefont {Pan}}]{cai_experimental_2013}%
  \BibitemOpen
  \bibfield  {author} {\bibinfo {author} {\bibfnamefont {X.-D.}\ \bibnamefont {Cai}}, \bibinfo {author} {\bibfnamefont {C.}~\bibnamefont {Weedbrook}}, \bibinfo {author} {\bibfnamefont {Z.-E.}\ \bibnamefont {Su}}, \bibinfo {author} {\bibfnamefont {M.-C.}\ \bibnamefont {Chen}}, \bibinfo {author} {\bibfnamefont {M.}~\bibnamefont {Gu}}, \bibinfo {author} {\bibfnamefont {M.-J.}\ \bibnamefont {Zhu}}, \bibinfo {author} {\bibfnamefont {L.}~\bibnamefont {Li}}, \bibinfo {author} {\bibfnamefont {N.-L.}\ \bibnamefont {Liu}}, \bibinfo {author} {\bibfnamefont {C.-Y.}\ \bibnamefont {Lu}}, \ and\ \bibinfo {author} {\bibfnamefont {J.-W.}\ \bibnamefont {Pan}},\ }\href {https://journals.aps.org/prl/abstract/10.1103/PhysRevLett.110.230501} {\bibfield  {journal} {\bibinfo  {journal} {Physical Review Letters}\ }\textbf {\bibinfo {volume} {110}},\ \bibinfo {pages} {230501} (\bibinfo {year} {2013})}\BibitemShut {NoStop}%
\bibitem [{\citenamefont {Barz}\ \emph {et~al.}(2014)\citenamefont {Barz}, \citenamefont {Kassal}, \citenamefont {Ringbauer}, \citenamefont {Lipp}, \citenamefont {Daki{\'c}}, \citenamefont {Aspuru-Guzik},\ and\ \citenamefont {Walther}}]{barz_solving_2013}%
  \BibitemOpen
  \bibfield  {author} {\bibinfo {author} {\bibfnamefont {S.}~\bibnamefont {Barz}}, \bibinfo {author} {\bibfnamefont {I.}~\bibnamefont {Kassal}}, \bibinfo {author} {\bibfnamefont {M.}~\bibnamefont {Ringbauer}}, \bibinfo {author} {\bibfnamefont {Y.~O.}\ \bibnamefont {Lipp}}, \bibinfo {author} {\bibfnamefont {B.}~\bibnamefont {Daki{\'c}}}, \bibinfo {author} {\bibfnamefont {A.}~\bibnamefont {Aspuru-Guzik}}, \ and\ \bibinfo {author} {\bibfnamefont {P.}~\bibnamefont {Walther}},\ }\href {https://www.nature.com/articles/srep06115} {\bibfield  {journal} {\bibinfo  {journal} {Scientific reports}\ }\textbf {\bibinfo {volume} {4}},\ \bibinfo {pages} {6115} (\bibinfo {year} {2014})}\BibitemShut {NoStop}%
\bibitem [{\citenamefont {Pan}\ \emph {et~al.}(2014)\citenamefont {Pan}, \citenamefont {Cao}, \citenamefont {Yao}, \citenamefont {Li}, \citenamefont {Ju}, \citenamefont {Chen}, \citenamefont {Peng}, \citenamefont {Kais},\ and\ \citenamefont {Du}}]{pan_experimental_2014}%
  \BibitemOpen
  \bibfield  {author} {\bibinfo {author} {\bibfnamefont {J.}~\bibnamefont {Pan}}, \bibinfo {author} {\bibfnamefont {Y.}~\bibnamefont {Cao}}, \bibinfo {author} {\bibfnamefont {X.}~\bibnamefont {Yao}}, \bibinfo {author} {\bibfnamefont {Z.}~\bibnamefont {Li}}, \bibinfo {author} {\bibfnamefont {C.}~\bibnamefont {Ju}}, \bibinfo {author} {\bibfnamefont {H.}~\bibnamefont {Chen}}, \bibinfo {author} {\bibfnamefont {X.}~\bibnamefont {Peng}}, \bibinfo {author} {\bibfnamefont {S.}~\bibnamefont {Kais}}, \ and\ \bibinfo {author} {\bibfnamefont {J.}~\bibnamefont {Du}},\ }\href {https://journals.aps.org/pra/abstract/10.1103/PhysRevA.89.022313} {\bibfield  {journal} {\bibinfo  {journal} {Physical Review A}\ }\textbf {\bibinfo {volume} {89}},\ \bibinfo {pages} {022313} (\bibinfo {year} {2014})}\BibitemShut {NoStop}%
\bibitem [{\citenamefont {Dervovic}\ \emph {et~al.}(2018)\citenamefont {Dervovic}, \citenamefont {Herbster}, \citenamefont {Mountney}, \citenamefont {Severini}, \citenamefont {Usher},\ and\ \citenamefont {Wossnig}}]{dervovic_quantum_2018}%
  \BibitemOpen
  \bibfield  {author} {\bibinfo {author} {\bibfnamefont {D.}~\bibnamefont {Dervovic}}, \bibinfo {author} {\bibfnamefont {M.}~\bibnamefont {Herbster}}, \bibinfo {author} {\bibfnamefont {P.}~\bibnamefont {Mountney}}, \bibinfo {author} {\bibfnamefont {S.}~\bibnamefont {Severini}}, \bibinfo {author} {\bibfnamefont {N.}~\bibnamefont {Usher}}, \ and\ \bibinfo {author} {\bibfnamefont {L.}~\bibnamefont {Wossnig}},\ }\href {https://arxiv.org/abs/1802.08227} {\enquote {\bibinfo {title} {Quantum linear systems algorithms: A primer},}\ } (\bibinfo {year} {2018}),\ \Eprint {http://arxiv.org/abs/1802.08227} {arXiv:1802.08227 [quant-ph]} \BibitemShut {NoStop}%
\bibitem [{\citenamefont {Cleve}\ \emph {et~al.}(1997)\citenamefont {Cleve}, \citenamefont {Ekert}, \citenamefont {Macchiavello},\ and\ \citenamefont {Mosca}}]{cleve_quantum_1997}%
  \BibitemOpen
  \bibfield  {author} {\bibinfo {author} {\bibfnamefont {R.}~\bibnamefont {Cleve}}, \bibinfo {author} {\bibfnamefont {A.}~\bibnamefont {Ekert}}, \bibinfo {author} {\bibfnamefont {C.}~\bibnamefont {Macchiavello}}, \ and\ \bibinfo {author} {\bibfnamefont {M.}~\bibnamefont {Mosca}},\ }\href {\doibase 10.1098/rspa.1998.0164} {\bibfield  {journal} {\bibinfo  {journal} {Proceedings of the Royal Society A: Mathematical, Physical and Engineering Sciences}\ }\textbf {\bibinfo {volume} {454}} (\bibinfo {year} {1997}),\ 10.1098/rspa.1998.0164}\BibitemShut {NoStop}%
\bibitem [{\citenamefont {Brassard}\ \emph {et~al.}(2002)\citenamefont {Brassard}, \citenamefont {Hoyer}, \citenamefont {Mosca},\ and\ \citenamefont {Tapp}}]{brassard2002quantum}%
  \BibitemOpen
  \bibfield  {author} {\bibinfo {author} {\bibfnamefont {G.}~\bibnamefont {Brassard}}, \bibinfo {author} {\bibfnamefont {P.}~\bibnamefont {Hoyer}}, \bibinfo {author} {\bibfnamefont {M.}~\bibnamefont {Mosca}}, \ and\ \bibinfo {author} {\bibfnamefont {A.}~\bibnamefont {Tapp}},\ }\href@noop {} {\bibfield  {journal} {\bibinfo  {journal} {Contemporary Mathematics}\ }\textbf {\bibinfo {volume} {305}},\ \bibinfo {pages} {53} (\bibinfo {year} {2002})}\BibitemShut {NoStop}%
\bibitem [{\citenamefont {Low}(2019)}]{Low2019hamsim}%
  \BibitemOpen
  \bibfield  {author} {\bibinfo {author} {\bibfnamefont {G.~H.}\ \bibnamefont {Low}},\ }in\ \href {\doibase 10.1145/3313276.3316386} {\emph {\bibinfo {booktitle} {Proceedings of the 51st Annual ACM SIGACT Symposium on Theory of Computing}}},\ \bibinfo {series and number} {STOC 2019}\ (\bibinfo  {publisher} {Association for Computing Machinery},\ \bibinfo {address} {New York, NY, USA},\ \bibinfo {year} {2019})\ p.\ \bibinfo {pages} {491–502}\BibitemShut {NoStop}%
\bibitem [{\citenamefont {Shewchuk}(1994)}]{shewchuk_introduction_1994}%
  \BibitemOpen
  \bibfield  {author} {\bibinfo {author} {\bibfnamefont {J.}~\bibnamefont {Shewchuk}},\ }\href {https://axon.cs.byu.edu/~martinez/classes/678/Papers/painless-conjugate-gradient.pdf} {\emph {\bibinfo {title} {An Introduction to the Conjugate Gradient Method Without the Agonizing Pain}}}\ (\bibinfo  {publisher} {Carnegie Mellon University, Department of Computer Science},\ \bibinfo {year} {1994})\BibitemShut {NoStop}%
\bibitem [{\citenamefont {Lin}(2022)}]{lin2022lecture}%
  \BibitemOpen
  \bibfield  {author} {\bibinfo {author} {\bibfnamefont {L.}~\bibnamefont {Lin}},\ }\href {https://arxiv.org/abs/2201.08309} {\bibfield  {journal} {\bibinfo  {journal} {arXiv preprint arXiv:2201.08309}\ } (\bibinfo {year} {2022})}\BibitemShut {NoStop}%
\bibitem [{\citenamefont {Sanders}\ \emph {et~al.}(2019)\citenamefont {Sanders}, \citenamefont {Low}, \citenamefont {Scherer},\ and\ \citenamefont {Berry}}]{sanders2012blackbox}%
  \BibitemOpen
  \bibfield  {author} {\bibinfo {author} {\bibfnamefont {Y.~R.}\ \bibnamefont {Sanders}}, \bibinfo {author} {\bibfnamefont {G.~H.}\ \bibnamefont {Low}}, \bibinfo {author} {\bibfnamefont {A.}~\bibnamefont {Scherer}}, \ and\ \bibinfo {author} {\bibfnamefont {D.~W.}\ \bibnamefont {Berry}},\ }\href {\doibase 10.1103/PhysRevLett.122.020502} {\bibfield  {journal} {\bibinfo  {journal} {Physical Review Letters}\ }\textbf {\bibinfo {volume} {122}},\ \bibinfo {pages} {020502} (\bibinfo {year} {2019})}\BibitemShut {NoStop}%
\bibitem [{\citenamefont {Berry}\ \emph {et~al.}(2007)\citenamefont {Berry}, \citenamefont {Ahokas}, \citenamefont {Cleve},\ and\ \citenamefont {Sanders}}]{berry2007efficient}%
  \BibitemOpen
  \bibfield  {author} {\bibinfo {author} {\bibfnamefont {D.~W.}\ \bibnamefont {Berry}}, \bibinfo {author} {\bibfnamefont {G.}~\bibnamefont {Ahokas}}, \bibinfo {author} {\bibfnamefont {R.}~\bibnamefont {Cleve}}, \ and\ \bibinfo {author} {\bibfnamefont {B.~C.}\ \bibnamefont {Sanders}},\ }\href {https://link.springer.com/article/10.1007/s00220-006-0150-x} {\bibfield  {journal} {\bibinfo  {journal} {Communications in Mathematical Physics}\ }\textbf {\bibinfo {volume} {270}},\ \bibinfo {pages} {359} (\bibinfo {year} {2007})}\BibitemShut {NoStop}%
\bibitem [{\citenamefont {Yoder}\ \emph {et~al.}(2014)\citenamefont {Yoder}, \citenamefont {Low},\ and\ \citenamefont {Chuang}}]{Yoder2014fixed}%
  \BibitemOpen
  \bibfield  {author} {\bibinfo {author} {\bibfnamefont {T.~J.}\ \bibnamefont {Yoder}}, \bibinfo {author} {\bibfnamefont {G.~H.}\ \bibnamefont {Low}}, \ and\ \bibinfo {author} {\bibfnamefont {I.~L.}\ \bibnamefont {Chuang}},\ }\href {\doibase 10.1103/physrevlett.113.210501} {\bibfield  {journal} {\bibinfo  {journal} {Physical Review Letters}\ }\textbf {\bibinfo {volume} {113}} (\bibinfo {year} {2014}),\ 10.1103/physrevlett.113.210501}\BibitemShut {NoStop}%
\bibitem [{\citenamefont {Berry}\ \emph {et~al.}(2015)\citenamefont {Berry}, \citenamefont {Childs},\ and\ \citenamefont {Kothari}}]{berry_optimal_2015}%
  \BibitemOpen
  \bibfield  {author} {\bibinfo {author} {\bibfnamefont {D.~W.}\ \bibnamefont {Berry}}, \bibinfo {author} {\bibfnamefont {A.~M.}\ \bibnamefont {Childs}}, \ and\ \bibinfo {author} {\bibfnamefont {R.}~\bibnamefont {Kothari}},\ }in\ \href {\doibase 10.1109/FOCS.2015.54} {\emph {\bibinfo {booktitle} {2015 IEEE 56th Annual Symposium on Foundations of Computer Science}}}\ (\bibinfo {year} {2015})\ pp.\ \bibinfo {pages} {792--809}\BibitemShut {NoStop}%
\bibitem [{\citenamefont {Low}\ and\ \citenamefont {Chuang}(2017)}]{low2017optimal}%
  \BibitemOpen
  \bibfield  {author} {\bibinfo {author} {\bibfnamefont {G.~H.}\ \bibnamefont {Low}}\ and\ \bibinfo {author} {\bibfnamefont {I.~L.}\ \bibnamefont {Chuang}},\ }\href {https://journals.aps.org/prl/abstract/10.1103/PhysRevLett.118.010501} {\bibfield  {journal} {\bibinfo  {journal} {Physical Review Letters}\ }\textbf {\bibinfo {volume} {118}},\ \bibinfo {pages} {010501} (\bibinfo {year} {2017})}\BibitemShut {NoStop}%
\bibitem [{\citenamefont {Martyn}\ \emph {et~al.}(2021)\citenamefont {Martyn}, \citenamefont {Rossi}, \citenamefont {Tan},\ and\ \citenamefont {Chuang}}]{martyn2021grand}%
  \BibitemOpen
  \bibfield  {author} {\bibinfo {author} {\bibfnamefont {J.~M.}\ \bibnamefont {Martyn}}, \bibinfo {author} {\bibfnamefont {Z.~M.}\ \bibnamefont {Rossi}}, \bibinfo {author} {\bibfnamefont {A.~K.}\ \bibnamefont {Tan}}, \ and\ \bibinfo {author} {\bibfnamefont {I.~L.}\ \bibnamefont {Chuang}},\ }\href {https://link.aps.org/doi/10.1103/PRXQuantum.2.040203} {\bibfield  {journal} {\bibinfo  {journal} {PRX Quantum}\ }\textbf {\bibinfo {volume} {2}},\ \bibinfo {pages} {040203} (\bibinfo {year} {2021})}\BibitemShut {NoStop}%
\bibitem [{\citenamefont {An}\ \emph {et~al.}(2023{\natexlab{a}})\citenamefont {An}, \citenamefont {Liu},\ and\ \citenamefont {Lin}}]{an2023linear}%
  \BibitemOpen
  \bibfield  {author} {\bibinfo {author} {\bibfnamefont {D.}~\bibnamefont {An}}, \bibinfo {author} {\bibfnamefont {J.-P.}\ \bibnamefont {Liu}}, \ and\ \bibinfo {author} {\bibfnamefont {L.}~\bibnamefont {Lin}},\ }\href {https://journals.aps.org/prl/abstract/10.1103/PhysRevLett.131.150603} {\bibfield  {journal} {\bibinfo  {journal} {Physical Review Letters}\ }\textbf {\bibinfo {volume} {131}},\ \bibinfo {pages} {150603} (\bibinfo {year} {2023}{\natexlab{a}})}\BibitemShut {NoStop}%
\bibitem [{\citenamefont {An}\ \emph {et~al.}(2023{\natexlab{b}})\citenamefont {An}, \citenamefont {Childs},\ and\ \citenamefont {Lin}}]{an2023quantum}%
  \BibitemOpen
  \bibfield  {author} {\bibinfo {author} {\bibfnamefont {D.}~\bibnamefont {An}}, \bibinfo {author} {\bibfnamefont {A.~M.}\ \bibnamefont {Childs}}, \ and\ \bibinfo {author} {\bibfnamefont {L.}~\bibnamefont {Lin}},\ }\href {https://arxiv.org/abs/2312.03916} {\bibfield  {journal} {\bibinfo  {journal} {arXiv preprint arXiv:2312.03916}\ } (\bibinfo {year} {2023}{\natexlab{b}})}\BibitemShut {NoStop}%
\bibitem [{\citenamefont {Berry}\ and\ \citenamefont {Childs}(2012)}]{berry_blackbox_2012}%
  \BibitemOpen
  \bibfield  {author} {\bibinfo {author} {\bibfnamefont {D.~W.}\ \bibnamefont {Berry}}\ and\ \bibinfo {author} {\bibfnamefont {A.~M.}\ \bibnamefont {Childs}},\ }\href {https://doi.org/10.26421/QIC12.1-2-4} {\bibfield  {journal} {\bibinfo  {journal} {Quantum Information \& Computation}\ }\textbf {\bibinfo {volume} {12}},\ \bibinfo {pages} {29} (\bibinfo {year} {2012})}\BibitemShut {NoStop}%
\bibitem [{\citenamefont {Low}\ and\ \citenamefont {Su}(2024{\natexlab{a}})}]{low2024quantumlinearalgorithmoptimal}%
  \BibitemOpen
  \bibfield  {author} {\bibinfo {author} {\bibfnamefont {G.~H.}\ \bibnamefont {Low}}\ and\ \bibinfo {author} {\bibfnamefont {Y.}~\bibnamefont {Su}},\ }\href {https://arxiv.org/abs/2410.18178} {\enquote {\bibinfo {title} {Quantum linear system algorithm with optimal queries to initial state preparation},}\ } (\bibinfo {year} {2024}{\natexlab{a}}),\ \Eprint {http://arxiv.org/abs/2410.18178} {arXiv:2410.18178 [quant-ph]} \BibitemShut {NoStop}%
\bibitem [{\citenamefont {Tang}\ and\ \citenamefont {Tian}(2024)}]{tang2024cs}%
  \BibitemOpen
  \bibfield  {author} {\bibinfo {author} {\bibfnamefont {E.}~\bibnamefont {Tang}}\ and\ \bibinfo {author} {\bibfnamefont {K.}~\bibnamefont {Tian}},\ }in\ \href {https://epubs.siam.org/doi/10.1137/1.9781611977936.13} {\emph {\bibinfo {booktitle} {2024 Symposium on Simplicity in Algorithms (SOSA)}}}\ (\bibinfo {organization} {SIAM},\ \bibinfo {year} {2024})\ pp.\ \bibinfo {pages} {121--143}\BibitemShut {NoStop}%
\bibitem [{\citenamefont {Ying}(2022)}]{ying2022stable}%
  \BibitemOpen
  \bibfield  {author} {\bibinfo {author} {\bibfnamefont {L.}~\bibnamefont {Ying}},\ }\href {\doibase 10.22331/q-2022-10-20-842} {\bibfield  {journal} {\bibinfo  {journal} {Quantum}\ }\textbf {\bibinfo {volume} {6}},\ \bibinfo {pages} {842} (\bibinfo {year} {2022})}\BibitemShut {NoStop}%
\bibitem [{\citenamefont {Boixo}\ \emph {et~al.}(2009)\citenamefont {Boixo}, \citenamefont {Knill},\ and\ \citenamefont {Somma}}]{Boixo2009eigenpath}%
  \BibitemOpen
  \bibfield  {author} {\bibinfo {author} {\bibfnamefont {S.}~\bibnamefont {Boixo}}, \bibinfo {author} {\bibfnamefont {E.}~\bibnamefont {Knill}}, \ and\ \bibinfo {author} {\bibfnamefont {R.}~\bibnamefont {Somma}},\ }\href {\doibase 10.26421/QIC9.9-10-7} {\bibfield  {journal} {\bibinfo  {journal} {Quantum Information and Computation}\ }\textbf {\bibinfo {volume} {9}} (\bibinfo {year} {2009}),\ 10.26421/QIC9.9-10-7}\BibitemShut {NoStop}%
\bibitem [{\citenamefont {Somma}\ and\ \citenamefont {Boixo}(2013)}]{Somma2013gapamplification}%
  \BibitemOpen
  \bibfield  {author} {\bibinfo {author} {\bibfnamefont {R.~D.}\ \bibnamefont {Somma}}\ and\ \bibinfo {author} {\bibfnamefont {S.}~\bibnamefont {Boixo}},\ }\href {\doibase 10.1137/120871997} {\bibfield  {journal} {\bibinfo  {journal} {SIAM Journal on Computing}\ }\textbf {\bibinfo {volume} {42}},\ \bibinfo {pages} {593} (\bibinfo {year} {2013})},\ \Eprint {http://arxiv.org/abs/https://doi.org/10.1137/120871997} {https://doi.org/10.1137/120871997} \BibitemShut {NoStop}%
\bibitem [{\citenamefont {{Hao Low}}\ and\ \citenamefont {{Wiebe}}(2018)}]{Low2018interaction}%
  \BibitemOpen
  \bibfield  {author} {\bibinfo {author} {\bibfnamefont {G.}~\bibnamefont {{Hao Low}}}\ and\ \bibinfo {author} {\bibfnamefont {N.}~\bibnamefont {{Wiebe}}},\ }\href {\doibase 10.48550/arXiv.1805.00675} {\bibfield  {journal} {\bibinfo  {journal} {arXiv e-prints}\ ,\ \bibinfo {eid} {arXiv:1805.00675}} (\bibinfo {year} {2018})},\ \Eprint {http://arxiv.org/abs/1805.00675} {arXiv:1805.00675 [quant-ph]} \BibitemShut {NoStop}%
\bibitem [{\citenamefont {Atia}\ and\ \citenamefont {Aharonov}(2017)}]{Atia2017}%
  \BibitemOpen
  \bibfield  {author} {\bibinfo {author} {\bibfnamefont {Y.}~\bibnamefont {Atia}}\ and\ \bibinfo {author} {\bibfnamefont {D.}~\bibnamefont {Aharonov}},\ }\href {\doibase 10.1038/s41467-017-01637-7} {\bibfield  {journal} {\bibinfo  {journal} {Nature Communications}\ }\textbf {\bibinfo {volume} {8}},\ \bibinfo {pages} {1572} (\bibinfo {year} {2017})}\BibitemShut {NoStop}%
\bibitem [{\citenamefont {Haah}\ \emph {et~al.}(2018)\citenamefont {Haah}, \citenamefont {Hastings}, \citenamefont {Kothari},\ and\ \citenamefont {Low}}]{Haah2018}%
  \BibitemOpen
  \bibfield  {author} {\bibinfo {author} {\bibfnamefont {J.}~\bibnamefont {Haah}}, \bibinfo {author} {\bibfnamefont {M.}~\bibnamefont {Hastings}}, \bibinfo {author} {\bibfnamefont {R.}~\bibnamefont {Kothari}}, \ and\ \bibinfo {author} {\bibfnamefont {G.~H.}\ \bibnamefont {Low}},\ }in\ \href {\doibase 10.1109/FOCS.2018.00041} {\emph {\bibinfo {booktitle} {2018 IEEE 59th Annual Symposium on Foundations of Computer Science (FOCS)}}}\ (\bibinfo {year} {2018})\ pp.\ \bibinfo {pages} {350--360}\BibitemShut {NoStop}%
\bibitem [{\citenamefont {Jansen}\ \emph {et~al.}(2007)\citenamefont {Jansen}, \citenamefont {Ruskai},\ and\ \citenamefont {Seiler}}]{jansen2007bounds}%
  \BibitemOpen
  \bibfield  {author} {\bibinfo {author} {\bibfnamefont {S.}~\bibnamefont {Jansen}}, \bibinfo {author} {\bibfnamefont {M.-B.}\ \bibnamefont {Ruskai}}, \ and\ \bibinfo {author} {\bibfnamefont {R.}~\bibnamefont {Seiler}},\ }\href {https://pubs.aip.org/aip/jmp/article/48/10/102111/379272/Bounds-for-the-adiabatic-approximation-with} {\bibfield  {journal} {\bibinfo  {journal} {Journal of Mathematical Physics}\ }\textbf {\bibinfo {volume} {48}} (\bibinfo {year} {2007})}\BibitemShut {NoStop}%
\bibitem [{\citenamefont {Grover}(2005)}]{grover2005fixed}%
  \BibitemOpen
  \bibfield  {author} {\bibinfo {author} {\bibfnamefont {L.~K.}\ \bibnamefont {Grover}},\ }\href {\doibase 10.1103/PhysRevLett.95.150501} {\bibfield  {journal} {\bibinfo  {journal} {Phys. Rev. Lett.}\ }\textbf {\bibinfo {volume} {95}},\ \bibinfo {pages} {150501} (\bibinfo {year} {2005})}\BibitemShut {NoStop}%
\bibitem [{\citenamefont {Szegedy}(2004)}]{szegedy2004quantum}%
  \BibitemOpen
  \bibfield  {author} {\bibinfo {author} {\bibfnamefont {M.}~\bibnamefont {Szegedy}},\ }in\ \href {https://doi.org/10.1109/FOCS.2004.53} {\emph {\bibinfo {booktitle} {Proceedings of the 45th Annual IEEE Symposium on Foundations of Computer Science}}},\ \bibinfo {series and number} {FOCS '04}\ (\bibinfo  {publisher} {IEEE Computer Society},\ \bibinfo {address} {USA},\ \bibinfo {year} {2004})\ p.\ \bibinfo {pages} {32–41}\BibitemShut {NoStop}%
\bibitem [{\citenamefont {Berry}\ \emph {et~al.}(2018)\citenamefont {Berry}, \citenamefont {Kieferov{\'a}}, \citenamefont {Scherer}, \citenamefont {Sanders}, \citenamefont {Low}, \citenamefont {Wiebe}, \citenamefont {Gidney},\ and\ \citenamefont {Babbush}}]{berry2018improved}%
  \BibitemOpen
  \bibfield  {author} {\bibinfo {author} {\bibfnamefont {D.~W.}\ \bibnamefont {Berry}}, \bibinfo {author} {\bibfnamefont {M.}~\bibnamefont {Kieferov{\'a}}}, \bibinfo {author} {\bibfnamefont {A.}~\bibnamefont {Scherer}}, \bibinfo {author} {\bibfnamefont {Y.~R.}\ \bibnamefont {Sanders}}, \bibinfo {author} {\bibfnamefont {G.~H.}\ \bibnamefont {Low}}, \bibinfo {author} {\bibfnamefont {N.}~\bibnamefont {Wiebe}}, \bibinfo {author} {\bibfnamefont {C.}~\bibnamefont {Gidney}}, \ and\ \bibinfo {author} {\bibfnamefont {R.}~\bibnamefont {Babbush}},\ }\href {https://www.nature.com/articles/s41534-018-0071-5} {\bibfield  {journal} {\bibinfo  {journal} {npj Quantum Information}\ }\textbf {\bibinfo {volume} {4}},\ \bibinfo {pages} {22} (\bibinfo {year} {2018})}\BibitemShut {NoStop}%
\bibitem [{\citenamefont {Kothari}\ and\ \citenamefont {O'Donnell}(2023)}]{kothari2023estimation}%
  \BibitemOpen
  \bibfield  {author} {\bibinfo {author} {\bibfnamefont {R.}~\bibnamefont {Kothari}}\ and\ \bibinfo {author} {\bibfnamefont {R.}~\bibnamefont {O'Donnell}},\ }in\ \href {\doibase 10.1137/1.9781611977554.ch44} {\emph {\bibinfo {booktitle} {Proceedings of the 2023 Annual ACM-SIAM Symposium on Discrete Algorithms (SODA)}}}\ (\bibinfo {year} {2023})\ pp.\ \bibinfo {pages} {1186--1215},\ \Eprint {http://arxiv.org/abs/https://epubs.siam.org/doi/pdf/10.1137/1.9781611977554.ch44} {https://epubs.siam.org/doi/pdf/10.1137/1.9781611977554.ch44} \BibitemShut {NoStop}%
\bibitem [{\citenamefont {Giovannetti}\ \emph {et~al.}(2008)\citenamefont {Giovannetti}, \citenamefont {Lloyd},\ and\ \citenamefont {Maccone}}]{giovannetti2008quantum}%
  \BibitemOpen
  \bibfield  {author} {\bibinfo {author} {\bibfnamefont {V.}~\bibnamefont {Giovannetti}}, \bibinfo {author} {\bibfnamefont {S.}~\bibnamefont {Lloyd}}, \ and\ \bibinfo {author} {\bibfnamefont {L.}~\bibnamefont {Maccone}},\ }\href {https://journals.aps.org/prl/abstract/10.1103/PhysRevLett.100.160501} {\bibfield  {journal} {\bibinfo  {journal} {Physical Review Letters}\ }\textbf {\bibinfo {volume} {100}},\ \bibinfo {pages} {160501} (\bibinfo {year} {2008})}\BibitemShut {NoStop}%
\bibitem [{\citenamefont {Aharonov}\ and\ \citenamefont {Ta-Shma}(2003)}]{aharonov2003adiabaticquantumstategeneration}%
  \BibitemOpen
  \bibfield  {author} {\bibinfo {author} {\bibfnamefont {D.}~\bibnamefont {Aharonov}}\ and\ \bibinfo {author} {\bibfnamefont {A.}~\bibnamefont {Ta-Shma}},\ }\href {https://arxiv.org/abs/quant-ph/0301023} {\enquote {\bibinfo {title} {Adiabatic quantum state generation and statistical zero knowledge},}\ } (\bibinfo {year} {2003}),\ \Eprint {http://arxiv.org/abs/quant-ph/0301023} {arXiv:quant-ph/0301023 [quant-ph]} \BibitemShut {NoStop}%
\bibitem [{\citenamefont {Childs}\ and\ \citenamefont {Kothari}(2011)}]{childs_simulating_2011}%
  \BibitemOpen
  \bibfield  {author} {\bibinfo {author} {\bibfnamefont {A.~M.}\ \bibnamefont {Childs}}\ and\ \bibinfo {author} {\bibfnamefont {R.}~\bibnamefont {Kothari}},\ }\enquote {\bibinfo {title} {Simulating sparse hamiltonians with star decompositions},}\ in\ \href {\doibase 10.1007/978-3-642-18073-6_8} {\emph {\bibinfo {booktitle} {Theory of Quantum Computation, Communication, and Cryptography}}}\ (\bibinfo  {publisher} {Springer Berlin Heidelberg},\ \bibinfo {year} {2011})\ pp.\ \bibinfo {pages} {94--103}\BibitemShut {NoStop}%
\bibitem [{\citenamefont {Low}\ \emph {et~al.}(2016)\citenamefont {Low}, \citenamefont {Yoder},\ and\ \citenamefont {Chuang}}]{low2016methodology}%
  \BibitemOpen
  \bibfield  {author} {\bibinfo {author} {\bibfnamefont {G.~H.}\ \bibnamefont {Low}}, \bibinfo {author} {\bibfnamefont {T.~J.}\ \bibnamefont {Yoder}}, \ and\ \bibinfo {author} {\bibfnamefont {I.~L.}\ \bibnamefont {Chuang}},\ }\href@noop {} {\bibfield  {journal} {\bibinfo  {journal} {Physical Review X}\ }\textbf {\bibinfo {volume} {6}},\ \bibinfo {pages} {041067} (\bibinfo {year} {2016})}\BibitemShut {NoStop}%
\bibitem [{\citenamefont {Costa}\ \emph {et~al.}(2024)\citenamefont {Costa}, \citenamefont {An}, \citenamefont {Babbush},\ and\ \citenamefont {Berry}}]{costa2024discreteadiabaticquantumlinear}%
  \BibitemOpen
  \bibfield  {author} {\bibinfo {author} {\bibfnamefont {P.~C.~S.}\ \bibnamefont {Costa}}, \bibinfo {author} {\bibfnamefont {D.}~\bibnamefont {An}}, \bibinfo {author} {\bibfnamefont {R.}~\bibnamefont {Babbush}}, \ and\ \bibinfo {author} {\bibfnamefont {D.}~\bibnamefont {Berry}},\ }\href {https://arxiv.org/abs/2312.07690} {\enquote {\bibinfo {title} {The discrete adiabatic quantum linear system solver has lower constant factors than the randomized adiabatic solver},}\ } (\bibinfo {year} {2024}),\ \Eprint {http://arxiv.org/abs/2312.07690} {arXiv:2312.07690 [quant-ph]} \BibitemShut {NoStop}%
\bibitem [{\citenamefont {Bagherimehrab}\ \emph {et~al.}(2023)\citenamefont {Bagherimehrab}, \citenamefont {Nakaji}, \citenamefont {Wiebe},\ and\ \citenamefont {Aspuru-Guzik}}]{bagherimehrab2023fast}%
  \BibitemOpen
  \bibfield  {author} {\bibinfo {author} {\bibfnamefont {M.}~\bibnamefont {Bagherimehrab}}, \bibinfo {author} {\bibfnamefont {K.}~\bibnamefont {Nakaji}}, \bibinfo {author} {\bibfnamefont {N.}~\bibnamefont {Wiebe}}, \ and\ \bibinfo {author} {\bibfnamefont {A.}~\bibnamefont {Aspuru-Guzik}},\ }\href {https://arxiv.org/abs/2306.11802} {\enquote {\bibinfo {title} {Fast quantum algorithm for differential equations},}\ } (\bibinfo {year} {2023}),\ \Eprint {http://arxiv.org/abs/2306.11802} {arXiv:2306.11802 [quant-ph]} \BibitemShut {NoStop}%
\bibitem [{\citenamefont {Tong}\ \emph {et~al.}(2021)\citenamefont {Tong}, \citenamefont {An}, \citenamefont {Wiebe},\ and\ \citenamefont {Lin}}]{tong_fast_2021}%
  \BibitemOpen
  \bibfield  {author} {\bibinfo {author} {\bibfnamefont {Y.}~\bibnamefont {Tong}}, \bibinfo {author} {\bibfnamefont {D.}~\bibnamefont {An}}, \bibinfo {author} {\bibfnamefont {N.}~\bibnamefont {Wiebe}}, \ and\ \bibinfo {author} {\bibfnamefont {L.}~\bibnamefont {Lin}},\ }\href@noop {} {\bibfield  {journal} {\bibinfo  {journal} {Physical Review A}\ }\textbf {\bibinfo {volume} {104}},\ \bibinfo {pages} {032422} (\bibinfo {year} {2021})}\BibitemShut {NoStop}%
\bibitem [{\citenamefont {Clader}\ \emph {et~al.}(2013)\citenamefont {Clader}, \citenamefont {Jacobs},\ and\ \citenamefont {Sprouse}}]{clader2013preconditioned}%
  \BibitemOpen
  \bibfield  {author} {\bibinfo {author} {\bibfnamefont {B.~D.}\ \bibnamefont {Clader}}, \bibinfo {author} {\bibfnamefont {B.~C.}\ \bibnamefont {Jacobs}}, \ and\ \bibinfo {author} {\bibfnamefont {C.~R.}\ \bibnamefont {Sprouse}},\ }\href {\doibase 10.1103/physrevlett.110.250504} {\bibfield  {journal} {\bibinfo  {journal} {Physical Review Letters}\ }\textbf {\bibinfo {volume} {110}} (\bibinfo {year} {2013}),\ 10.1103/physrevlett.110.250504}\BibitemShut {NoStop}%
\bibitem [{\citenamefont {Knill}\ \emph {et~al.}(2007)\citenamefont {Knill}, \citenamefont {Ortiz},\ and\ \citenamefont {Somma}}]{knill2007optimal}%
  \BibitemOpen
  \bibfield  {author} {\bibinfo {author} {\bibfnamefont {E.}~\bibnamefont {Knill}}, \bibinfo {author} {\bibfnamefont {G.}~\bibnamefont {Ortiz}}, \ and\ \bibinfo {author} {\bibfnamefont {R.~D.}\ \bibnamefont {Somma}},\ }\href {\doibase 10.1103/PhysRevA.75.012328} {\bibfield  {journal} {\bibinfo  {journal} {Phys. Rev. A}\ }\textbf {\bibinfo {volume} {75}},\ \bibinfo {pages} {012328} (\bibinfo {year} {2007})}\BibitemShut {NoStop}%
\bibitem [{\citenamefont {Alase}\ \emph {et~al.}(2022)\citenamefont {Alase}, \citenamefont {Nerem}, \citenamefont {Bagherimehrab}, \citenamefont {Høyer},\ and\ \citenamefont {Sanders}}]{alase_tight_2022}%
  \BibitemOpen
  \bibfield  {author} {\bibinfo {author} {\bibfnamefont {A.}~\bibnamefont {Alase}}, \bibinfo {author} {\bibfnamefont {R.~R.}\ \bibnamefont {Nerem}}, \bibinfo {author} {\bibfnamefont {M.}~\bibnamefont {Bagherimehrab}}, \bibinfo {author} {\bibfnamefont {P.}~\bibnamefont {Høyer}}, \ and\ \bibinfo {author} {\bibfnamefont {B.~C.}\ \bibnamefont {Sanders}},\ }\href@noop {} {\bibfield  {journal} {\bibinfo  {journal} {Physical Review Research}\ }\textbf {\bibinfo {volume} {4}},\ \bibinfo {pages} {023237} (\bibinfo {year} {2022})}\BibitemShut {NoStop}%
\bibitem [{\citenamefont {Huggins}\ \emph {et~al.}(2022)\citenamefont {Huggins}, \citenamefont {Wan}, \citenamefont {McClean}, \citenamefont {O'Brien}, \citenamefont {Wiebe},\ and\ \citenamefont {Babbush}}]{huggins2022nearlyoptimal}%
  \BibitemOpen
  \bibfield  {author} {\bibinfo {author} {\bibfnamefont {W.~J.}\ \bibnamefont {Huggins}}, \bibinfo {author} {\bibfnamefont {K.}~\bibnamefont {Wan}}, \bibinfo {author} {\bibfnamefont {J.}~\bibnamefont {McClean}}, \bibinfo {author} {\bibfnamefont {T.~E.}\ \bibnamefont {O'Brien}}, \bibinfo {author} {\bibfnamefont {N.}~\bibnamefont {Wiebe}}, \ and\ \bibinfo {author} {\bibfnamefont {R.}~\bibnamefont {Babbush}},\ }\href {\doibase 10.1103/PhysRevLett.129.240501} {\bibfield  {journal} {\bibinfo  {journal} {Phys. Rev. Lett.}\ }\textbf {\bibinfo {volume} {129}},\ \bibinfo {pages} {240501} (\bibinfo {year} {2022})}\BibitemShut {NoStop}%
\bibitem [{Note1()}]{Note1}%
  \BibitemOpen
  \bibinfo {note} {Although we do not go into the details of complexity theory, for completeness we briefly comment on complexity classes. $\protect \mathsf {BQP}$ and $\protect \mathsf {PSPACE}$ correspond to classes of decision problems, i.e., subsets of the set of all bitstrings $\{0,1\}^*$. Given such set $L$, an algorithm must decide if some particular bitstring $z$ is in $L$ or not. The class $\protect \mathsf {BQP}$ corresponds to problems that can be decided in polynomial time by quantum algorithms and $\protect \mathsf {PSPACE}$ corresponds to problems decidable in polynomial space. A more detailed exposition on these definitions and background on theoretical computer science can be found in Ref.~\cite {Arora2009}}\BibitemShut {NoStop}%
\bibitem [{\citenamefont {Harrow}\ and\ \citenamefont {Kothari}(2021)}]{harrow_inpreparation_2021}%
  \BibitemOpen
  \bibfield  {author} {\bibinfo {author} {\bibfnamefont {A.~W.}\ \bibnamefont {Harrow}}\ and\ \bibinfo {author} {\bibfnamefont {R.}~\bibnamefont {Kothari}},\ }\href@noop {} {} (\bibinfo {year} {2021}),\ \bibinfo {note} {in preparation}\BibitemShut {NoStop}%
\bibitem [{\citenamefont {Orsucci}\ and\ \citenamefont {Dunjko}(2021)}]{Orsucci2021solvingclassesof}%
  \BibitemOpen
  \bibfield  {author} {\bibinfo {author} {\bibfnamefont {D.}~\bibnamefont {Orsucci}}\ and\ \bibinfo {author} {\bibfnamefont {V.}~\bibnamefont {Dunjko}},\ }\href {\doibase 10.22331/q-2021-11-08-573} {\bibfield  {journal} {\bibinfo  {journal} {Quantum}\ }\textbf {\bibinfo {volume} {5}},\ \bibinfo {pages} {573} (\bibinfo {year} {2021})}\BibitemShut {NoStop}%
\bibitem [{\citenamefont {Wang}\ and\ \citenamefont {Zhang}(2024)}]{wang_tight_2024}%
  \BibitemOpen
  \bibfield  {author} {\bibinfo {author} {\bibfnamefont {Q.}~\bibnamefont {Wang}}\ and\ \bibinfo {author} {\bibfnamefont {Z.}~\bibnamefont {Zhang}},\ }\href {\doibase 10.1103/physreva.110.012422} {\bibfield  {journal} {\bibinfo  {journal} {Physical Review A}\ }\textbf {\bibinfo {volume} {110}} (\bibinfo {year} {2024}),\ 10.1103/physreva.110.012422}\BibitemShut {NoStop}%
\bibitem [{\citenamefont {Jennings}\ \emph {et~al.}(2023)\citenamefont {Jennings}, \citenamefont {Lostaglio}, \citenamefont {Pallister}, \citenamefont {Sornborger},\ and\ \citenamefont {Subaşı}}]{jennings_efficientquantumlinearsolver_2023}%
  \BibitemOpen
  \bibfield  {author} {\bibinfo {author} {\bibfnamefont {D.}~\bibnamefont {Jennings}}, \bibinfo {author} {\bibfnamefont {M.}~\bibnamefont {Lostaglio}}, \bibinfo {author} {\bibfnamefont {S.}~\bibnamefont {Pallister}}, \bibinfo {author} {\bibfnamefont {A.~T.}\ \bibnamefont {Sornborger}}, \ and\ \bibinfo {author} {\bibfnamefont {Y.}~\bibnamefont {Subaşı}},\ }\href {https://arxiv.org/abs/2305.11352} {\enquote {\bibinfo {title} {Efficient quantum linear solver algorithm with detailed running costs},}\ } (\bibinfo {year} {2023}),\ \Eprint {http://arxiv.org/abs/2305.11352} {arXiv:2305.11352 [quant-ph]} \BibitemShut {NoStop}%
\bibitem [{\citenamefont {Preskill}(2018)}]{preskill_quantum_2018}%
  \BibitemOpen
  \bibfield  {author} {\bibinfo {author} {\bibfnamefont {J.}~\bibnamefont {Preskill}},\ }\href {\doibase 10.22331/q-2018-08-06-79} {\bibfield  {journal} {\bibinfo  {journal} {Quantum}\ }\textbf {\bibinfo {volume} {2}},\ \bibinfo {pages} {79} (\bibinfo {year} {2018})}\BibitemShut {NoStop}%
\bibitem [{\citenamefont {Bharti}\ \emph {et~al.}(2022)\citenamefont {Bharti}, \citenamefont {Cervera-Lierta}, \citenamefont {Kyaw}, \citenamefont {Haug}, \citenamefont {Alperin-Lea}, \citenamefont {Anand}, \citenamefont {Degroote}, \citenamefont {Heimonen}, \citenamefont {Kottmann}, \citenamefont {Menke}, \citenamefont {Mok}, \citenamefont {Sim}, \citenamefont {Kwek},\ and\ \citenamefont {Aspuru-Guzik}}]{bharti_noisyintermediatealgorithms_2022}%
  \BibitemOpen
  \bibfield  {author} {\bibinfo {author} {\bibfnamefont {K.}~\bibnamefont {Bharti}}, \bibinfo {author} {\bibfnamefont {A.}~\bibnamefont {Cervera-Lierta}}, \bibinfo {author} {\bibfnamefont {T.~H.}\ \bibnamefont {Kyaw}}, \bibinfo {author} {\bibfnamefont {T.}~\bibnamefont {Haug}}, \bibinfo {author} {\bibfnamefont {S.}~\bibnamefont {Alperin-Lea}}, \bibinfo {author} {\bibfnamefont {A.}~\bibnamefont {Anand}}, \bibinfo {author} {\bibfnamefont {M.}~\bibnamefont {Degroote}}, \bibinfo {author} {\bibfnamefont {H.}~\bibnamefont {Heimonen}}, \bibinfo {author} {\bibfnamefont {J.~S.}\ \bibnamefont {Kottmann}}, \bibinfo {author} {\bibfnamefont {T.}~\bibnamefont {Menke}}, \bibinfo {author} {\bibfnamefont {W.-K.}\ \bibnamefont {Mok}}, \bibinfo {author} {\bibfnamefont {S.}~\bibnamefont {Sim}}, \bibinfo {author} {\bibfnamefont {L.-C.}\ \bibnamefont {Kwek}}, \ and\ \bibinfo {author} {\bibfnamefont {A.}~\bibnamefont {Aspuru-Guzik}},\ }\href {\doibase 10.1103/revmodphys.94.015004} {\bibfield  {journal} {\bibinfo  {journal} {Reviews
  of Modern Physics}\ }\textbf {\bibinfo {volume} {94}} (\bibinfo {year} {2022}),\ 10.1103/revmodphys.94.015004}\BibitemShut {NoStop}%
\bibitem [{\citenamefont {Katabarwa}\ \emph {et~al.}(2024)\citenamefont {Katabarwa}, \citenamefont {Gratsea}, \citenamefont {Caesura},\ and\ \citenamefont {Johnson}}]{katabarwa_early_2023}%
  \BibitemOpen
  \bibfield  {author} {\bibinfo {author} {\bibfnamefont {A.}~\bibnamefont {Katabarwa}}, \bibinfo {author} {\bibfnamefont {K.}~\bibnamefont {Gratsea}}, \bibinfo {author} {\bibfnamefont {A.}~\bibnamefont {Caesura}}, \ and\ \bibinfo {author} {\bibfnamefont {P.~D.}\ \bibnamefont {Johnson}},\ }\href {https://journals.aps.org/prxquantum/abstract/10.1103/PRXQuantum.5.020101} {\bibfield  {journal} {\bibinfo  {journal} {PRX Quantum}\ }\textbf {\bibinfo {volume} {5}},\ \bibinfo {pages} {020101} (\bibinfo {year} {2024})}\BibitemShut {NoStop}%
\bibitem [{\citenamefont {Liang}\ \emph {et~al.}(2024)\citenamefont {Liang}, \citenamefont {Zhou}, \citenamefont {Dalal},\ and\ \citenamefont {Johnson}}]{liang_modeling_2024}%
  \BibitemOpen
  \bibfield  {author} {\bibinfo {author} {\bibfnamefont {Q.}~\bibnamefont {Liang}}, \bibinfo {author} {\bibfnamefont {Y.}~\bibnamefont {Zhou}}, \bibinfo {author} {\bibfnamefont {A.}~\bibnamefont {Dalal}}, \ and\ \bibinfo {author} {\bibfnamefont {P.}~\bibnamefont {Johnson}},\ }\href {\doibase 10.1103/PhysRevResearch.6.023118} {\bibfield  {journal} {\bibinfo  {journal} {Phys. Rev. Res.}\ }\textbf {\bibinfo {volume} {6}},\ \bibinfo {pages} {023118} (\bibinfo {year} {2024})}\BibitemShut {NoStop}%
\bibitem [{\citenamefont {Ni}\ \emph {et~al.}(2023)\citenamefont {Ni}, \citenamefont {Li},\ and\ \citenamefont {Ying}}]{ni_lowdepth_2023}%
  \BibitemOpen
  \bibfield  {author} {\bibinfo {author} {\bibfnamefont {H.}~\bibnamefont {Ni}}, \bibinfo {author} {\bibfnamefont {H.}~\bibnamefont {Li}}, \ and\ \bibinfo {author} {\bibfnamefont {L.}~\bibnamefont {Ying}},\ }\href {\doibase 10.22331/q-2023-11-06-1165} {\bibfield  {journal} {\bibinfo  {journal} {Quantum}\ }\textbf {\bibinfo {volume} {7}},\ \bibinfo {pages} {1165} (\bibinfo {year} {2023})}\BibitemShut {NoStop}%
\bibitem [{\citenamefont {Huang}\ \emph {et~al.}(2021)\citenamefont {Huang}, \citenamefont {Bharti},\ and\ \citenamefont {Rebentrost}}]{huang_near_2021}%
  \BibitemOpen
  \bibfield  {author} {\bibinfo {author} {\bibfnamefont {H.-Y.}\ \bibnamefont {Huang}}, \bibinfo {author} {\bibfnamefont {K.}~\bibnamefont {Bharti}}, \ and\ \bibinfo {author} {\bibfnamefont {P.}~\bibnamefont {Rebentrost}},\ }\href {\doibase 10.1088/1367-2630/ac325f} {\bibfield  {journal} {\bibinfo  {journal} {New Journal of Physics}\ }\textbf {\bibinfo {volume} {23}},\ \bibinfo {pages} {113021} (\bibinfo {year} {2021})}\BibitemShut {NoStop}%
\bibitem [{\citenamefont {Bravo-Prieto}\ \emph {et~al.}(2023)\citenamefont {Bravo-Prieto}, \citenamefont {LaRose}, \citenamefont {Cerezo}, \citenamefont {Subasi}, \citenamefont {Cincio},\ and\ \citenamefont {Coles}}]{BravoPrieto_variational_2023}%
  \BibitemOpen
  \bibfield  {author} {\bibinfo {author} {\bibfnamefont {C.}~\bibnamefont {Bravo-Prieto}}, \bibinfo {author} {\bibfnamefont {R.}~\bibnamefont {LaRose}}, \bibinfo {author} {\bibfnamefont {M.}~\bibnamefont {Cerezo}}, \bibinfo {author} {\bibfnamefont {Y.}~\bibnamefont {Subasi}}, \bibinfo {author} {\bibfnamefont {L.}~\bibnamefont {Cincio}}, \ and\ \bibinfo {author} {\bibfnamefont {P.~J.}\ \bibnamefont {Coles}},\ }\href {\doibase 10.22331/q-2023-11-22-1188} {\bibfield  {journal} {\bibinfo  {journal} {{Quantum}}\ }\textbf {\bibinfo {volume} {7}},\ \bibinfo {pages} {1188} (\bibinfo {year} {2023})}\BibitemShut {NoStop}%
\bibitem [{\citenamefont {Xu}\ \emph {et~al.}(2021)\citenamefont {Xu}, \citenamefont {Sun}, \citenamefont {Endo}, \citenamefont {Li}, \citenamefont {Benjamin},\ and\ \citenamefont {Yuan}}]{xu_variational_2021}%
  \BibitemOpen
  \bibfield  {author} {\bibinfo {author} {\bibfnamefont {X.}~\bibnamefont {Xu}}, \bibinfo {author} {\bibfnamefont {J.}~\bibnamefont {Sun}}, \bibinfo {author} {\bibfnamefont {S.}~\bibnamefont {Endo}}, \bibinfo {author} {\bibfnamefont {Y.}~\bibnamefont {Li}}, \bibinfo {author} {\bibfnamefont {S.~C.}\ \bibnamefont {Benjamin}}, \ and\ \bibinfo {author} {\bibfnamefont {X.}~\bibnamefont {Yuan}},\ }\href {\doibase 10.1016/j.scib.2021.06.023} {\bibfield  {journal} {\bibinfo  {journal} {Science Bulletin}\ }\textbf {\bibinfo {volume} {66}},\ \bibinfo {pages} {2181–2188} (\bibinfo {year} {2021})}\BibitemShut {NoStop}%
\bibitem [{\citenamefont {Perelshtein}\ \emph {et~al.}(2022)\citenamefont {Perelshtein}, \citenamefont {Pakhomchik}, \citenamefont {Melnikov}, \citenamefont {Novikov}, \citenamefont {Glatz}, \citenamefont {Paraoanu}, \citenamefont {Vinokur},\ and\ \citenamefont {Lesovik}}]{perelshtein_solving_2022}%
  \BibitemOpen
  \bibfield  {author} {\bibinfo {author} {\bibfnamefont {M.~R.}\ \bibnamefont {Perelshtein}}, \bibinfo {author} {\bibfnamefont {A.~I.}\ \bibnamefont {Pakhomchik}}, \bibinfo {author} {\bibfnamefont {A.~A.}\ \bibnamefont {Melnikov}}, \bibinfo {author} {\bibfnamefont {A.~A.}\ \bibnamefont {Novikov}}, \bibinfo {author} {\bibfnamefont {A.}~\bibnamefont {Glatz}}, \bibinfo {author} {\bibfnamefont {G.~S.}\ \bibnamefont {Paraoanu}}, \bibinfo {author} {\bibfnamefont {V.~M.}\ \bibnamefont {Vinokur}}, \ and\ \bibinfo {author} {\bibfnamefont {G.~B.}\ \bibnamefont {Lesovik}},\ }\href {\doibase 10.1002/andp.202200082} {\bibfield  {journal} {\bibinfo  {journal} {Annalen der Physik}\ }\textbf {\bibinfo {volume} {534}} (\bibinfo {year} {2022}),\ 10.1002/andp.202200082}\BibitemShut {NoStop}%
\bibitem [{\citenamefont {Chen}\ \emph {et~al.}(2024)\citenamefont {Chen}, \citenamefont {Ma}, \citenamefont {Ye}, \citenamefont {Xu}, \citenamefont {Tan}, \citenamefont {Zhuang}, \citenamefont {Xu}, \citenamefont {Wang}, \citenamefont {Sun}, \citenamefont {Chen}, \citenamefont {Du}, \citenamefont {Guo}, \citenamefont {Zhang}, \citenamefont {Tao}, \citenamefont {Wang}, \citenamefont {Yang}, \citenamefont {Zhao}, \citenamefont {Wang}, \citenamefont {Zhang}, \citenamefont {Zhang}, \citenamefont {Zhao}, \citenamefont {Jia}, \citenamefont {Kong}, \citenamefont {Dou}, \citenamefont {Wang}, \citenamefont {Liu}, \citenamefont {Xue}, \citenamefont {Zhang}, \citenamefont {Huang}, \citenamefont {Duan}, \citenamefont {Wu},\ and\ \citenamefont {Guo}}]{chen2024enablinglargescalehighprecisionfluid}%
  \BibitemOpen
  \bibfield  {author} {\bibinfo {author} {\bibfnamefont {Z.-Y.}\ \bibnamefont {Chen}}, \bibinfo {author} {\bibfnamefont {T.-Y.}\ \bibnamefont {Ma}}, \bibinfo {author} {\bibfnamefont {C.-C.}\ \bibnamefont {Ye}}, \bibinfo {author} {\bibfnamefont {L.}~\bibnamefont {Xu}}, \bibinfo {author} {\bibfnamefont {M.-Y.}\ \bibnamefont {Tan}}, \bibinfo {author} {\bibfnamefont {X.-N.}\ \bibnamefont {Zhuang}}, \bibinfo {author} {\bibfnamefont {X.-F.}\ \bibnamefont {Xu}}, \bibinfo {author} {\bibfnamefont {Y.-J.}\ \bibnamefont {Wang}}, \bibinfo {author} {\bibfnamefont {T.-P.}\ \bibnamefont {Sun}}, \bibinfo {author} {\bibfnamefont {Y.}~\bibnamefont {Chen}}, \bibinfo {author} {\bibfnamefont {L.}~\bibnamefont {Du}}, \bibinfo {author} {\bibfnamefont {L.-L.}\ \bibnamefont {Guo}}, \bibinfo {author} {\bibfnamefont {H.-F.}\ \bibnamefont {Zhang}}, \bibinfo {author} {\bibfnamefont {H.-R.}\ \bibnamefont {Tao}}, \bibinfo {author} {\bibfnamefont {T.-L.}\ \bibnamefont {Wang}}, \bibinfo {author} {\bibfnamefont {X.-Y.}\ \bibnamefont {Yang}},
  \bibinfo {author} {\bibfnamefont {Z.-A.}\ \bibnamefont {Zhao}}, \bibinfo {author} {\bibfnamefont {P.}~\bibnamefont {Wang}}, \bibinfo {author} {\bibfnamefont {S.}~\bibnamefont {Zhang}}, \bibinfo {author} {\bibfnamefont {C.}~\bibnamefont {Zhang}}, \bibinfo {author} {\bibfnamefont {R.-Z.}\ \bibnamefont {Zhao}}, \bibinfo {author} {\bibfnamefont {Z.-L.}\ \bibnamefont {Jia}}, \bibinfo {author} {\bibfnamefont {W.-C.}\ \bibnamefont {Kong}}, \bibinfo {author} {\bibfnamefont {M.-H.}\ \bibnamefont {Dou}}, \bibinfo {author} {\bibfnamefont {J.-C.}\ \bibnamefont {Wang}}, \bibinfo {author} {\bibfnamefont {H.-Y.}\ \bibnamefont {Liu}}, \bibinfo {author} {\bibfnamefont {C.}~\bibnamefont {Xue}}, \bibinfo {author} {\bibfnamefont {P.-J.-Y.}\ \bibnamefont {Zhang}}, \bibinfo {author} {\bibfnamefont {S.-H.}\ \bibnamefont {Huang}}, \bibinfo {author} {\bibfnamefont {P.}~\bibnamefont {Duan}}, \bibinfo {author} {\bibfnamefont {Y.-C.}\ \bibnamefont {Wu}}, \ and\ \bibinfo {author} {\bibfnamefont {G.-P.}\ \bibnamefont {Guo}},\ }\href
  {https://arxiv.org/abs/2406.06063} {\enquote {\bibinfo {title} {Enabling large-scale and high-precision fluid simulations on near-term quantum computers},}\ } (\bibinfo {year} {2024}),\ \Eprint {http://arxiv.org/abs/2406.06063} {arXiv:2406.06063 [physics.comp-ph]} \BibitemShut {NoStop}%
\bibitem [{\citenamefont {Pellow-Jarman}\ \emph {et~al.}(2023)\citenamefont {Pellow-Jarman}, \citenamefont {Sinayskiy}, \citenamefont {Pillay},\ and\ \citenamefont {Petruccione}}]{pellowjarman_nearterm_2023}%
  \BibitemOpen
  \bibfield  {author} {\bibinfo {author} {\bibfnamefont {A.}~\bibnamefont {Pellow-Jarman}}, \bibinfo {author} {\bibfnamefont {I.}~\bibnamefont {Sinayskiy}}, \bibinfo {author} {\bibfnamefont {A.}~\bibnamefont {Pillay}}, \ and\ \bibinfo {author} {\bibfnamefont {F.}~\bibnamefont {Petruccione}},\ }\href {\doibase 10.1007/s11128-023-04020-2} {\bibfield  {journal} {\bibinfo  {journal} {Quantum Information Processing}\ }\textbf {\bibinfo {volume} {22}},\ \bibinfo {pages} {258} (\bibinfo {year} {2023})}\BibitemShut {NoStop}%
\bibitem [{\citenamefont {O'Malley}\ \emph {et~al.}(2024)\citenamefont {O'Malley}, \citenamefont {Henderson}, \citenamefont {Pelofske}, \citenamefont {Greer}, \citenamefont {Subasi}, \citenamefont {Golden}, \citenamefont {Lowrie},\ and\ \citenamefont {Eidenbenz}}]{omalley_neartermquantumalgorithmsolving_2024}%
  \BibitemOpen
  \bibfield  {author} {\bibinfo {author} {\bibfnamefont {D.}~\bibnamefont {O'Malley}}, \bibinfo {author} {\bibfnamefont {J.~M.}\ \bibnamefont {Henderson}}, \bibinfo {author} {\bibfnamefont {E.}~\bibnamefont {Pelofske}}, \bibinfo {author} {\bibfnamefont {S.}~\bibnamefont {Greer}}, \bibinfo {author} {\bibfnamefont {Y.}~\bibnamefont {Subasi}}, \bibinfo {author} {\bibfnamefont {J.~K.}\ \bibnamefont {Golden}}, \bibinfo {author} {\bibfnamefont {R.}~\bibnamefont {Lowrie}}, \ and\ \bibinfo {author} {\bibfnamefont {S.}~\bibnamefont {Eidenbenz}},\ }\href {https://arxiv.org/abs/2205.00645} {\enquote {\bibinfo {title} {A near-term quantum algorithm for solving linear systems of equations based on the woodbury identity},}\ } (\bibinfo {year} {2024}),\ \Eprint {http://arxiv.org/abs/2205.00645} {arXiv:2205.00645 [quant-ph]} \BibitemShut {NoStop}%
\bibitem [{\citenamefont {Ghisoni}\ \emph {et~al.}(2024)\citenamefont {Ghisoni}, \citenamefont {Scala}, \citenamefont {Bajoni},\ and\ \citenamefont {Gerace}}]{ghisoni2024shadowquantumlinearsolver}%
  \BibitemOpen
  \bibfield  {author} {\bibinfo {author} {\bibfnamefont {F.}~\bibnamefont {Ghisoni}}, \bibinfo {author} {\bibfnamefont {F.}~\bibnamefont {Scala}}, \bibinfo {author} {\bibfnamefont {D.}~\bibnamefont {Bajoni}}, \ and\ \bibinfo {author} {\bibfnamefont {D.}~\bibnamefont {Gerace}},\ }\href {https://arxiv.org/abs/2409.08929} {\enquote {\bibinfo {title} {Shadow quantum linear solver: A resource efficient quantum algorithm for linear systems of equations},}\ } (\bibinfo {year} {2024}),\ \Eprint {http://arxiv.org/abs/2409.08929} {arXiv:2409.08929 [quant-ph]} \BibitemShut {NoStop}%
\bibitem [{\citenamefont {Larocca}\ \emph {et~al.}(2024)\citenamefont {Larocca}, \citenamefont {Thanasilp}, \citenamefont {Wang}, \citenamefont {Sharma}, \citenamefont {Biamonte}, \citenamefont {Coles}, \citenamefont {Cincio}, \citenamefont {McClean}, \citenamefont {Holmes},\ and\ \citenamefont {Cerezo}}]{larocca2024reviewbarrenplateausvariational}%
  \BibitemOpen
  \bibfield  {author} {\bibinfo {author} {\bibfnamefont {M.}~\bibnamefont {Larocca}}, \bibinfo {author} {\bibfnamefont {S.}~\bibnamefont {Thanasilp}}, \bibinfo {author} {\bibfnamefont {S.}~\bibnamefont {Wang}}, \bibinfo {author} {\bibfnamefont {K.}~\bibnamefont {Sharma}}, \bibinfo {author} {\bibfnamefont {J.}~\bibnamefont {Biamonte}}, \bibinfo {author} {\bibfnamefont {P.~J.}\ \bibnamefont {Coles}}, \bibinfo {author} {\bibfnamefont {L.}~\bibnamefont {Cincio}}, \bibinfo {author} {\bibfnamefont {J.~R.}\ \bibnamefont {McClean}}, \bibinfo {author} {\bibfnamefont {Z.}~\bibnamefont {Holmes}}, \ and\ \bibinfo {author} {\bibfnamefont {M.}~\bibnamefont {Cerezo}},\ }\href {https://arxiv.org/abs/2405.00781} {\bibfield  {journal} {\bibinfo  {journal} {arXiv preprint arXiv:2405.00781}\ } (\bibinfo {year} {2024})},\ \Eprint {http://arxiv.org/abs/2405.00781} {arXiv:2405.00781 [quant-ph]} \BibitemShut {NoStop}%
\bibitem [{\citenamefont {Grimsley}\ \emph {et~al.}(2019)\citenamefont {Grimsley}, \citenamefont {Economou}, \citenamefont {Barnes},\ and\ \citenamefont {Mayhall}}]{grimsley_adaptive_2019}%
  \BibitemOpen
  \bibfield  {author} {\bibinfo {author} {\bibfnamefont {H.~R.}\ \bibnamefont {Grimsley}}, \bibinfo {author} {\bibfnamefont {S.~E.}\ \bibnamefont {Economou}}, \bibinfo {author} {\bibfnamefont {E.}~\bibnamefont {Barnes}}, \ and\ \bibinfo {author} {\bibfnamefont {N.~J.}\ \bibnamefont {Mayhall}},\ }\href {\doibase 10.1038/s41467-019-10988-2} {\bibfield  {journal} {\bibinfo  {journal} {Nature Communications}\ }\textbf {\bibinfo {volume} {10}} (\bibinfo {year} {2019}),\ 10.1038/s41467-019-10988-2}\BibitemShut {NoStop}%
\bibitem [{\citenamefont {Low}\ and\ \citenamefont {Su}(2024{\natexlab{b}})}]{low2024quantumeigenvalueprocessing}%
  \BibitemOpen
  \bibfield  {author} {\bibinfo {author} {\bibfnamefont {G.~H.}\ \bibnamefont {Low}}\ and\ \bibinfo {author} {\bibfnamefont {Y.}~\bibnamefont {Su}},\ }\href {https://arxiv.org/abs/2401.06240} {\enquote {\bibinfo {title} {Quantum eigenvalue processing},}\ } (\bibinfo {year} {2024}{\natexlab{b}}),\ \Eprint {http://arxiv.org/abs/2401.06240} {arXiv:2401.06240 [quant-ph]} \BibitemShut {NoStop}%
\bibitem [{\citenamefont {Augustino}\ \emph {et~al.}(2023)\citenamefont {Augustino}, \citenamefont {Nannicini}, \citenamefont {Terlaky},\ and\ \citenamefont {Zuluaga}}]{Augustino_2023}%
  \BibitemOpen
  \bibfield  {author} {\bibinfo {author} {\bibfnamefont {B.}~\bibnamefont {Augustino}}, \bibinfo {author} {\bibfnamefont {G.}~\bibnamefont {Nannicini}}, \bibinfo {author} {\bibfnamefont {T.}~\bibnamefont {Terlaky}}, \ and\ \bibinfo {author} {\bibfnamefont {L.~F.}\ \bibnamefont {Zuluaga}},\ }\href {\doibase 10.22331/q-2023-09-11-1110} {\bibfield  {journal} {\bibinfo  {journal} {Quantum}\ }\textbf {\bibinfo {volume} {7}},\ \bibinfo {pages} {1110} (\bibinfo {year} {2023})}\BibitemShut {NoStop}%
\bibitem [{\citenamefont {Butcher}(2016)}]{butcher2016classicalodesbook}%
  \BibitemOpen
  \bibfield  {author} {\bibinfo {author} {\bibfnamefont {J.}~\bibnamefont {Butcher}},\ }\href {https://books.google.ca/books?id=JlSvDAAAQBAJ} {\emph {\bibinfo {title} {Numerical Methods for Ordinary Differential Equations}}}\ (\bibinfo  {publisher} {Wiley},\ \bibinfo {year} {2016})\BibitemShut {NoStop}%
\bibitem [{\citenamefont {Ames}(2014)}]{ames2014classicalpdesbook1}%
  \BibitemOpen
  \bibfield  {author} {\bibinfo {author} {\bibfnamefont {W.}~\bibnamefont {Ames}},\ }\href {https://books.google.ca/books?id=KmjiBQAAQBAJ} {\emph {\bibinfo {title} {Numerical Methods for Partial Differential Equations}}},\ Computer Science and Scientific Computing\ (\bibinfo  {publisher} {Elsevier Science},\ \bibinfo {year} {2014})\BibitemShut {NoStop}%
\bibitem [{\citenamefont {Evans}\ \emph {et~al.}(2012)\citenamefont {Evans}, \citenamefont {Blackledge},\ and\ \citenamefont {Yardley}}]{evans2012classicalpdesbook2}%
  \BibitemOpen
  \bibfield  {author} {\bibinfo {author} {\bibfnamefont {G.}~\bibnamefont {Evans}}, \bibinfo {author} {\bibfnamefont {J.}~\bibnamefont {Blackledge}}, \ and\ \bibinfo {author} {\bibfnamefont {P.}~\bibnamefont {Yardley}},\ }\href {https://books.google.ca/books?id=O1f1BwAAQBAJ} {\emph {\bibinfo {title} {Numerical Methods for Partial Differential Equations}}},\ Springer Undergraduate Mathematics Series\ (\bibinfo  {publisher} {Springer London},\ \bibinfo {year} {2012})\BibitemShut {NoStop}%
\bibitem [{\citenamefont {Cockburn}\ \emph {et~al.}(2012)\citenamefont {Cockburn}, \citenamefont {Karniadakis},\ and\ \citenamefont {Shu}}]{cockburn2012classicaldgbook}%
  \BibitemOpen
  \bibfield  {author} {\bibinfo {author} {\bibfnamefont {B.}~\bibnamefont {Cockburn}}, \bibinfo {author} {\bibfnamefont {G.}~\bibnamefont {Karniadakis}}, \ and\ \bibinfo {author} {\bibfnamefont {C.}~\bibnamefont {Shu}},\ }\href {https://books.google.ca/books?id=uOHrCAAAQBAJ} {\emph {\bibinfo {title} {Discontinuous Galerkin Methods: Theory, Computation and Applications}}},\ Lecture Notes in Computational Science and Engineering\ (\bibinfo  {publisher} {Springer Berlin Heidelberg},\ \bibinfo {year} {2012})\BibitemShut {NoStop}%
\bibitem [{\citenamefont {Dahmen}\ \emph {et~al.}(1997)\citenamefont {Dahmen}, \citenamefont {Kurdila},\ and\ \citenamefont {Oswald}}]{dahmen1997classicalmultiscalebook}%
  \BibitemOpen
  \bibfield  {author} {\bibinfo {author} {\bibfnamefont {W.}~\bibnamefont {Dahmen}}, \bibinfo {author} {\bibfnamefont {A.}~\bibnamefont {Kurdila}}, \ and\ \bibinfo {author} {\bibfnamefont {P.}~\bibnamefont {Oswald}},\ }\href {https://books.google.ca/books?id=mgXxybT83jwC} {\emph {\bibinfo {title} {Multiscale Wavelet Methods for Partial Differential Equations}}},\ ISSN\ (\bibinfo  {publisher} {Elsevier Science},\ \bibinfo {year} {1997})\BibitemShut {NoStop}%
\bibitem [{\citenamefont {Cao}\ \emph {et~al.}(2013)\citenamefont {Cao}, \citenamefont {Papageorgiou}, \citenamefont {Petras}, \citenamefont {Traub},\ and\ \citenamefont {Kais}}]{cao2013quantum}%
  \BibitemOpen
  \bibfield  {author} {\bibinfo {author} {\bibfnamefont {Y.}~\bibnamefont {Cao}}, \bibinfo {author} {\bibfnamefont {A.}~\bibnamefont {Papageorgiou}}, \bibinfo {author} {\bibfnamefont {I.}~\bibnamefont {Petras}}, \bibinfo {author} {\bibfnamefont {J.}~\bibnamefont {Traub}}, \ and\ \bibinfo {author} {\bibfnamefont {S.}~\bibnamefont {Kais}},\ }\href {\doibase 10.1088/1367-2630/15/1/013021} {\bibfield  {journal} {\bibinfo  {journal} {New Journal of Physics}\ }\textbf {\bibinfo {volume} {15}},\ \bibinfo {pages} {013021} (\bibinfo {year} {2013})}\BibitemShut {NoStop}%
\bibitem [{\citenamefont {Childs}\ \emph {et~al.}(2020)\citenamefont {Childs}, \citenamefont {Liu},\ and\ \citenamefont {Ostrander}}]{childs_highprecision_2020}%
  \BibitemOpen
  \bibfield  {author} {\bibinfo {author} {\bibfnamefont {A.~M.}\ \bibnamefont {Childs}}, \bibinfo {author} {\bibfnamefont {J.-P.}\ \bibnamefont {Liu}}, \ and\ \bibinfo {author} {\bibfnamefont {A.}~\bibnamefont {Ostrander}},\ }\href@noop {} {\enquote {\bibinfo {title} {High-precision quantum algorithms for partial differential equations},}\ } (\bibinfo {year} {2020}),\ \Eprint {http://arxiv.org/abs/2002.07868} {arXiv:2002.07868 [quant-ph]} \BibitemShut {NoStop}%
\bibitem [{\citenamefont {Childs}\ and\ \citenamefont {Liu}(2020)}]{childs2020quantum}%
  \BibitemOpen
  \bibfield  {author} {\bibinfo {author} {\bibfnamefont {A.~M.}\ \bibnamefont {Childs}}\ and\ \bibinfo {author} {\bibfnamefont {J.-P.}\ \bibnamefont {Liu}},\ }\href {https://link.springer.com/article/10.1007/s00220-020-03699-z} {\bibfield  {journal} {\bibinfo  {journal} {Communications in Mathematical Physics}\ }\textbf {\bibinfo {volume} {375}},\ \bibinfo {pages} {1427} (\bibinfo {year} {2020})}\BibitemShut {NoStop}%
\bibitem [{\citenamefont {Linden}\ \emph {et~al.}(2020)\citenamefont {Linden}, \citenamefont {Montanaro},\ and\ \citenamefont {Shao}}]{linden2020quantumvsclassicalalgorithms}%
  \BibitemOpen
  \bibfield  {author} {\bibinfo {author} {\bibfnamefont {N.}~\bibnamefont {Linden}}, \bibinfo {author} {\bibfnamefont {A.}~\bibnamefont {Montanaro}}, \ and\ \bibinfo {author} {\bibfnamefont {C.}~\bibnamefont {Shao}},\ }\href {https://arxiv.org/abs/2004.06516} {\bibfield  {journal} {\bibinfo  {journal} {arxiv}\ } (\bibinfo {year} {2020})},\ \Eprint {http://arxiv.org/abs/2004.06516} {arXiv:2004.06516 [quant-ph]} \BibitemShut {NoStop}%
\bibitem [{\citenamefont {Mardal}\ and\ \citenamefont {Winther}(2011)}]{mardal2011classicalpreconditioningbook}%
  \BibitemOpen
  \bibfield  {author} {\bibinfo {author} {\bibfnamefont {K.-A.}\ \bibnamefont {Mardal}}\ and\ \bibinfo {author} {\bibfnamefont {R.}~\bibnamefont {Winther}},\ }\href {\doibase https://doi.org/10.1002/nla.716} {\bibfield  {journal} {\bibinfo  {journal} {Numerical Linear Algebra with Applications}\ }\textbf {\bibinfo {volume} {18}},\ \bibinfo {pages} {1} (\bibinfo {year} {2011})},\ \Eprint {http://arxiv.org/abs/https://onlinelibrary.wiley.com/doi/pdf/10.1002/nla.716} {https://onlinelibrary.wiley.com/doi/pdf/10.1002/nla.716} \BibitemShut {NoStop}%
\bibitem [{\citenamefont {Fang}\ \emph {et~al.}(2023)\citenamefont {Fang}, \citenamefont {Lin},\ and\ \citenamefont {Tong}}]{fang2023time}%
  \BibitemOpen
  \bibfield  {author} {\bibinfo {author} {\bibfnamefont {D.}~\bibnamefont {Fang}}, \bibinfo {author} {\bibfnamefont {L.}~\bibnamefont {Lin}}, \ and\ \bibinfo {author} {\bibfnamefont {Y.}~\bibnamefont {Tong}},\ }\href {https://quantum-journal.org/papers/q-2023-03-20-955/} {\bibfield  {journal} {\bibinfo  {journal} {Quantum}\ }\textbf {\bibinfo {volume} {7}},\ \bibinfo {pages} {955} (\bibinfo {year} {2023})}\BibitemShut {NoStop}%
\bibitem [{\citenamefont {Costa}\ \emph {et~al.}(2019)\citenamefont {Costa}, \citenamefont {Jordan},\ and\ \citenamefont {Ostrander}}]{costa2019quantum}%
  \BibitemOpen
  \bibfield  {author} {\bibinfo {author} {\bibfnamefont {P.~C.}\ \bibnamefont {Costa}}, \bibinfo {author} {\bibfnamefont {S.}~\bibnamefont {Jordan}}, \ and\ \bibinfo {author} {\bibfnamefont {A.}~\bibnamefont {Ostrander}},\ }\href {https://journals.aps.org/pra/abstract/10.1103/PhysRevA.99.012323} {\bibfield  {journal} {\bibinfo  {journal} {Physical Review A}\ }\textbf {\bibinfo {volume} {99}},\ \bibinfo {pages} {012323} (\bibinfo {year} {2019})}\BibitemShut {NoStop}%
\bibitem [{\citenamefont {Babbush}\ \emph {et~al.}(2023)\citenamefont {Babbush}, \citenamefont {Berry}, \citenamefont {Kothari}, \citenamefont {Somma},\ and\ \citenamefont {Wiebe}}]{babbush2023exponential}%
  \BibitemOpen
  \bibfield  {author} {\bibinfo {author} {\bibfnamefont {R.}~\bibnamefont {Babbush}}, \bibinfo {author} {\bibfnamefont {D.~W.}\ \bibnamefont {Berry}}, \bibinfo {author} {\bibfnamefont {R.}~\bibnamefont {Kothari}}, \bibinfo {author} {\bibfnamefont {R.~D.}\ \bibnamefont {Somma}}, \ and\ \bibinfo {author} {\bibfnamefont {N.}~\bibnamefont {Wiebe}},\ }\href {https://journals.aps.org/prx/abstract/10.1103/PhysRevX.13.041041} {\bibfield  {journal} {\bibinfo  {journal} {Physical Review X}\ }\textbf {\bibinfo {volume} {13}},\ \bibinfo {pages} {041041} (\bibinfo {year} {2023})}\BibitemShut {NoStop}%
\bibitem [{\citenamefont {Jin}\ \emph {et~al.}(2022)\citenamefont {Jin}, \citenamefont {Liu},\ and\ \citenamefont {Yu}}]{jin2022quantumsimulationpartialdifferential}%
  \BibitemOpen
  \bibfield  {author} {\bibinfo {author} {\bibfnamefont {S.}~\bibnamefont {Jin}}, \bibinfo {author} {\bibfnamefont {N.}~\bibnamefont {Liu}}, \ and\ \bibinfo {author} {\bibfnamefont {Y.}~\bibnamefont {Yu}},\ }\href {https://arxiv.org/abs/2212.13969} {\enquote {\bibinfo {title} {Quantum simulation of partial differential equations via schrödingerisation},}\ } (\bibinfo {year} {2022}),\ \Eprint {http://arxiv.org/abs/2212.13969} {arXiv:2212.13969 [quant-ph]} \BibitemShut {NoStop}%
\bibitem [{\citenamefont {Feynman}(1982)}]{feynman1982simulating}%
  \BibitemOpen
  \bibfield  {author} {\bibinfo {author} {\bibfnamefont {R.~P.}\ \bibnamefont {Feynman}},\ }\href {\doibase 10.1007/BF02650179} {\bibfield  {journal} {\bibinfo  {journal} {International Journal of Theoretical Physics}\ }\textbf {\bibinfo {volume} {21}},\ \bibinfo {pages} {467} (\bibinfo {year} {1982})}\BibitemShut {NoStop}%
\bibitem [{\citenamefont {Feynman}(1986)}]{feynman1986quantum}%
  \BibitemOpen
  \bibfield  {author} {\bibinfo {author} {\bibfnamefont {R.~P.}\ \bibnamefont {Feynman}},\ }\href@noop {} {\bibfield  {journal} {\bibinfo  {journal} {Found. Phys.}\ }\textbf {\bibinfo {volume} {16}},\ \bibinfo {pages} {507} (\bibinfo {year} {1986})}\BibitemShut {NoStop}%
\bibitem [{\citenamefont {Berry}(2014)}]{berry_highorder_2014}%
  \BibitemOpen
  \bibfield  {author} {\bibinfo {author} {\bibfnamefont {D.~W.}\ \bibnamefont {Berry}},\ }\href {\doibase 10.1088/1751-8113/47/10/105301} {\bibfield  {journal} {\bibinfo  {journal} {Journal of Physics A: Mathematical and Theoretical}\ }\textbf {\bibinfo {volume} {47}},\ \bibinfo {pages} {105301} (\bibinfo {year} {2014})}\BibitemShut {NoStop}%
\bibitem [{\citenamefont {Berry}\ \emph {et~al.}(2017)\citenamefont {Berry}, \citenamefont {Childs}, \citenamefont {Ostrander},\ and\ \citenamefont {Wang}}]{berry2017quantum}%
  \BibitemOpen
  \bibfield  {author} {\bibinfo {author} {\bibfnamefont {D.~W.}\ \bibnamefont {Berry}}, \bibinfo {author} {\bibfnamefont {A.~M.}\ \bibnamefont {Childs}}, \bibinfo {author} {\bibfnamefont {A.}~\bibnamefont {Ostrander}}, \ and\ \bibinfo {author} {\bibfnamefont {G.}~\bibnamefont {Wang}},\ }\href {https://link.springer.com/article/10.1007/s00220-017-3002-y} {\bibfield  {journal} {\bibinfo  {journal} {Communications in Mathematical Physics}\ }\textbf {\bibinfo {volume} {356}},\ \bibinfo {pages} {1057} (\bibinfo {year} {2017})}\BibitemShut {NoStop}%
\bibitem [{\citenamefont {Krovi}(2023)}]{krovi2023improved}%
  \BibitemOpen
  \bibfield  {author} {\bibinfo {author} {\bibfnamefont {H.}~\bibnamefont {Krovi}},\ }\href {https://quantum-journal.org/papers/q-2023-02-02-913/} {\bibfield  {journal} {\bibinfo  {journal} {Quantum}\ }\textbf {\bibinfo {volume} {7}},\ \bibinfo {pages} {913} (\bibinfo {year} {2023})}\BibitemShut {NoStop}%
\bibitem [{\citenamefont {Berry}\ and\ \citenamefont {Costa}(2024)}]{berry2022quantum}%
  \BibitemOpen
  \bibfield  {author} {\bibinfo {author} {\bibfnamefont {D.~W.}\ \bibnamefont {Berry}}\ and\ \bibinfo {author} {\bibfnamefont {P.~C.}\ \bibnamefont {Costa}},\ }\href {https://quantum-journal.org/papers/q-2024-06-13-1369/} {\bibfield  {journal} {\bibinfo  {journal} {Quantum}\ }\textbf {\bibinfo {volume} {8}},\ \bibinfo {pages} {1369} (\bibinfo {year} {2024})}\BibitemShut {NoStop}%
\bibitem [{\citenamefont {Jennings}\ \emph {et~al.}(2024)\citenamefont {Jennings}, \citenamefont {Lostaglio}, \citenamefont {Lowrie}, \citenamefont {Pallister},\ and\ \citenamefont {Sornborger}}]{jennings2024costsolvinglineardifferential}%
  \BibitemOpen
  \bibfield  {author} {\bibinfo {author} {\bibfnamefont {D.}~\bibnamefont {Jennings}}, \bibinfo {author} {\bibfnamefont {M.}~\bibnamefont {Lostaglio}}, \bibinfo {author} {\bibfnamefont {R.~B.}\ \bibnamefont {Lowrie}}, \bibinfo {author} {\bibfnamefont {S.}~\bibnamefont {Pallister}}, \ and\ \bibinfo {author} {\bibfnamefont {A.~T.}\ \bibnamefont {Sornborger}},\ }\href {https://arxiv.org/abs/2309.07881} {\enquote {\bibinfo {title} {The cost of solving linear differential equations on a quantum computer: Fast-forwarding to explicit resource counts},}\ } (\bibinfo {year} {2024}),\ \Eprint {http://arxiv.org/abs/2309.07881} {arXiv:2309.07881 [quant-ph]} \BibitemShut {NoStop}%
\bibitem [{\citenamefont {An}\ \emph {et~al.}(2024)\citenamefont {An}, \citenamefont {Onwunta},\ and\ \citenamefont {Yang}}]{an2024fastforwardingquantumalgorithmslinear}%
  \BibitemOpen
  \bibfield  {author} {\bibinfo {author} {\bibfnamefont {D.}~\bibnamefont {An}}, \bibinfo {author} {\bibfnamefont {A.}~\bibnamefont {Onwunta}}, \ and\ \bibinfo {author} {\bibfnamefont {G.}~\bibnamefont {Yang}},\ }\href {https://arxiv.org/abs/2410.13189} {\enquote {\bibinfo {title} {Fast-forwarding quantum algorithms for linear dissipative differential equations},}\ } (\bibinfo {year} {2024}),\ \Eprint {http://arxiv.org/abs/2410.13189} {arXiv:2410.13189 [quant-ph]} \BibitemShut {NoStop}%
\bibitem [{\citenamefont {Leyton}\ and\ \citenamefont {Osborne}(2008)}]{leyton_quantum_2008}%
  \BibitemOpen
  \bibfield  {author} {\bibinfo {author} {\bibfnamefont {S.~K.}\ \bibnamefont {Leyton}}\ and\ \bibinfo {author} {\bibfnamefont {T.~J.}\ \bibnamefont {Osborne}},\ }\href {https://arxiv.org/abs/0812.4423} {\bibfield  {journal} {\bibinfo  {journal} {arXiv preprint 0812.4423}\ } (\bibinfo {year} {2008})}\BibitemShut {NoStop}%
\bibitem [{\citenamefont {Carleman}(1932)}]{carleman1932application}%
  \BibitemOpen
  \bibfield  {author} {\bibinfo {author} {\bibfnamefont {T.}~\bibnamefont {Carleman}},\ }\href {https://doi.org/10.1007/BF02546499} {\bibfield  {journal} {\bibinfo  {journal} {Acta Mathematica}\ }\textbf {\bibinfo {volume} {59}},\ \bibinfo {pages} {63} (\bibinfo {year} {1932})}\BibitemShut {NoStop}%
\bibitem [{\citenamefont {Liu}\ \emph {et~al.}(2021)\citenamefont {Liu}, \citenamefont {Kolden}, \citenamefont {Krovi}, \citenamefont {Loureiro}, \citenamefont {Trivisa},\ and\ \citenamefont {Childs}}]{liu2021efficient}%
  \BibitemOpen
  \bibfield  {author} {\bibinfo {author} {\bibfnamefont {J.-P.}\ \bibnamefont {Liu}}, \bibinfo {author} {\bibfnamefont {H.~{\O}.}\ \bibnamefont {Kolden}}, \bibinfo {author} {\bibfnamefont {H.~K.}\ \bibnamefont {Krovi}}, \bibinfo {author} {\bibfnamefont {N.~F.}\ \bibnamefont {Loureiro}}, \bibinfo {author} {\bibfnamefont {K.}~\bibnamefont {Trivisa}}, \ and\ \bibinfo {author} {\bibfnamefont {A.~M.}\ \bibnamefont {Childs}},\ }\href {https://www.pnas.org/doi/abs/10.1073/pnas.2026805118} {\bibfield  {journal} {\bibinfo  {journal} {Proceedings of the National Academy of Sciences}\ }\textbf {\bibinfo {volume} {118}},\ \bibinfo {pages} {e2026805118} (\bibinfo {year} {2021})}\BibitemShut {NoStop}%
\bibitem [{\citenamefont {Liu}\ \emph {et~al.}(2023)\citenamefont {Liu}, \citenamefont {An}, \citenamefont {Fang}, \citenamefont {Wang}, \citenamefont {Low},\ and\ \citenamefont {Jordan}}]{liu2023efficient}%
  \BibitemOpen
  \bibfield  {author} {\bibinfo {author} {\bibfnamefont {J.-P.}\ \bibnamefont {Liu}}, \bibinfo {author} {\bibfnamefont {D.}~\bibnamefont {An}}, \bibinfo {author} {\bibfnamefont {D.}~\bibnamefont {Fang}}, \bibinfo {author} {\bibfnamefont {J.}~\bibnamefont {Wang}}, \bibinfo {author} {\bibfnamefont {G.~H.}\ \bibnamefont {Low}}, \ and\ \bibinfo {author} {\bibfnamefont {S.}~\bibnamefont {Jordan}},\ }\href {https://link.springer.com/article/10.1007/s00220-023-04857-9} {\bibfield  {journal} {\bibinfo  {journal} {Communications in Mathematical Physics}\ }\textbf {\bibinfo {volume} {404}},\ \bibinfo {pages} {963} (\bibinfo {year} {2023})}\BibitemShut {NoStop}%
\bibitem [{\citenamefont {Costa}\ \emph {et~al.}(2023)\citenamefont {Costa}, \citenamefont {Schleich}, \citenamefont {Morales},\ and\ \citenamefont {Berry}}]{costa2023further}%
  \BibitemOpen
  \bibfield  {author} {\bibinfo {author} {\bibfnamefont {P.}~\bibnamefont {Costa}}, \bibinfo {author} {\bibfnamefont {P.}~\bibnamefont {Schleich}}, \bibinfo {author} {\bibfnamefont {M.~E.}\ \bibnamefont {Morales}}, \ and\ \bibinfo {author} {\bibfnamefont {D.~W.}\ \bibnamefont {Berry}},\ }\href {https://arxiv.org/abs/2312.09518} {\bibfield  {journal} {\bibinfo  {journal} {arXiv preprint arXiv:2312.09518}\ } (\bibinfo {year} {2023})}\BibitemShut {NoStop}%
\bibitem [{\citenamefont {Xue}\ \emph {et~al.}(2021)\citenamefont {Xue}, \citenamefont {Wu},\ and\ \citenamefont {Guo}}]{xue2021quantum}%
  \BibitemOpen
  \bibfield  {author} {\bibinfo {author} {\bibfnamefont {C.}~\bibnamefont {Xue}}, \bibinfo {author} {\bibfnamefont {Y.-C.}\ \bibnamefont {Wu}}, \ and\ \bibinfo {author} {\bibfnamefont {G.-P.}\ \bibnamefont {Guo}},\ }\href {https://iopscience.iop.org/article/10.1088/1367-2630/ac3eff} {\bibfield  {journal} {\bibinfo  {journal} {New Journal of Physics}\ }\textbf {\bibinfo {volume} {23}},\ \bibinfo {pages} {123035} (\bibinfo {year} {2021})}\BibitemShut {NoStop}%
\bibitem [{\citenamefont {Forets}\ and\ \citenamefont {Pouly}(2017)}]{forets2017explicit}%
  \BibitemOpen
  \bibfield  {author} {\bibinfo {author} {\bibfnamefont {M.}~\bibnamefont {Forets}}\ and\ \bibinfo {author} {\bibfnamefont {A.}~\bibnamefont {Pouly}},\ }\href {https://arxiv.org/abs/1711.02552} {\bibfield  {journal} {\bibinfo  {journal} {arXiv preprint arXiv:1711.02552}\ } (\bibinfo {year} {2017})}\BibitemShut {NoStop}%
\bibitem [{\citenamefont {Li}\ \emph {et~al.}(2023)\citenamefont {Li}, \citenamefont {Yin}, \citenamefont {Wiebe}, \citenamefont {Chun}, \citenamefont {Schenter}, \citenamefont {Cheung},\ and\ \citenamefont {M{\"u}lmenst{\"a}dt}}]{li2023potential}%
  \BibitemOpen
  \bibfield  {author} {\bibinfo {author} {\bibfnamefont {X.}~\bibnamefont {Li}}, \bibinfo {author} {\bibfnamefont {X.}~\bibnamefont {Yin}}, \bibinfo {author} {\bibfnamefont {N.}~\bibnamefont {Wiebe}}, \bibinfo {author} {\bibfnamefont {J.}~\bibnamefont {Chun}}, \bibinfo {author} {\bibfnamefont {G.~K.}\ \bibnamefont {Schenter}}, \bibinfo {author} {\bibfnamefont {M.~S.}\ \bibnamefont {Cheung}}, \ and\ \bibinfo {author} {\bibfnamefont {J.}~\bibnamefont {M{\"u}lmenst{\"a}dt}},\ }\href {https://arxiv.org/abs/2303.16550} {\bibfield  {journal} {\bibinfo  {journal} {arXiv preprint arXiv:2303.16550}\ } (\bibinfo {year} {2023})}\BibitemShut {NoStop}%
\bibitem [{\citenamefont {Mitchell}(1997)}]{mitchell_machine_1997}%
  \BibitemOpen
  \bibfield  {author} {\bibinfo {author} {\bibfnamefont {T.~M.}\ \bibnamefont {Mitchell}},\ }\href@noop {} {\emph {\bibinfo {title} {Machine Learning}}},\ \bibinfo {edition} {1st}\ ed.\ (\bibinfo  {publisher} {McGraw-Hill, Inc.},\ \bibinfo {address} {USA},\ \bibinfo {year} {1997})\BibitemShut {NoStop}%
\bibitem [{\citenamefont {Goodfellow}\ \emph {et~al.}(2016)\citenamefont {Goodfellow}, \citenamefont {Bengio},\ and\ \citenamefont {Courville}}]{goodfellow_deep_2016}%
  \BibitemOpen
  \bibfield  {author} {\bibinfo {author} {\bibfnamefont {I.}~\bibnamefont {Goodfellow}}, \bibinfo {author} {\bibfnamefont {Y.}~\bibnamefont {Bengio}}, \ and\ \bibinfo {author} {\bibfnamefont {A.}~\bibnamefont {Courville}},\ }\href@noop {} {\emph {\bibinfo {title} {Deep Learning}}}\ (\bibinfo  {publisher} {MIT Press},\ \bibinfo {year} {2016})\ \bibinfo {note} {\url{http://www.deeplearningbook.org}}\BibitemShut {NoStop}%
\bibitem [{\citenamefont {Schuld}\ and\ \citenamefont {Petruccione}(2021)}]{schuld_machine_2021}%
  \BibitemOpen
  \bibfield  {author} {\bibinfo {author} {\bibfnamefont {M.}~\bibnamefont {Schuld}}\ and\ \bibinfo {author} {\bibfnamefont {F.}~\bibnamefont {Petruccione}},\ }\href {https://books.google.com.au/books?id=-N5IEAAAQBAJ} {\emph {\bibinfo {title} {Machine Learning with Quantum Computers}}},\ Quantum Science and Technology\ (\bibinfo  {publisher} {Springer International Publishing},\ \bibinfo {year} {2021})\BibitemShut {NoStop}%
\bibitem [{\citenamefont {Adcock}\ \emph {et~al.}(2015)\citenamefont {Adcock}, \citenamefont {Allen}, \citenamefont {Day}, \citenamefont {Frick}, \citenamefont {Hinchliff}, \citenamefont {Johnson}, \citenamefont {Morley-Short}, \citenamefont {Pallister}, \citenamefont {Price},\ and\ \citenamefont {Stanisic}}]{adcock_advances_2015}%
  \BibitemOpen
  \bibfield  {author} {\bibinfo {author} {\bibfnamefont {J.}~\bibnamefont {Adcock}}, \bibinfo {author} {\bibfnamefont {E.}~\bibnamefont {Allen}}, \bibinfo {author} {\bibfnamefont {M.}~\bibnamefont {Day}}, \bibinfo {author} {\bibfnamefont {S.}~\bibnamefont {Frick}}, \bibinfo {author} {\bibfnamefont {J.}~\bibnamefont {Hinchliff}}, \bibinfo {author} {\bibfnamefont {M.}~\bibnamefont {Johnson}}, \bibinfo {author} {\bibfnamefont {S.}~\bibnamefont {Morley-Short}}, \bibinfo {author} {\bibfnamefont {S.}~\bibnamefont {Pallister}}, \bibinfo {author} {\bibfnamefont {A.}~\bibnamefont {Price}}, \ and\ \bibinfo {author} {\bibfnamefont {S.}~\bibnamefont {Stanisic}},\ }\href {https://arxiv.org/abs/1512.02900} {\bibfield  {journal} {\bibinfo  {journal} {arXiv:1512.02900}\ } (\bibinfo {year} {2015})}\BibitemShut {NoStop}%
\bibitem [{\citenamefont {Duan}\ \emph {et~al.}(2020)\citenamefont {Duan}, \citenamefont {Yuan}, \citenamefont {Yu}, \citenamefont {Huang},\ and\ \citenamefont {Hsieh}}]{duan_survey_2020}%
  \BibitemOpen
  \bibfield  {author} {\bibinfo {author} {\bibfnamefont {B.}~\bibnamefont {Duan}}, \bibinfo {author} {\bibfnamefont {J.}~\bibnamefont {Yuan}}, \bibinfo {author} {\bibfnamefont {C.-H.}\ \bibnamefont {Yu}}, \bibinfo {author} {\bibfnamefont {J.}~\bibnamefont {Huang}}, \ and\ \bibinfo {author} {\bibfnamefont {C.-Y.}\ \bibnamefont {Hsieh}},\ }\href {\doibase https://doi.org/10.1016/j.physleta.2020.126595} {\bibfield  {journal} {\bibinfo  {journal} {Physics Letters A}\ }\textbf {\bibinfo {volume} {384}},\ \bibinfo {pages} {126595} (\bibinfo {year} {2020})}\BibitemShut {NoStop}%
\bibitem [{\citenamefont {Rebentrost}\ \emph {et~al.}(2014)\citenamefont {Rebentrost}, \citenamefont {Mohseni},\ and\ \citenamefont {Lloyd}}]{rebentrost_quantum_2014}%
  \BibitemOpen
  \bibfield  {author} {\bibinfo {author} {\bibfnamefont {P.}~\bibnamefont {Rebentrost}}, \bibinfo {author} {\bibfnamefont {M.}~\bibnamefont {Mohseni}}, \ and\ \bibinfo {author} {\bibfnamefont {S.}~\bibnamefont {Lloyd}},\ }\href {https://journals.aps.org/prl/abstract/10.1103/PhysRevLett.113.130503} {\bibfield  {journal} {\bibinfo  {journal} {Physical Review Letters}\ }\textbf {\bibinfo {volume} {113}},\ \bibinfo {pages} {130503} (\bibinfo {year} {2014})}\BibitemShut {NoStop}%
\bibitem [{\citenamefont {Wiebe}\ \emph {et~al.}(2012)\citenamefont {Wiebe}, \citenamefont {Braun},\ and\ \citenamefont {Lloyd}}]{wiebe_quantum_2012}%
  \BibitemOpen
  \bibfield  {author} {\bibinfo {author} {\bibfnamefont {N.}~\bibnamefont {Wiebe}}, \bibinfo {author} {\bibfnamefont {D.}~\bibnamefont {Braun}}, \ and\ \bibinfo {author} {\bibfnamefont {S.}~\bibnamefont {Lloyd}},\ }\href {https://journals.aps.org/prl/abstract/10.1103/PhysRevLett.109.050505} {\bibfield  {journal} {\bibinfo  {journal} {Physical Review Letters}\ }\textbf {\bibinfo {volume} {109}},\ \bibinfo {pages} {050505} (\bibinfo {year} {2012})}\BibitemShut {NoStop}%
\bibitem [{\citenamefont {Liu}\ and\ \citenamefont {Zhang}(2017)}]{liu_fast_2017}%
  \BibitemOpen
  \bibfield  {author} {\bibinfo {author} {\bibfnamefont {Y.}~\bibnamefont {Liu}}\ and\ \bibinfo {author} {\bibfnamefont {S.}~\bibnamefont {Zhang}},\ }\href {https://www.sciencedirect.com/science/article/pii/S0304397516302304} {\bibfield  {journal} {\bibinfo  {journal} {Theoretical Computer Science}\ }\textbf {\bibinfo {volume} {657}},\ \bibinfo {pages} {38} (\bibinfo {year} {2017})}\BibitemShut {NoStop}%
\bibitem [{\citenamefont {Yu}\ \emph {et~al.}(2019)\citenamefont {Yu}, \citenamefont {Gao},\ and\ \citenamefont {Wen}}]{yu_improved_2019}%
  \BibitemOpen
  \bibfield  {author} {\bibinfo {author} {\bibfnamefont {C.-H.}\ \bibnamefont {Yu}}, \bibinfo {author} {\bibfnamefont {F.}~\bibnamefont {Gao}}, \ and\ \bibinfo {author} {\bibfnamefont {Q.-Y.}\ \bibnamefont {Wen}},\ }\href {https://ieeexplore.ieee.org/document/8812924} {\bibfield  {journal} {\bibinfo  {journal} {IEEE Transactions on Knowledge and Data Engineering}\ }\textbf {\bibinfo {volume} {33}},\ \bibinfo {pages} {858} (\bibinfo {year} {2019})}\BibitemShut {NoStop}%
\bibitem [{\citenamefont {Schuld}\ \emph {et~al.}(2016)\citenamefont {Schuld}, \citenamefont {Sinayskiy},\ and\ \citenamefont {Petruccione}}]{schuld_prediction_2016}%
  \BibitemOpen
  \bibfield  {author} {\bibinfo {author} {\bibfnamefont {M.}~\bibnamefont {Schuld}}, \bibinfo {author} {\bibfnamefont {I.}~\bibnamefont {Sinayskiy}}, \ and\ \bibinfo {author} {\bibfnamefont {F.}~\bibnamefont {Petruccione}},\ }\href {https://doi.org/10.1103/PhysRevA.94.022342} {\bibfield  {journal} {\bibinfo  {journal} {Physical Review A}\ }\textbf {\bibinfo {volume} {94}},\ \bibinfo {pages} {022342} (\bibinfo {year} {2016})}\BibitemShut {NoStop}%
\bibitem [{\citenamefont {Date}\ and\ \citenamefont {Potok}(2021)}]{date_adiabatic_2021}%
  \BibitemOpen
  \bibfield  {author} {\bibinfo {author} {\bibfnamefont {P.}~\bibnamefont {Date}}\ and\ \bibinfo {author} {\bibfnamefont {T.}~\bibnamefont {Potok}},\ }\href {\doibase 10.1038/s41598-021-01445-6} {\bibfield  {journal} {\bibinfo  {journal} {Scientific Reports}\ }\textbf {\bibinfo {volume} {11}} (\bibinfo {year} {2021}),\ 10.1038/s41598-021-01445-6}\BibitemShut {NoStop}%
\bibitem [{\citenamefont {A{\"\i}meur}\ \emph {et~al.}(2013)\citenamefont {A{\"\i}meur}, \citenamefont {Brassard},\ and\ \citenamefont {Gambs}}]{aimeur_2013_quantum}%
  \BibitemOpen
  \bibfield  {author} {\bibinfo {author} {\bibfnamefont {E.}~\bibnamefont {A{\"\i}meur}}, \bibinfo {author} {\bibfnamefont {G.}~\bibnamefont {Brassard}}, \ and\ \bibinfo {author} {\bibfnamefont {S.}~\bibnamefont {Gambs}},\ }\href {https://link.springer.com/article/10.1007/s10994-012-5316-5} {\bibfield  {journal} {\bibinfo  {journal} {Machine Learning}\ }\textbf {\bibinfo {volume} {90}},\ \bibinfo {pages} {261} (\bibinfo {year} {2013})}\BibitemShut {NoStop}%
\bibitem [{\citenamefont {Wiebe}\ \emph {et~al.}(2015)\citenamefont {Wiebe}, \citenamefont {Kapoor},\ and\ \citenamefont {Svore}}]{wiebe_quantum_2014}%
  \BibitemOpen
  \bibfield  {author} {\bibinfo {author} {\bibfnamefont {N.}~\bibnamefont {Wiebe}}, \bibinfo {author} {\bibfnamefont {A.}~\bibnamefont {Kapoor}}, \ and\ \bibinfo {author} {\bibfnamefont {K.~M.}\ \bibnamefont {Svore}},\ }\href {\doibase https://doi.org/10.26421/QIC16.7-8-1} {\bibfield  {journal} {\bibinfo  {journal} {Quantum Info. Comput.}\ }\textbf {\bibinfo {volume} {15}},\ \bibinfo {pages} {318–358} (\bibinfo {year} {2015})}\BibitemShut {NoStop}%
\bibitem [{\citenamefont {Lloyd}\ \emph {et~al.}(2014)\citenamefont {Lloyd}, \citenamefont {Mohseni},\ and\ \citenamefont {Rebentrost}}]{lloyd_quantum_2014}%
  \BibitemOpen
  \bibfield  {author} {\bibinfo {author} {\bibfnamefont {S.}~\bibnamefont {Lloyd}}, \bibinfo {author} {\bibfnamefont {M.}~\bibnamefont {Mohseni}}, \ and\ \bibinfo {author} {\bibfnamefont {P.}~\bibnamefont {Rebentrost}},\ }\href {\doibase 10.1038/nphys3029} {\bibfield  {journal} {\bibinfo  {journal} {Nature Physics}\ }\textbf {\bibinfo {volume} {10}},\ \bibinfo {pages} {631–633} (\bibinfo {year} {2014})}\BibitemShut {NoStop}%
\bibitem [{\citenamefont {Tang}(2021)}]{tang_quantum_2021}%
  \BibitemOpen
  \bibfield  {author} {\bibinfo {author} {\bibfnamefont {E.}~\bibnamefont {Tang}},\ }\href {\doibase 10.1103/PhysRevLett.127.060503} {\bibfield  {journal} {\bibinfo  {journal} {Phys. Rev. Lett.}\ }\textbf {\bibinfo {volume} {127}},\ \bibinfo {pages} {060503} (\bibinfo {year} {2021})}\BibitemShut {NoStop}%
\bibitem [{\citenamefont {Kerenidis}\ and\ \citenamefont {Prakash}(2016)}]{kerenidis_quantum_2016}%
  \BibitemOpen
  \bibfield  {author} {\bibinfo {author} {\bibfnamefont {I.}~\bibnamefont {Kerenidis}}\ and\ \bibinfo {author} {\bibfnamefont {A.}~\bibnamefont {Prakash}},\ }\href {https://arxiv.org/abs/1603.08675} {\bibfield  {journal} {\bibinfo  {journal} {arXiv preprint arXiv:1603.08675}\ } (\bibinfo {year} {2016})}\BibitemShut {NoStop}%
\bibitem [{\citenamefont {Rebentrost}\ \emph {et~al.}(2018)\citenamefont {Rebentrost}, \citenamefont {Bromley}, \citenamefont {Weedbrook},\ and\ \citenamefont {Lloyd}}]{rebentrost_quantumhop_2018}%
  \BibitemOpen
  \bibfield  {author} {\bibinfo {author} {\bibfnamefont {P.}~\bibnamefont {Rebentrost}}, \bibinfo {author} {\bibfnamefont {T.~R.}\ \bibnamefont {Bromley}}, \bibinfo {author} {\bibfnamefont {C.}~\bibnamefont {Weedbrook}}, \ and\ \bibinfo {author} {\bibfnamefont {S.}~\bibnamefont {Lloyd}},\ }\href {https://journals.aps.org/pra/abstract/10.1103/PhysRevA.98.042308} {\bibfield  {journal} {\bibinfo  {journal} {Physical Review A}\ }\textbf {\bibinfo {volume} {98}},\ \bibinfo {pages} {042308} (\bibinfo {year} {2018})}\BibitemShut {NoStop}%
\bibitem [{\citenamefont {Wossnig}\ \emph {et~al.}(2018)\citenamefont {Wossnig}, \citenamefont {Zhao},\ and\ \citenamefont {Prakash}}]{wossnig_quantum_2018}%
  \BibitemOpen
  \bibfield  {author} {\bibinfo {author} {\bibfnamefont {L.}~\bibnamefont {Wossnig}}, \bibinfo {author} {\bibfnamefont {Z.}~\bibnamefont {Zhao}}, \ and\ \bibinfo {author} {\bibfnamefont {A.}~\bibnamefont {Prakash}},\ }\href {\doibase 10.1103/physrevlett.120.050502} {\bibfield  {journal} {\bibinfo  {journal} {Physical Review Letters}\ }\textbf {\bibinfo {volume} {120}},\ \bibinfo {pages} {050502} (\bibinfo {year} {2018})}\BibitemShut {NoStop}%
\bibitem [{\citenamefont {Farhi}\ \emph {et~al.}(2014)\citenamefont {Farhi}, \citenamefont {Goldstone},\ and\ \citenamefont {Gutmann}}]{farhi_quantum_2014}%
  \BibitemOpen
  \bibfield  {author} {\bibinfo {author} {\bibfnamefont {E.}~\bibnamefont {Farhi}}, \bibinfo {author} {\bibfnamefont {J.}~\bibnamefont {Goldstone}}, \ and\ \bibinfo {author} {\bibfnamefont {S.}~\bibnamefont {Gutmann}},\ }\href@noop {} {\enquote {\bibinfo {title} {A quantum approximate optimization algorithm},}\ } (\bibinfo {year} {2014}),\ \Eprint {http://arxiv.org/abs/1411.4028} {arXiv:1411.4028 [quant-ph]} \BibitemShut {NoStop}%
\bibitem [{\citenamefont {Tang}(2019)}]{tang_quantum_2019}%
  \BibitemOpen
  \bibfield  {author} {\bibinfo {author} {\bibfnamefont {E.}~\bibnamefont {Tang}},\ }in\ \href {\doibase 10.1145/3313276.3316310} {\emph {\bibinfo {booktitle} {Proceedings of the 51st Annual ACM SIGACT Symposium on Theory of Computing}}},\ \bibinfo {series and number} {STOC 2019}\ (\bibinfo  {publisher} {Association for Computing Machinery},\ \bibinfo {address} {New York, NY, USA},\ \bibinfo {year} {2019})\ p.\ \bibinfo {pages} {217–228}\BibitemShut {NoStop}%
\bibitem [{\citenamefont {Tang}(2022)}]{tang_dequantizing_2022}%
  \BibitemOpen
  \bibfield  {author} {\bibinfo {author} {\bibfnamefont {E.}~\bibnamefont {Tang}},\ }\href {\doibase 10.1038/s42254-022-00511-w} {\bibfield  {journal} {\bibinfo  {journal} {Nature Reviews Physics}\ }\textbf {\bibinfo {volume} {4}},\ \bibinfo {pages} {692} (\bibinfo {year} {2022})}\BibitemShut {NoStop}%
\bibitem [{\citenamefont {Baskaran}\ \emph {et~al.}(2023)\citenamefont {Baskaran}, \citenamefont {Rawat}, \citenamefont {Jayashankar}, \citenamefont {Chakravarti}, \citenamefont {Sugisaki}, \citenamefont {Roy}, \citenamefont {Mandal}, \citenamefont {Mukherjee},\ and\ \citenamefont {Prasannaa}}]{baskaran2023adapting}%
  \BibitemOpen
  \bibfield  {author} {\bibinfo {author} {\bibfnamefont {N.}~\bibnamefont {Baskaran}}, \bibinfo {author} {\bibfnamefont {A.~S.}\ \bibnamefont {Rawat}}, \bibinfo {author} {\bibfnamefont {A.}~\bibnamefont {Jayashankar}}, \bibinfo {author} {\bibfnamefont {D.}~\bibnamefont {Chakravarti}}, \bibinfo {author} {\bibfnamefont {K.}~\bibnamefont {Sugisaki}}, \bibinfo {author} {\bibfnamefont {S.}~\bibnamefont {Roy}}, \bibinfo {author} {\bibfnamefont {S.~B.}\ \bibnamefont {Mandal}}, \bibinfo {author} {\bibfnamefont {D.}~\bibnamefont {Mukherjee}}, \ and\ \bibinfo {author} {\bibfnamefont {V.}~\bibnamefont {Prasannaa}},\ }\href {https://journals.aps.org/prresearch/abstract/10.1103/PhysRevResearch.5.043113} {\bibfield  {journal} {\bibinfo  {journal} {Physical Review Research}\ }\textbf {\bibinfo {volume} {5}},\ \bibinfo {pages} {043113} (\bibinfo {year} {2023})}\BibitemShut {NoStop}%
\bibitem [{\citenamefont {Tsemo}\ \emph {et~al.}(2024)\citenamefont {Tsemo}, \citenamefont {Jayashankar}, \citenamefont {Sugisaki}, \citenamefont {Baskaran}, \citenamefont {Chakraborty},\ and\ \citenamefont {Prasannaa}}]{tsemo2024enhancing}%
  \BibitemOpen
  \bibfield  {author} {\bibinfo {author} {\bibfnamefont {P.~B.}\ \bibnamefont {Tsemo}}, \bibinfo {author} {\bibfnamefont {A.}~\bibnamefont {Jayashankar}}, \bibinfo {author} {\bibfnamefont {K.}~\bibnamefont {Sugisaki}}, \bibinfo {author} {\bibfnamefont {N.}~\bibnamefont {Baskaran}}, \bibinfo {author} {\bibfnamefont {S.}~\bibnamefont {Chakraborty}}, \ and\ \bibinfo {author} {\bibfnamefont {V.}~\bibnamefont {Prasannaa}},\ }\href {https://arxiv.org/abs/2407.21641} {\bibfield  {journal} {\bibinfo  {journal} {arXiv preprint arXiv:2407.21641}\ } (\bibinfo {year} {2024})}\BibitemShut {NoStop}%
\bibitem [{\citenamefont {Arora}\ and\ \citenamefont {Barak}(2009)}]{Arora2009}%
  \BibitemOpen
  \bibfield  {author} {\bibinfo {author} {\bibfnamefont {S.}~\bibnamefont {Arora}}\ and\ \bibinfo {author} {\bibfnamefont {B.}~\bibnamefont {Barak}},\ }\href {https://www.cambridge.org/core/books/computational-complexity/3453CAFDEB0B4820B186FE69A64E1086} {\emph {\bibinfo {title} {Computational Complexity}}}\ (\bibinfo  {publisher} {Cambridge University Press},\ \bibinfo {address} {Cambridge, England},\ \bibinfo {year} {2009})\BibitemShut {NoStop}%
\end{thebibliography}%

\end{document}